\documentclass{aa}

\usepackage{verbatim}
\usepackage{graphicx}
\usepackage{epstopdf}
\usepackage{subfigure}
\usepackage{amsmath}
\usepackage{hyperref}
\usepackage{tabularx}
\usepackage{longtable}

\title{The Upper Atmospheres of Terrestrial Planets: Carbon Dioxide Cooling and the Earth's Thermospheric Evolution}

\titlerunning{Thermal and Chemical Structures of Planetary Atmospheres}

\author{C. P. Johnstone\inst{\ref{vienna}} \and M. G\"{u}del\inst{\ref{vienna}} \and H. Lammer\inst{\ref{graz}} \and K. G. Kislyakova\inst{\ref{graz}}}
\institute{
University of Vienna, Department of Astrophysics, T\"{u}rkenschanzstrasse 17, 1180 Vienna, Austria \label{vienna}
\and
Space Research Institute, Austrian Academy of Sciences, Graz, Austria \label{graz}
}

\abstract{
The thermal and chemical structures of the upper atmospheres of planets crucially influence losses to space and must be understood to constrain the effects of losses on atmospheric evolution.
}{
We develop a 1D first-principles hydrodynamic atmosphere model that calculates atmospheric thermal and chemical structures for arbitrary planetary parameters, chemical compositions, and stellar inputs.
We apply the model to study the reaction of the Earth's upper atmosphere to large changes in the CO$_2$ abundance and to changes in the input solar XUV field due to the Sun's activity evolution from 3~Gyr in the past to 2.5~Gyr in the future.
}{
For the thermal atmosphere structure, we consider heating from the absorption of stellar X-ray, UV, and IR radiation, heating from exothermic chemical reactions, electron heating from collisions with non-thermal photoelectrons, Joule heating, cooling from IR emission by several species, thermal conduction, and energy exchanges between the neutral, ion, and electron gases.
For the chemical structure, we consider $\sim$500 chemical reactions, including 56 photoreactions, eddy and molecular diffusion, and advection.
In addition, we calculate the atmospheric structure by solving the hydrodynamic equations. 
To solve the equations in our model, we develop the Kompot code and provide detailed descriptions of the numerical methods used in the appendices.
}{
We verify our model by calculating the structures of the upper atmospheres of the modern Earth and Venus.
By varying the CO$_2$ abundances at the lower boundary (65~km) of our Earth model, we show that the atmospheric thermal structure is significantly altered. 
Increasing the CO$_2$ abundances leads to massive reduction in thermospheric temperature, contraction of the atmosphere, and reductions in the ion densities indicating that CO$_2$ can significantly influence atmospheric erosion.
Our models for the evolution of the Earth's upper atmosphere indicate that the thermospheric structure has not changed significantly in the last 2~Gyr and is unlikely to change signficantly in the next few Gyr.
The largest changes that we see take place between 3~Gyr and 2~Gyr ago, with even larger changes expected at even earlier times.
}{}

\begin{document} 

\maketitle


\section{Introduction} \label{sect:intro}

Planetary atmospheres evolve due to interactions with the planet's surface and losses into space. 
At the surface, gas can be removed from the atmosphere by several processes, such as subduction (\citealt{Marty03}), and added to the atmosphere by other processes, such as outgassing during magma ocean solidification (\citealt{Noack14}).
At the top of the atmosphere, gases are lost to space, which over time can lead to significant atmospheric erosion (\citealt{Lammer14}; \citealt{Luger15}).
Atmospheric loss into space takes place by a large number of different mechanisms (e.g. \citealt{Lammer08}).
One factor that is common to almost all of these processes is the fact that the loss rates depend strongly on the thermal and chemical structure of the upper atmosphere. 
Atmospheres that are hotter and more expanded have higher loss rates by essentially all mechanisms (e.g. \citealt{Lichtenegger10}).

Much recent work has studied hydrodynamic losses of atmospheres.
Many of these studies concentrate mostly on atmospheres composed primarily of H and He (e.g. \citealt{Lammer14}; \citealt{Shaikhislamov14}; \citealt{Luger15}; \citealt{Khodachenko15}; \citealt{OwenMohanty16}).
Such atmospheres can experience very high hydrodynamic losses, largely due to the small average molecular masses of the gas (\citealt{Erkaev13}).
Atmospheres composed of water vapor, such as the possible early atmosphere of Venus, likely also undergo hydrodynamic escape as the dissociation of H$_2$O creates large amounts of atomic H (\citealt{Lichtenegger16}).
Generally more interesting for planetary habitability are atmospheres dominated by heavier molecules, such as CO$_2$, N$_2$, and O$_2$.
The physical processes in these atmospheres are very complex (e.g. \citealt{Kulikov07}; \citealt{Tian08a}), and detailed models are needed to understand their structures.
Such atmospheres are less likely to undergo hydrodynamic losses due to their higher molecular masses, and other atmospheric loss processes must be taken into account, such as polar ion outflows (\citealt{Glocer12}; \citealt{Airapetian17}) and pick-up of exospheric gas by the stellar wind (\citealt{Kislyakova14}) and coronal mass ejections (\citealt{Khodachenko07}; \citealt{Lammer07}).
In all cases, the specific atmospheric composition is critically important for the detailed physics of the upper atmosphere (\citealt{Kulikov07}). 

The most important input into the upper atmospheres of planets is the irradiation by the central star, especially in X-ray and ultraviolet (together `XUV')\footnotemark~wavelengths, though IR photons can also be important.
The absorption causes dissociation and ionization, and significant heating.
The energy gained by this heating is mostly lost by cooling due to IR emission from several molecules, most notably CO$_2$.
In the upper thermosphere of the Earth, the local heating is much stronger than the local cooling, and the excess energy is transported into the lower thermosphere by thermal conduction.
The chemical structure of the upper atmosphere is determined by the composition of the lower atmosphere, chemical/photochemical reactions, and diffusion. 
Sophisticated models that take into account all of these processes have been applied for solar system planets for decades (e.g. \citealt{FoxBougher91}; \citealt{Roble95}; \citealt{ridley2006global}), but only a few studies have applied such models to planetary atmospheres under very different conditions to those of the current solar system terrestrial planets (\citealt{Tian08a}; \citealt{Tian09}).

\footnotetext{
Several meanings of the abbreviation `XUV' are used in the literature. 
In this paper, we use the term to refer to the X-ray and UV spectrum from 10 to 4000~\AA. 
}

The need for sophisticated first principles upper atmosphere models is clear when considering the range of atmospheric conditions that exist.
In addition to different atmospheric compositions, the distribution of planets spans the entire range of possible masses and orbital distances from their host stars (\citealt{LopezMorales16}).
Furthermore, different planets are exposed to very different conditions from the central star.
Observations of young solar analogues have shown that the Sun was much more active in X-rays and UV than it currently is (\citealt{Guedel97}; \citealt{Ribas05}).
Recently, \citet{Tu15} showed that the early evolution of the Sun's activity depended sensitively on its early rotation rate; this is important since we do not know how rapidly the Sun was rotating, and different evolutionary tracks for XUV can lead to different atmospheric evolution scenarios (\citealt{Johnstone15}).
Stellar activity evolution depends also on the star's mass, with lower mass stars remaining highly active for longer amounts of time (\citealt{West08}). 
In addition, the exact shape of a star's XUV spectrum depends on its spectral type and activity (\citealt{Telleschi05}; \citealt{JohnstoneGuedel15}; \citealt{Fontenla16}).

The aim of this paper is to develop and validate a first principles physical model for the upper atmospheres of planets and to apply it to the Earth to understand how the atmosphere reacts to changes in the CO$_2$ abundances and the solar XUV spectrum.
This physical model will be used as an important component in future studies on the evolution of terrestrial atmospheres.
In Section~\ref{sect:model}, we present the complete physical model. 
In Section~\ref{sect:validation}, we validate the model by calculating the atmospheric structures of Earth and Venus.
In Section~\ref{sect:results}, we study the effects of enhanced CO$_2$ abundances and the effects of the solar XUV evolution between 3~Gyr in the past and 2.5~Gyr in the future on the structure of the Earth's upper atmosphere.
In Section~\ref{sect:discussion}, we summarise and discuss our results.
To solve the physical model presented in this paper, we have developed The Kompot Code, which we describe in the appendices.\footnotemark
In the appendices, we describe in detail the numerical methods used to solve the equations described in Section~\ref{sect:model}.

\footnotetext{
We will make The Kompot Code publicly available in the near future, and it will be obtainable by contacting the authors directly. 
}

\section{Model} \label{sect:model}

\subsection{Model Overview} \label{sect:modeloverview}

The purpose of our model is to calculate the atmospheric properties as a function of altitude for arbitrary planetary atmospheres.
The input parameters are the planetary mass and radius, the atmospheric properties at the base of the simulation, and the stellar radiation spectrum at the top of the atmosphere.
Our computational domain is 1D and points radially outwards from the planet's centre, extending between the lower boundary at an arbitrary altitude in the middle atmosphere to the upper boundary at the exobase. 
In the description of the state of the atmosphere, we make a few basic assumptions. 
Firstly, we assume that the gas has one bulk advection speed shared by the entire gas, though different chemical species have different diffusion speeds.
Secondly, we assume that the neutrals, ions, and electrons have their own temperatures that evolve separately.
Thirdly, we assume quasineutrality, meaning that the electron density is equal to the total ion density everywhere.
The two stellar inputs are the XUV (i.e. X-ray and ultraviolet) field between 10 and 4000~\AA, and the infrared field between 1 and 20~$\mu$m.

In this model, we break the gas down into components in two separate ways: in Eqn.~\ref{eqn:main_speciescontinuity} the gas is broken down by different chemical species (e.g. N$_2$, O$_2$, CO$_2$, etc.), and in Eqns.~\ref{eqn:main_neutralenergy}--\ref{eqn:main_electronenergy}, the gas is broken down into neutrals, ions, and electrons.
In the rest of the paper, we define the `components' of the gas as the neutral, ion, and electron gases, and are referred to using the subscripts n, i, and e.
Unless otherwise stated, when we discuss the electrons, we are referring to the thermal electron gas, and not the non-thermal electrons produced in photoionization reactions. 

The main physical processes taken into account in this model are
\begin{itemize}
\item
atmospheric expansion/contraction in response to changes in the gas temperature and composition,
\item
the transfer of X-ray, ultraviolet, and infrared radiation through the atmosphere, including the production of non-thermal electrons by photoionization reactions,
\item
atmospheric chemistry, including photochemistry and reactions driven by impacts with non-thermal electrons,
\item
molecular and eddy diffusion,
\item
neutral heating by stellar XUV and IR radiation,
\item
electron heating by impacts with non-thermal electrons,
\item
infrared cooling, particularly by CO$_2$ molecules,
\item
heat conduction for each gas component,
\item
and energy exchange between the components.
\end{itemize}
The equations that describe the changes of the atmosphere due to these processes are
\begin{equation} \label{eqn:main_speciescontinuity}
\frac{\partial n_j}{ \partial t} 
+ \frac{1}{r^2} \frac{\partial \left[ r^2 ( n_j v + \Phi_{\mathrm{d},j}) \right] }{\partial r} 
=
S_j ,
\end{equation}
\begin{equation} \label{eqn:main_momentum}
\frac{\partial ( \rho v ) }{ \partial t} 
+ \frac{1}{r^2} \frac{\partial \left[ r^2 \left( \rho v^2 + p \right) \right] }{\partial r} 
=
- \rho g 
+ \frac{2 p }{r} ,
\end{equation}
\begin{equation} \label{eqn:main_neutralenergy}
\begin{aligned}
\frac{\partial e_\mathrm{n} }{ \partial t} 
+ & \frac{1}{r^2} \frac{\partial \left[ r^2 v \left( e_\mathrm{n} + p_\mathrm{n} \right) \right] }{\partial r} 
= - \rho_\mathrm{n} v g  \\
& 
+ \left( Q_{\mathrm{h,n}} - Q_{\mathrm{c,n}}
- Q_\mathrm{in} - Q_\mathrm{en} \right) \\
 & + \frac{1}{r^2} \frac{\partial}{\partial r} \left[ r^2 \kappa_\mathrm{mol} \frac{\partial T_\mathrm{n}}{\partial r} + r^2 \kappa_\mathrm{eddy} \left( \frac{\partial T_\mathrm{n}}{\partial r} + \frac{g}{c_\mathrm{P}} \right) \right] ,
\end{aligned} 
\end{equation}
\begin{equation} \label{eqn:main_ionenergy}
\begin{aligned}
\frac{\partial e_\mathrm{i} }{ \partial t} &
+  \frac{1}{r^2} \frac{\partial \left[ r^2 v \left( e_\mathrm{i} + p_\mathrm{i} \right) \right] }{\partial r} 
= - \rho_\mathrm{i} v g \\
&  
+ \left( Q_{\mathrm{h,i}} - Q_{\mathrm{c,i}}
- Q_\mathrm{ei} + Q_\mathrm{in} \right)
+ \frac{1}{r^2} \frac{\partial}{\partial r} \left[ r^2 \kappa_\mathrm{i} \frac{\partial T_\mathrm{i}}{\partial r}  \right],
\end{aligned}
\end{equation}
\begin{equation} \label{eqn:main_electronenergy}
\begin{aligned}
\frac{\partial e_\mathrm{e} }{ \partial t} &
+  \frac{1}{r^2} \frac{\partial \left[ r^2 v \left( e_\mathrm{e} + p_\mathrm{e} \right) \right] }{\partial r} 
= - \rho_\mathrm{e} v g \\
&  
+ \left( Q_{\mathrm{h,e}} - Q_{\mathrm{c,e}}
+ Q_\mathrm{ei} + Q_\mathrm{en} \right)
+ \frac{1}{r^2} \frac{\partial}{\partial r} \left[ r^2 \kappa_\mathrm{e} \frac{\partial T_\mathrm{e}}{\partial r}  \right],
\end{aligned}
\end{equation}
where
$r$ is the radius,
$n_j$ is the number density of the $j$th species,
$\rho$ is the total mass density,
$v$ is the bulk advection speed,
\mbox{$\rho v$} is the momentum density,
$\rho_k$, $e_k$, $p_k$ and $T_k$ are the mass density, energy density, thermal pressure, and temperature of the $k$th component of the gas,
$\Phi_{\mathrm{d},j}$ and $S_j$ are the diffusive particle flux and chemical source term of the $j$th species,
$g$ is the gravitational acceleration,
$Q_{\mathrm{h},k}$ and $Q_{\mathrm{c},k}$ are the heating and cooling functions for the $k$th component,
$Q_\mathrm{ei}$, $Q_\mathrm{in}$, and $Q_\mathrm{en}$ are the electron-ion, ion-neutral, and electron-neutral heat exchange functions,
$\kappa_\mathrm{mol}$ and $\kappa_\mathrm{eddy}$ are the molecular and eddy thermal conductivities,
$\kappa_\mathrm{i}$ and $\kappa_\mathrm{e}$ are the ion and electron thermal conductivities, 
and $c_\mathrm{P}$ is the specific heat at constant pressure.
Since chemistry and diffusion do not change the total mass density of the gas, Eqn.~\ref{eqn:main_speciescontinuity} implies the standard mass continuity equation.
It is also important at times to calculate $\gamma$, i.e. the ratio of specific heats, for the neutral and ion gases, which are mixtures of species with different $\gamma$ values; for this we assume \mbox{$\gamma_j = 5/3$} for atomic species and \mbox{$\gamma_j = 7/5$} for molecular species\footnotemark.

\footnotetext{
To get $\gamma$ for a gas mixture, we first calculate $C_{\mathrm{V},j}$ and $C_{\mathrm{P},j}$ for each species using Mayer's relation and their individual $\gamma_j$ values, where the subscript $j$ refers to an individual species.
Then, we calculate the total heat capacities as the density weighted average heat capacities of the constituent gases.
Finally, $\gamma$ is calculated simply as the ratio~$C_\mathrm{P}/C_\mathrm{V}$.
}

The exobaseis assumed to be where the mean-free-path of particles becomes larger than the pressure scale height.
The mean free path is calculated from \mbox{$l_\mathrm{mfp} = 1 / (\sigma N)$}, where $\sigma$ is the total collision cross-section, and $N$ is the total number density.
In reality, different species have different $\sigma$; however, the values tend to be similar and our calculated exobase location is not sensitive to small changes in $\sigma$.
We therefore assume \mbox{$\sigma = 2\times10^{-15}$~cm$^{-2}$} always.

\subsection{Hydrodynamics} \label{sect:hydro}

Including gravity, the purely hydrodynamic parts of Eqns.~\ref{eqn:main_speciescontinuity}--\ref{eqn:main_electronenergy} are
\begin{equation} \label{eqn:hydro_mass}
\frac{\partial \rho}{ \partial t} 
+ \frac{1}{r^2} \frac{\partial \left( r^2 \rho v \right) }{\partial r} 
 = 0 ,
\end{equation}
\begin{equation} \label{eqn:hydro_momentum}
\frac{\partial ( \rho v ) }{ \partial t} 
+ \frac{1}{r^2} \frac{\partial \left[ r^2 \left( \rho v^2 + p \right) \right] }{\partial r} 
=
- \rho g 
+ \frac{2 p }{r} ,
\end{equation}
\begin{equation} \label{eqn:hydro_neutralenergy}
\frac{\partial e_\mathrm{n} }{ \partial t} 
+  \frac{1}{r^2} \frac{\partial \left[ r^2 u \left( e_\mathrm{n} + p_\mathrm{n} \right) \right] }{\partial r} 
= - \rho_\mathrm{n} u g ,
\end{equation}
where the ion and electron energy equations are identical to the neutral energy equations with the n subscript replaced with the i and e subscripts.
In Appendix~\ref{appendix:hydro}, we give an explicit method for solving these equations.
Explicitly solving the full set of hydrodynamic equations is undesirable when the atmosphere is static, or close to static.
Note that no atmosphere is ever fully hydrostatic since there is always some escape at the top of the atmosphere, meaning that there will always be a net upward flow of material.

In this paper, we use a method for solving the hydrodynamic equations given by \citet{Tian08a}. 
This method is not appropriate when the atmosphere is transonic, which is often the case for strongly irradiated planets.
The basic simplifying assumption of the method is that the mass and momentum density structures are in a steady state, such that \mbox{$ \partial \rho / \partial t = 0 $} and \mbox{$ \partial ( \rho v ) / \partial t = 0 $}.
Eqns.~\ref{eqn:hydro_mass} and \ref{eqn:hydro_momentum} can then be written
\begin{equation} \label{eqn:hydro_mass_steady}
2 r \rho v + r^2 v \frac{d\rho}{dr} + r^2 \rho \frac{dv}{dr} = 0 ,
\end{equation}
\begin{equation} \label{eqn:hydro_momentum_steady}
\frac{2 \rho v^2}{r} + 2 v \rho \frac{dv}{dr} + v^2 \frac{d\rho}{dr} + \frac{dp}{dr} = - \rho g .
\end{equation}
Assuming an ideal gas, the pressure is given by \mbox{$p = \bar{m}^{-1} \rho k_\mathrm{B} T$} and the radial derivative of $p$ is
\begin{equation} \label{eqn:}
\frac{d p}{dr} = - \frac{\rho}{\bar{m}} v_0^2 \frac{d\bar{m}}{dr} + v_0^2 \frac{d\rho}{dr} + \rho \frac{v_0^2}{T} \frac{dT}{dr} ,
\end{equation}
where \mbox{$v_0^2 = k_\mathrm{B} T / \bar{m}$} and $\bar{m}$ and $T$ are the average molecular mass and temperature of the entire gas. 
Putting these three equations together gives
\begin{equation} \label{eqn:semistatic1}
\frac{1}{v} \left( 1 - \frac{v^2}{v_0^2} \right) \frac{dv}{dr} = \frac{1}{T} \frac{dT}{dr} + \frac{g}{v_0^2} - \frac{1}{\bar{m}} \frac{d\bar{m}}{dr} - \frac{2}{r} ,
\end{equation}
\begin{equation} \label{eqn:semistatic2}
\frac{1}{\rho} \frac{d \rho}{dr} = - \frac{1}{T} \frac{dT}{dr} - \frac{g}{v_0^2} + \frac{1}{\bar{m}} \frac{d\bar{m}}{dr} - \frac{v}{v_0^2} \frac{dv}{dr}.
\end{equation}
When \mbox{$v=0$}, Eqn.~\ref{eqn:semistatic2} gives the density structure of a hydrostatic atmosphere.
As in \citet{Tian08a}, we solve these equations consecutively. 
Firstly, we update the energies using Eqns.~\ref{eqn:hydro_neutralenergy}--\ref{eqn:hydro_neutralenergy}.
With the new temperature structure and the already known structure of $\bar{m}$, we then recalculate the structure of $v$ by integrating from the exobase downwards through the grid using Eqn.~\ref{eqn:semistatic1}, assuming the outflow speed at the exobase is known. 
Finally, since the density at the lower boundary of the simulation is a fixed value and is therefore known, we calculate the structure of $\rho$ by integrating upwards to the exobase using Eqn.~\ref{eqn:semistatic2}.
For both the solution of the energy equations and the integration of $\rho$ and $v$, we use the implicit Crank-Nicolson scheme, as described in Appendix~\ref{appendix:energyequationimplicit}.

When the atmosphere is supersonic at the upper boundary, the material has escape velocity and simply flows away from the planet; in these cases, the appropriate boundary conditions are zero-gradient outflow conditions.
Specifically, the values for each quantity in the final grid cell are made equal to the values in the second to last grid cell.
When the atmosphere is subsonic at the upper boundary, we assume an outflow speed that is consistent with the Jeans escape rate.
We first calculate the Jeans mass escape rate, $\dot{M}_\mathrm{Jeans}$, using the expressions given in \citet{Luger15}, and then calculate the upper boundary velocity from \mbox{$v_\mathrm{exo} = \dot{M}_\mathrm{Jeans} / ( 4 \pi r_\mathrm{exo}^2 \rho_\mathrm{exo})$}.

When the bulk flow of the atmosphere is not negligible, the effects of advection on the species densities must be taken into account.
We do this using the advection scheme described in Appendix~\ref{appendix:hydro} by converting the calculated cell boundary mass fluxes into individual species particle fluxes.
When using the semi-static hydrodynamic approach, we use the advection scheme to calculate the mass fluxes only (i.e. Eqns.~\ref{eqn:muscl1}--\ref{eqn:muscl2}).
Since the total mass density structure is being calculated at every timestep assuming that it has already come to a steady state, the changes in the density are a result of advection that is not explicitly calculated in the model.
To take this into account, after updating the structure of $\rho$ using Eqn.~\ref{eqn:semistatic2}, we scale the species number densities by a species-independent factor at each location to ensure that \mbox{$\rho = \sum_j m_j n_j$}, where the sum is over all species.

\subsection{Stellar Radiation and Non-thermal Electrons}

\begin{figure}
\fbox{\includegraphics[trim = 2mm 6mm 3mm 25mm, clip=true,width=0.49\textwidth]{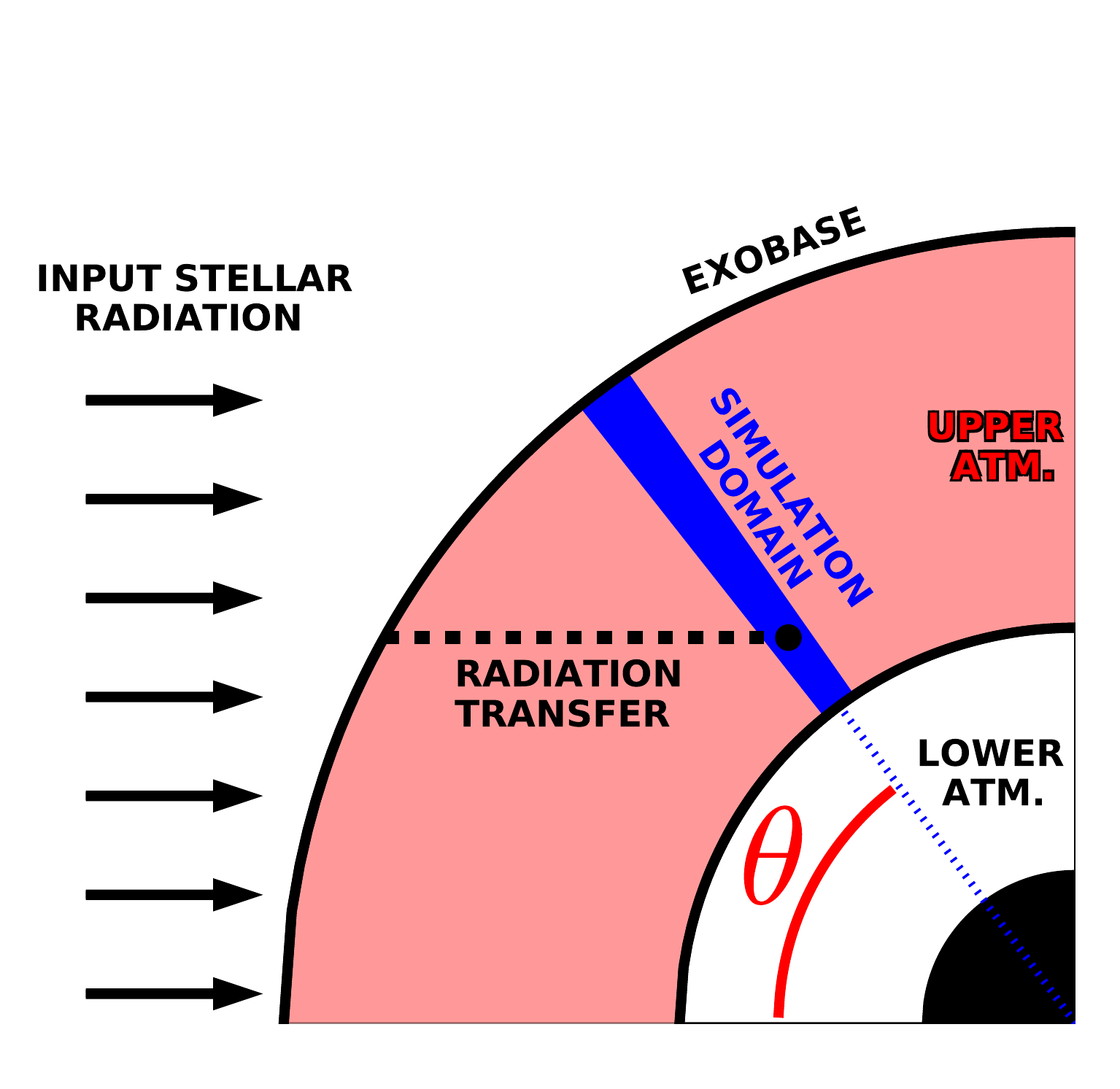}}
\caption{
Cartoon illustrating the geometry of the radiation transfer in our model. 
The bottom right of the cartoon is the centre of the planet and the incoming stellar radiation is travelling horizontally from left to right. 
The inner black region is the planet and the shaded region is the upper atmosphere. 
The simulation domain of our 1D atmosphere model, illustrated with the blue region, is a region that is pointing radially outwards from the planet centre.   
The angle $\theta$ is the zenith angle. 
To calculate the stellar XUV and IR spectra at any given point, we perform radiation transfer along the dashed black line. 
}
\label{fig:geometrycartoon}
\end{figure}

\subsubsection{X-ray and Ultraviolet radiation}

The most important external input into the upper atmosphere is the star's X-ray and ultraviolet (=`XUV') radiation. 
Its importance stems from the fact that atmospheric gases absorb radiation at XUV wavelengths very effectively.
This means that the XUV radiation is absorbed high in the atmosphere where the gas densities are low and relatively small energy inputs can lead to large temperature changes. 
The XUV spectrum also drives the most important chemical processes in the upper atmosphere, and is therefore essential for calculating the chemical structure of the atmosphere.

We irradiate the atmosphere with a stellar XUV spectrum between 10 and 4000~\AA.
The XUV spectrum is divided into 1000 energy bins and represented by the irradiance, $I_\nu$, which is the energy flux per unit frequency.
This input spectrum is assumed to be unattenuated at the exobase.
The radiation transfer through the atmosphere is then calculated based on the density structures of each absorbing species.

A weakness of 1D atmosphere models is that in reality the planet is being irradiated from one side only, which makes fully simulating the atmosphere at minimum a 2D problem.
In 1D models, approximate simplifying assumptions must be made. 
We assume that the computational domain is pointing in an arbitrary direction relative to the position of the star.
The angle between this direction and the direction that points directly at the star is the zenith angle, $\theta$.
We calculate the XUV spectrum at each point in the atmosphere by doing the radiation transfer from the exobase to each point separately.
This geometry is demonstrating in Fig.~\ref{fig:geometrycartoon}, where the dashed black line shows the path that the radiation takes through the atmosphere.
We assume that the state of the atmosphere at any given altitude is uniform over all latitudes and longitudes.
This means that when doing the radiation transfer, we get the densities of each species at each given point by taking the values at the point in our simulation that has the same altitude.  
In all simulations in this paper, except our Venus simulation in Section~\ref{sect:currentVenus}, we assume a zenith angle of 66$^\circ$.
We find in Section~\ref{sect:currentEarth} for the case of the Earth that this gives a decent representation of the atmosphere averaged over all longitudes and latitudes. 

\begin{figure}
\includegraphics[trim = 0mm 0mm 0mm 0mm, clip=true,width=0.49\textwidth]{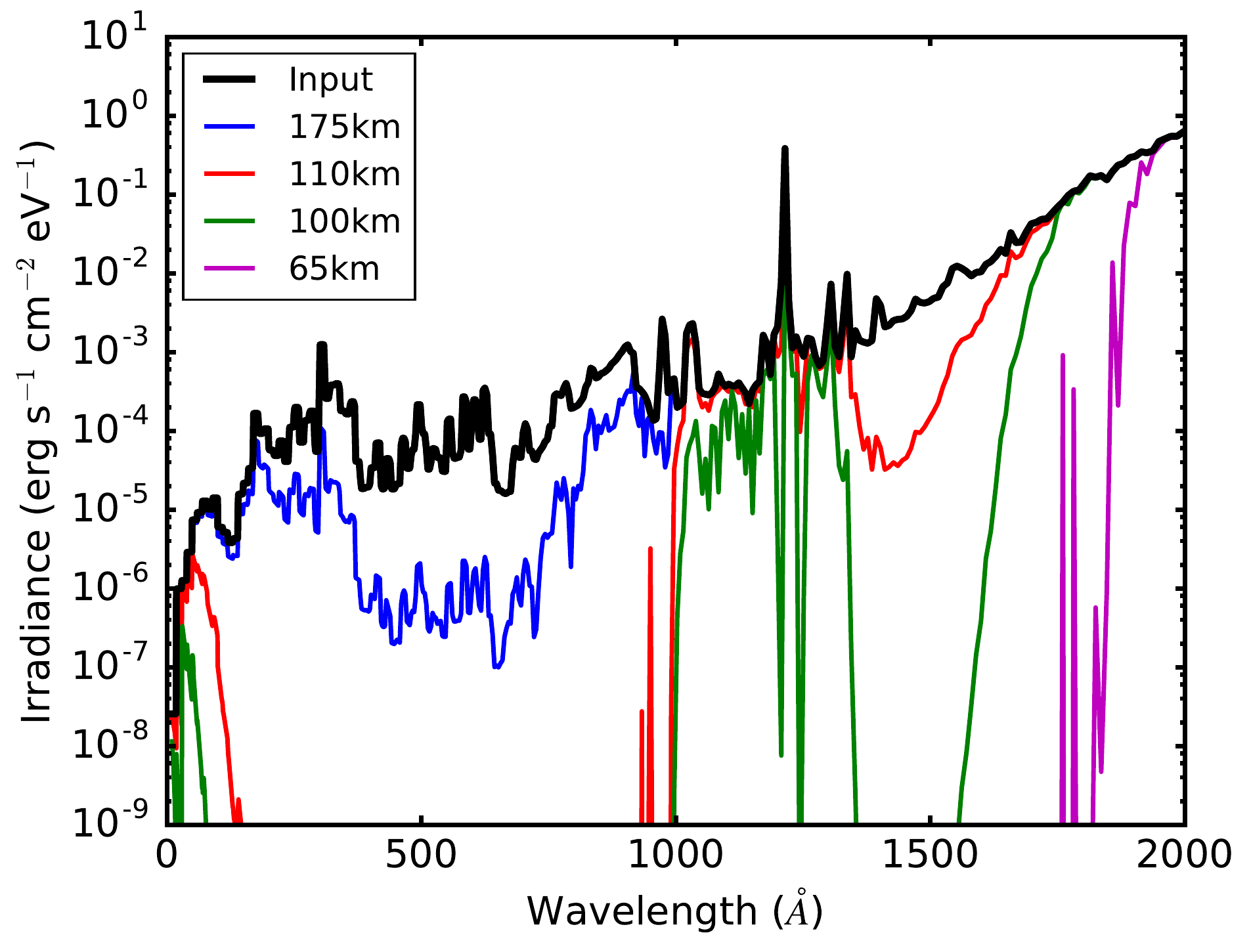}
\caption{
Figure showing the stellar XUV irradiance spectrum at different altitudes in the atmosphere of the Earth, calculated using our current Earth model presented in Section~\ref{sect:currentEarth}.
The black and purple lines show the spectrum at the top and bottom of our model respectively.  
}
\label{fig:XUVradtrans}
\end{figure}

To calculate the XUV spectrum at a given grid cell, we integrate along the dashed black line in Fig.~\ref{fig:geometrycartoon} for each energy bin using spatial steps with length \mbox{$\Delta s$} given by \mbox{$H/5$}, where \mbox{$H = N / (dN/dr)$} is the density scale height.
The change in the irradiance over a path \mbox{$\Delta s$} is given by
\begin{equation}
I_\nu ( s + \Delta s ) = I_\nu (s) \mathrm{e}^{-\Delta \tau_\nu} ,
\end{equation}
where \mbox{$\Delta \tau_\nu$} is the optical depth along the path length \mbox{$\Delta s$} and is given by
\begin{equation} \label{eqn:tauxuv}
\Delta \tau_\nu = \Delta s \sum\limits_j \sigma_{\nu,j} [R_j] ,
\end{equation}
where the sum is over all photoreactions in our chemical network, $\sigma_{\nu,j}$ is the cross-section of the $j$th photoreaction at frequency $\nu$, and $[R_j]$ is the number density of the reactant in the $j$th photoreaction.
By summing over individual photoreactions instead of using the total absorption cross-sections for each species, we ensure that the radiation transfer and the photochemistry are fully consistent.
The individual photoreaction cross-sections are discussed in Section~\ref{sect:chemistry}.
In Fig.~\ref{fig:XUVradtrans}, we show the XUV spectrum at several altitudes in our model for the current Earth. 

\subsubsection{Infrared radiation} \label{sect:stellarIR}

\begin{figure}
\includegraphics[trim = 0mm 0mm 0mm 0mm, clip=true,width=0.49\textwidth]{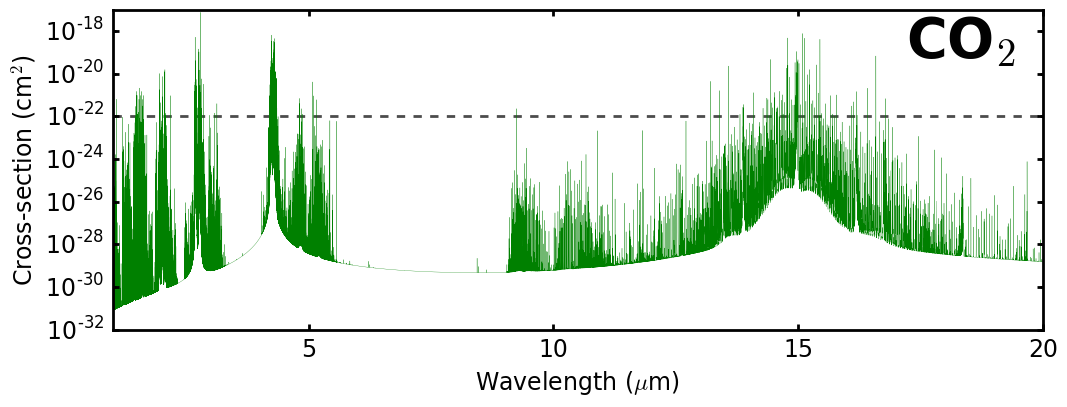}
\includegraphics[trim = 0mm 0mm 0mm 0mm, clip=true,width=0.49\textwidth]{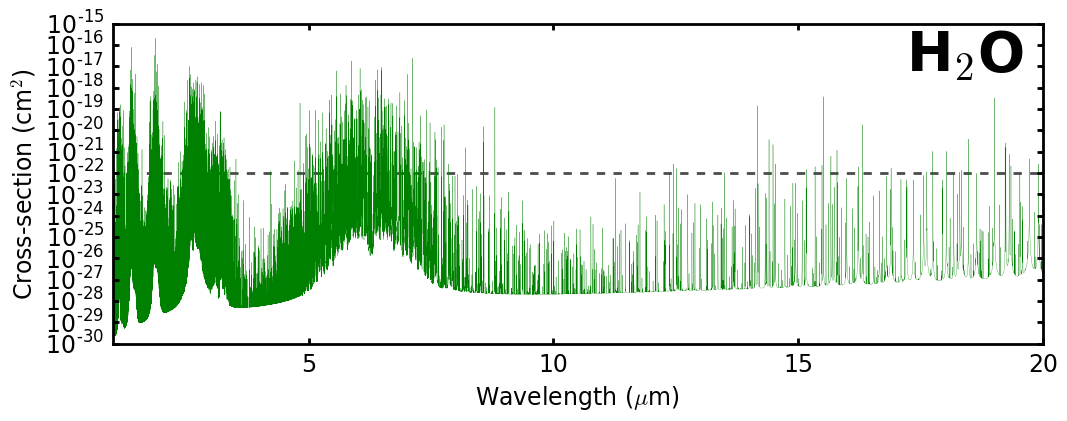}
\caption{
Figure showing the IR absorption spectra of CO$_2$ (\emph{upper-panel}) and H$_2$O (\emph{lower-panel}) as calculated by kspectrum (\citealt{Eymet16}). 
The gas is assumed to have a temperature of 200~K and a pressure of $10^{-2}$~mbar. 
The horizontal dashed line shows the cross-section of $10^{-22}$~cm$^{-3}$.
}
\label{fig:CO2IRspectrum}
\end{figure}

Another input into the atmosphere that can be important is the stellar infrared radiation.
Although this has a negligible effect on the Earth's upper atmosphere, it is a signficant source of heating for Mars and Venus (\citealt{BougherDickinson88}; \citealt{FoxBougher91}). 
This difference is due to the different abundances of CO$_2$, which is a strong absorber of IR radiation.
We calculate the transfer of the stellar IR spectrum between 1 and 20~$\mu$m through the atmosphere, and its effect on atmospheric heating. 
For the input stellar IR spectrum, we assume a simple blackbody spectrum with a temperature of 5777~K.
We make the same geometrical assumptions for the IR radiation transfer as we make for the XUV, as demonstrated in Fig.~\ref{fig:geometrycartoon}, and perform the integration from the exobase to each grid cell using the same method described for XUV.
For IR transfer, the sum in Eqn.~\ref{eqn:tauxuv} for the optical depth is over all considered absorbing species, where $\sigma_{\nu,j}$ and $[R_j]$ are the cross-sections and number densities of the $j$th absorbing species.  
We consider only absorption of IR radiation by CO$_2$ and H$_2$O molecules.
In future studies, the influences of other molecules will be included when necessary.

We calculate the absorption spectra of CO$_2$ and H$_2$O using the software package \emph{kspectrum} (\citealt{Eymet16}), which is an open-source code for calculating the high resolution absorption spectra of common atmospheric gases using the HITRAN~2008 and HITEMP~2010 molecular spectroscopic databases (\citealt{rothman2009hitran}; \citealt{rothman2010hitemp}). 
Although the absorption spectrum is temperature and pressure dependent, we calculate the cross-sections at 200~K and $10^{-2}$~mbar only and use these values everywhere in the atmosphere.
In order to resolve all features in the CO$_2$ absorption spectrum, a large number (\mbox{$\sim 10^6$}) of spectral bins are needed.
The wavelength-dependent absorption cross-sections for CO$_2$ and H$_2$O are shown in Fig.~\ref{fig:CO2IRspectrum}.
However, including so many energy bins is computationally too expensive for our model; instead, we only consider energy bins that have cross-sections above $10^{-22}$~cm$^{2}$ for at least one of the considered molecules.
Tests have shown that we get identical results using this threshold, while limiting the number of energy bins to something reasonable (\mbox{$\sim 10^4$}).

The heating of the atmosphere by the absorption of IR radiation is discussed in Section~\ref{sect:modelheating}.
We calculate the heating assuming that all energy removed from the radiation field by absorbtion is immediately added to the thermal energy reservoir of the neutral gas.
What actually happens is that the absorption of photons excites the molecules and the heating of the gas only takes place when they are then collisionally deexcited. 
Some of this energy will not in fact end up in heat, but will be reradiated back into space. 
To take this into account, we add an additional excitation term into the equations for 15~$\mu$m CO$_2$ cooling in Section~\ref{sect:cooling}.
We write the excitation rate due to stellar IR photons as
\begin{equation} \label{eqn:infraredCO2excitation}
S_\mathrm{IR} = \frac{ \int\limits_\nu \sigma_{\nu,\mathrm{co}_2} [\mathrm{CO}_2] I_\nu d\nu }{ (h\nu)_{15\mu \mathrm{m} } } ,
\end{equation}
where the integral is over all considered frequencies, $\sigma_{\nu,\mathrm{co}_2}$ and $[\mathrm{CO}_2]$ are the absorption cross-section and number density of CO$_2$ at frequency $\nu$, and \mbox{$(h\nu)_{15\mu \mathrm{m} }$} is the energy of a 15~\mbox{$\mu$m} photon.
The cgs units for $S_\mathrm{IR}$ are \mbox{excitations~s$^{-1}$~cm$^{-3}$}.
The numerator in Eqn.~\ref{eqn:infraredCO2excitation} gives the volumetric heating rate due to the absorbtion of IR photons by CO$_2$.  
The assumption here is that all energy absorbed from the IR field by CO$_2$ eventually contributes to the excitation of the 15~\mbox{$\mu$m} bending mode in CO$_2$ molecules.
The main absorption bands are at 15~\mbox{$\mu$m}, 4.3~\mbox{$\mu$m}, 2.7~\mbox{$\mu$m}, and 2.0~\mbox{$\mu$m}.
The latter two are combination bands, and photons absorbed in these bands cause multiple excitations in the 15~\mbox{$\mu$m} and 4.3~\mbox{$\mu$m} band transitions.
Our assumption in Eqn.~\ref{eqn:infraredCO2excitation} is reasonable if the majority of energy in the 4.3~\mbox{$\mu$m} vibrational state is transferred to the 15~\mbox{$\mu$m} vibrational state by vibrational-vibrational exchanges, as argued by \citet{taylor1969survey} (see the discussion in Section~2 of \citealt{Dickinson76}).

\subsubsection{Non-thermal electrons} \label{sect:photoelectron}

\begin{figure}
\includegraphics[trim = 0mm 0mm 0mm 0mm, clip=true,width=0.49\textwidth]{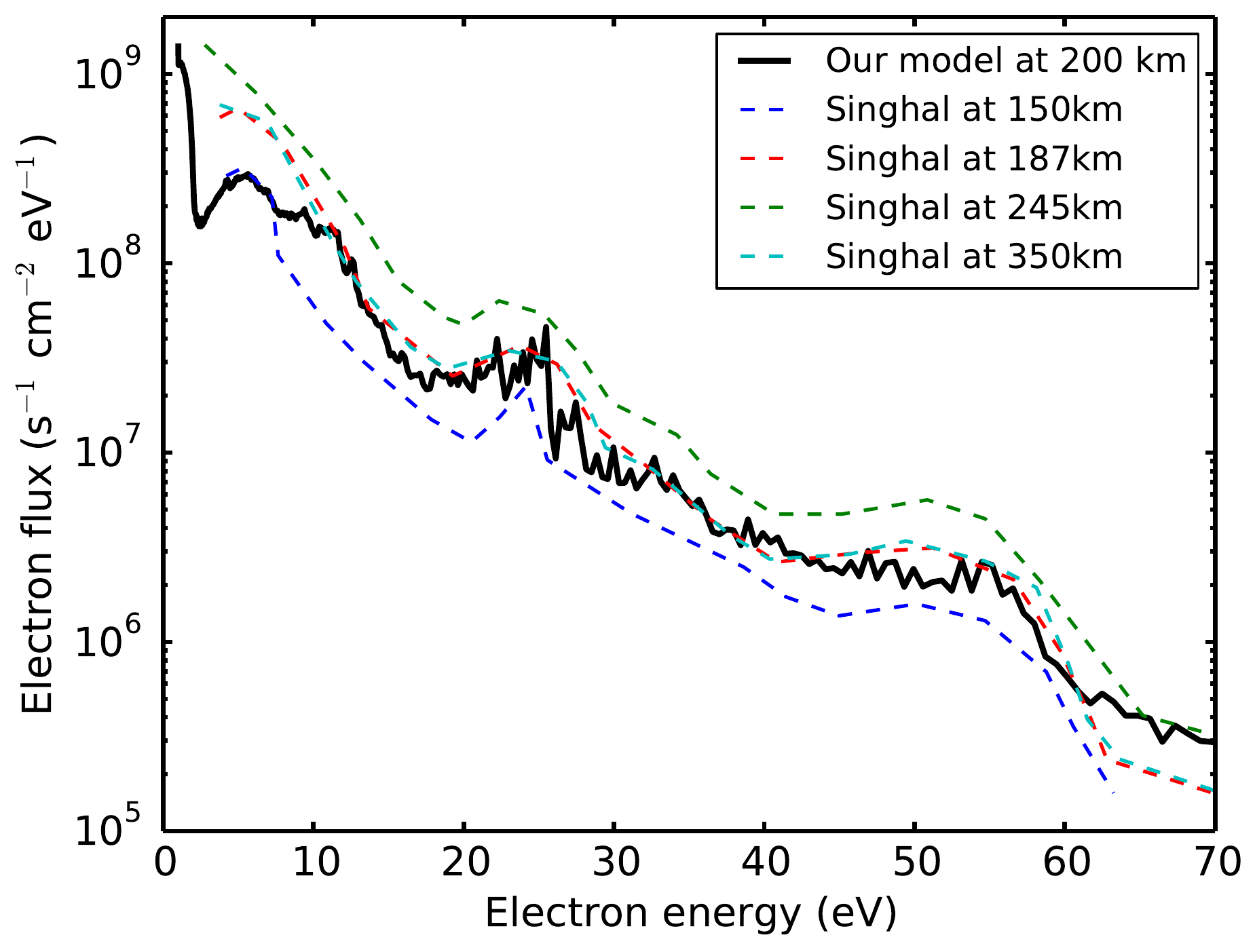}
\caption{
Figure showing our prediction for the non-thermal electron spectrum at 200~km for the current Earth. 
For comparison, the dashed lines show the spectra at several altitudes calculated by \citet{SinghalHaider84} (taken from their Fig.~3 and multiplied by 4$\pi$ to move the sr$^{-1}$ from the units).
The sudden increase in the flux at low electron energies is due to the thermal electron spectrum becoming dominant. 
}
\label{fig:earthpespectrum}
\end{figure}

Many of the photoionization reactions that take place in the upper atmosphere are caused by photons that contain significantly more energy than is needed simply to cause the ionization.
This additional energy is given to the produced photoelectrons in the form of kinetic energy, which results in a population of photoelectrons that have significantly larger energies than the thermal energy of the electron gas.
These high energy photoelectrons then lose their energy by collisions with other atmospheric particles.
For the thermal electrons, elastic collisions with non-thermal electrons is the main heat source and is the reason why the electron gas becomes hotter than the neutral and ion gases in the Earth's upper thermosphere (\citealt{SmithtroSolomon08}).
 
The two main assumptions in our model are that the photoelectrons lose their energy locally where they are created and that the non-thermal electron spectrum is in a steady state at each point.
Given the latter assumption, the spectrum can be calculated simply by balancing sources and sinks of electrons at each electron energy.
In future models, we will include also a more sophisticated electron transport model; this was shown to influence the heat deposition by \citet{Tian08b}.
For the effects of collisions, the situation is complicated by the fact that a given atmospheric species can interact with a non-thermal electron in multiple ways. 
Each of these different interactions has a different energy-dependent cross-section and takes a different amount of energy from the impacting electron.

At a given electron energy, $E_\mathrm{e}$, the two sources of electrons are photoionization reactions producing electrons with energy $E_\mathrm{e}$, and the degradation of more energetic electrons through collisions with the ambient gas.
The production spectrum for photoelectrons by photoionization reactions, $P_\mathrm{e} (E)$, is given by
\begin{equation} \label{eqn:PEproduction}
P_\mathrm{e} (E_\mathrm{e}) = \sum\limits_k  \frac{ I_\mathrm{xuv} (E_\mathrm{e} + \delta E_k) }{E_\mathrm{e}+\delta E_k} \sigma_k (E_\mathrm{e} + \delta E_k) [R_k] ,
\end{equation}
where the sum is over all photoionization reactions considered, $\sigma_k (E_\mathrm{e})$ is the energy dependent cross-section for the $k$th photoionization reaction, $[R_k]$ is the number density of the reactant in the $k$th photoionization reaction, and $\delta E_k$ is the energy required for the ionization to take place. 
We calculate the non-thermal electron spectrum using
\begin{equation} \label{eqn:PEspectrum}
\phi_\mathrm{e} (E_\mathrm{e}) 
= 
\frac{
P_\mathrm{e} (E_\mathrm{e}) + \sum\limits_i \sum\limits_j n_i \sigma_{ij} (E_\mathrm{e} + \delta E_{ij}) \phi_\mathrm{e}(E_\mathrm{e} + \delta E_{ij})
}{
\sum\limits_i \sum\limits_j n_i \sigma_{ij} (E_\mathrm{e})
},
\end{equation}
where \mbox{$\phi_\mathrm{e} (E_\mathrm{e})$} is the electron flux at energy $E_\mathrm{e}$ and $\sigma_{ij}$ is the cross-sections for electron impact interactions.
This expression is described in more detail by \citet{SchunkNagy78}.
In both the numerator and the denominator, the first sum is over all species that the electrons interact with and the second sum is over all possible interactions with that species. 
The second term in the numerator is the source term from the degradation of higher energy electrons.
The fact that photoelectrons can only lose energy, so that $\phi_\mathrm{e}$ at a given energy depends on the higher energies values of $\phi_\mathrm{e}$ only, makes solving Eqn.~\ref{eqn:PEspectrum} trivial.
To do this, we break the spectrum down into 100 discreet energy bins logarithmically spaced between 1 and 1000~eV.
We first calculate $\phi_\mathrm{e}$ at the bin with the highest energy assuming \mbox{$\phi_\mathrm{e} = 0$} for higher energies, and then iterate downwards through the spectrum, calculating $\phi_\mathrm{e}$ in each bin. 

The neutral interacting species that we consider are N$_2$, O$_2$, O, CO$_2$, CO, and He.
For N$_2$, we use the electron impact cross-sections given in \citet{GreenBarth65}.
For non-ionizing O$_2$ transitions, we use the cross-sections from \citet{Watson67}, and for O$_2$ ionizations we use cross-sections from \citet{Jackman77}.
For O, CO$_2$, and CO, we use cross-sections from \citet{Jackman77}.
For He, we use cross-sections from \citet{jusick1967electron}.

In Fig.~\ref{fig:earthpespectrum} we show our predicted non-thermal electron spectrum up to 70~eV for the current Earth's atmosphere at an altitude of 200~km.
This spectrum is an output of the Earth model presented in Section~\ref{sect:currentEarth}.
For comparison, we also show the spectra calculated for several altitudes by \citet{SinghalHaider84}, which they compared to other models and observations in their Fig.~3.
Our spectrum match theirs well for almost the entire energy range, indicating that our model calculates approximately realistic non-thermal electron spectra.

In our chemical network, we have included several ionization reactions due to impacts with non-thermal electrons.
For each of these reactions, the total cross-sections for use in Eqn.~\ref{eqn:PEreactionrate} are calculated by summing over the cross-sections for each corresponding ionization interaction.
For many of the transitions that are not direct ionizations, an ionization can still take place by autoionization.
We take these into account when calculating the total ionization cross-sections for O, CO$_2$, and CO by multiplying the cross-sections for these transitions by the autoionization factors given by \citet{Jackman77}. 
These reactions also remove energy from non-thermal photoelectrons, but the situation is complicated since the products of these reactions include two electrons.
The energy of the original non-thermal electron that is not used to cause the reaction is distributed between the two electrons.
For simplicity, we assume that one of the electrons gets this energy, and the other just becomes a normal thermal electron.

\subsection{Chemical Structure of the Atmosphere} 


\subsubsection{Chemistry} \label{sect:chemistry}

In this study, we attempt to construct a general chemical network that can be applied to a range of atmospheres with arbitrary compositions. 
This is difficult given the huge numbers of reactions and species that any network could consider and the uncertainties in the rate coefficients for the reactions, particularly at high temperatures.
We do this by combining the networks of several previously published atmospheric models.
The networks that we use are from \citet{FoxSung01}, \citet{Verronen02}, \citet{Yelle04}, \citet{Verronen05}, \citet{GarciaMunoz07}, \citet{Tian08a}, \citet{RichardsVoglozin11}, \citet{Fox15}, and \citet{Foxetal15}.
We include almost all reactions from these papers, with some reactions being excluded if they introduced species that we consider unimportant.
The rate coefficients are taken also from these studies in almost every case.
Where multiple papers give different rate coefficients for the same reaction, we take the values almost arbitrarily, or find the coefficients on the KIDA database (\citealt{Wakelam12}).
In addition, we add a few reactions from KIDA that are not in any of these networks when necessary to stop reactions from creating species that are not destroyed.  
For the photoreactions, we take all of the relevant reactions from the PHIDRATES database (\citealt{HuebnerMukherjee15}), which provides wavelength dependent cross-sections for the entire XUV spectrum. 
These cross-sections are all temperature independent, which is in many cases unrealistic and could lead to inaccuracies in our photochemistry (\citealt{Venot17}).
The few reactions involving non-thermal electrons produced in photoionization reactions are described in Section~\ref{sect:photoelectron}.
The resulting network, which is given in Appendix~\ref{appendix:chemicalnetwork}, contains 63 species, including 30 ion species, and 503 reactions, including 56 photoreactions and 7 photoelectron reactions. 

The reaction rate of the $k$th chemical reaction, $R_k$, is related to the rate coefficient, $k_k$, by
\begin{equation} \label{eqn:reactionrate}
R_k = k_k \prod\limits_i n_i ,
\end{equation}
where the RHS gives the product of the densities of all reactants.
The rate coefficients for normal reactions are typically functions of temperature and many of the reactions have temperature limits, both of which are listed in Table~\ref{table:chemicalnetwork}.
A difficulty in our model is that we calculate separate neutral, ion, and electron temperatures, and in many cases it is unclear which of these temperatures to use to calculate the rate coefficients.
For reactions that have only neutral reactants, we use the neutral temperature; for reactions that have a mixture of neutrals, ions, and electrons as reactants, we simply use the averages of the temperatures of the involved components (e.g. if a reaction has one neutral reactant and one ion reactant, we set \mbox{$T_\mathrm{gas} = ( T_\mathrm{n} + T_\mathrm{i} ) / 2$} in the equation for the rate coefficient).
For photoreactions, the rate coefficients depend on the XUV spectrum and the wavelength dependent cross-sections by
\begin{equation} \label{eqn:photoreactionrate}
k_k = \int\limits_{E_\mathrm{t}}^\infty \frac{ \sigma_k  I_E }{E} dE,
\end{equation}
where $E$ is the photon energy, $E_\mathrm{t}$ is the threshold energy for the reaction, $\sigma_k$ is the cross-section, and $I_E$ is the irradience in units of energy flux per unit energy (the quantity $I_\nu$ used elsewhere in this paper is the irradience in units of energy flux per unit frequency).
Similarly, the equation for the rate coefficients of reactions involving inelastic collisions with non-thermal electrons is
\begin{equation} \label{eqn:PEreactionrate}
k_k = \int\limits_{E_\mathrm{t}}^\infty \sigma_k \phi_\mathrm{e} dE.
\end{equation}
The result is a set of ordinary differential equations, one for each species, describing the rates of change of the species densities.
For the $j$th species, this can be written
\begin{equation} \label{eqn:chemicalrate}
\frac{d n_j}{dt} = \sum\limits_k R_k - \sum\limits_i R_i = S_j,
\end{equation}
where the first sum is over all reactions that create the $j$th species, and the second sum is over all reactions that destroy it.  
The $S_j$ term is the total source term for the $j$th species in the RHS of Eqn.~\ref{eqn:main_speciescontinuity}.
To evolve $n_j$ using Eqn.~\ref{eqn:chemicalrate}, we use an implicit Rosenbrock solver described in Appendix~\ref{appendix:chemicalnetwork}.

In our model, we break the gas down into neutral, ion, and electron components.
A difficulty in our model is that the chemical reactions cause the transfer of mass, momentum, and energy between the components simply due to the changes in the identities of atoms and molecules.
For example, consider the reaction \mbox{N$^+$ + O$_2$ $\rightarrow$ NO$^+$ + O}; this reaction transfers an O atom from the neutral gas to the ion gas.
The changes in the mass and momentum densities of the components are trivial to calculate, but the changes in the energy densities are not.
We avoid this problem by assuming that the temperatures are unaffected when updating the species densities due to chemistry.
The heating of the gas due to exothermic and endothermic chemical reactions, and the energy exchanges between the neutral, ion, and electron gases, are calculated separately, as described in Section~\ref{sect:thermalstructure}.

\subsubsection{Diffusion}

Many of the species considered in the simulation are created and destroyed slowly by chemical/photochemical reactions. 
For these species, a very important transport mechanism is diffusion. 
Our model takes into account both molecular and eddy diffusion.
Eddy diffusion evolves the density profiles so that they all follow the pressure scale height of the entire gas; molecular diffusion evolves the density profiles so that they all follow their own pressure scale heights.
In the homosphere, eddy diffusion dominates and the mixing ratios of the long-lived species are independent of altitude.
In the heterosphere, molecular diffusion dominates and the densities of heavy species decrease with increasing altitude faster than the densities of light species, meaning that light species become increasingly dominant at higher altitudes (this also happens due to the dissociation of heavy molecules). 
The equation that we use for the diffusive flux of the $j$th species, including both eddy and molecular diffusion, is
\begin{equation} \label{eqn:diffusionflux}
\Phi_{\mathrm{d},j} = n_j v_{\mathrm{d},j} ,
\end{equation}
where $v_{\mathrm{d},j}$ is the diffusion speed, given by
\begin{equation} \label{eqn:diffusionvel}
\begin{aligned}
v_{\mathrm{d},j} = & - D_j \left[ \frac{1}{n_j} \frac{dn_j}{dr} - \frac{1}{N} \frac{dN}{dr} + \left( 1 - \frac{m_j}{\bar{m}} \right) \frac{1}{p} \frac{dp}{dr} + \frac{\alpha_{\mathrm{T},j}}{T} \frac{dT}{dr} \right] \\
 & - K_\mathrm{E} \left[ \frac{1}{n_j} \frac{dn_j}{dr} -\frac{1}{N} \frac{dN}{dr} \right] ,
\end{aligned}
\end{equation}
where $D_j$ and $K_\mathrm{E}$ are the molecular and eddy diffusion coefficients, $n_j$ and $N$ are the particle number densities of the $j$th species and of the entire gas, $m_j$ and $\bar{m}$ are the molecular masses of the $j$th species and of the entire gas, $p$ and $T$ are the thermal pressure and temperature of the entire gas, and $\alpha_{\mathrm{T},j}$ is the thermal diffusion factor. 
To solve these equations, we use the implicit Crank-Nicolson method described in Appendix~\ref{appendix:diffusion}.

For molecular diffusion, the diffusion coefficient for a given species, $D_j$, depends on both the species itself, the composition of the background gas, and the temperature. 
For all diffusion coefficients, we use the relation
\begin{equation} \label{eqn:}
D_j = \frac{\alpha_j \times 10^{17} T^{s_j}}{N}.
\end{equation}
For H, H$_2$, He, CH$_4$, CO, Ar, CO$_2$, and O, we use values for $\alpha_j$ and $s_j$ given in Table~15.1 and Table~15.2 of \mbox{\citet{banks1973aeronomy}} assuming an N$_2$ background atmosphere, which is likely reasonable since the values are very similar for the other background atmospheres.
For simplicity, we assume \mbox{$\alpha_j = 1$} and \mbox{$s_j = 0.75$} for all other species. 
For H, H$_2$, and He, we assume \mbox{$\alpha_{\mathrm{T},j} = -0.38$} and for Ar, we assume \mbox{$\alpha_{\mathrm{T},j} = 0.17$} (\citealt{banks1973aeronomy}); for all other species, we assume \mbox{$\alpha_{\mathrm{T},j} = 0$}.

In models such as ours, the eddy diffusion coefficients as a function of altitude are free parameters; this is the only free parameter in our model.
We assume it is given by \mbox{$K_\mathrm{E} = A N^B$}, where $K_\mathrm{E}$ and $N$ have the units cm$^2$~s$^{-1}$ and cm$^{-3}$ respectively. 
This functional form is typically used for models of Venus and Mars (\citealt{zahn1980upper}; \citealt{FoxSung01}).
For Venus, we use \mbox{$A = 2 \times 10^{13}$} and \mbox{$B=-0.5$} and impose a maximum value for $K_\mathrm{E}$ of \mbox{$6 \times 10^{8}$~cm$^{2}$~s$^{-1}$}.
These values were used by \citet{Fox15} for the upper atmosphere of Mars, and are very similar to values found for Venus (\citealt{zahn1980upper}).
For the current Earth, we first fit $A$ and $B$ to the tabulated $K_\mathrm{E}$ values given by \citet{Roble95}, but we scale these values up by a factor of ten in order to fit the expected O$_2$ densities at high altitudes in our Earth model. 
This gives \mbox{$A = 10^{8}$} and \mbox{$B=-0.1$}.
Note that even without scaling up the eddy diffusion coefficients from \citet{Roble95}, we obtain good fits to the density profiles of all other species. 

For atmospheres that are close to hydrostatic, it is important to specify a diffusion flux at the exobase. 
We assume an outward diffusion flux that corresponds to Jeans escape.
This is only necessary for the lightest species, i.e. H and He, so for all species more massive than $4m_\mathrm{p}$, we assume a zero flux. 
In simulations where the gas at the exobase is supersonic, and therefore is faster than the escape velocity, we simply assume a zero diffusion flux at the exobase.

\subsection{Thermal Structure of the Atmosphere}  \label{sect:thermalstructure}

\subsubsection{Heating}  \label{sect:modelheating}

\begin{figure}
\fbox{\includegraphics[trim = 0mm 0mm 0mm 0mm, clip=true,width=0.49\textwidth]{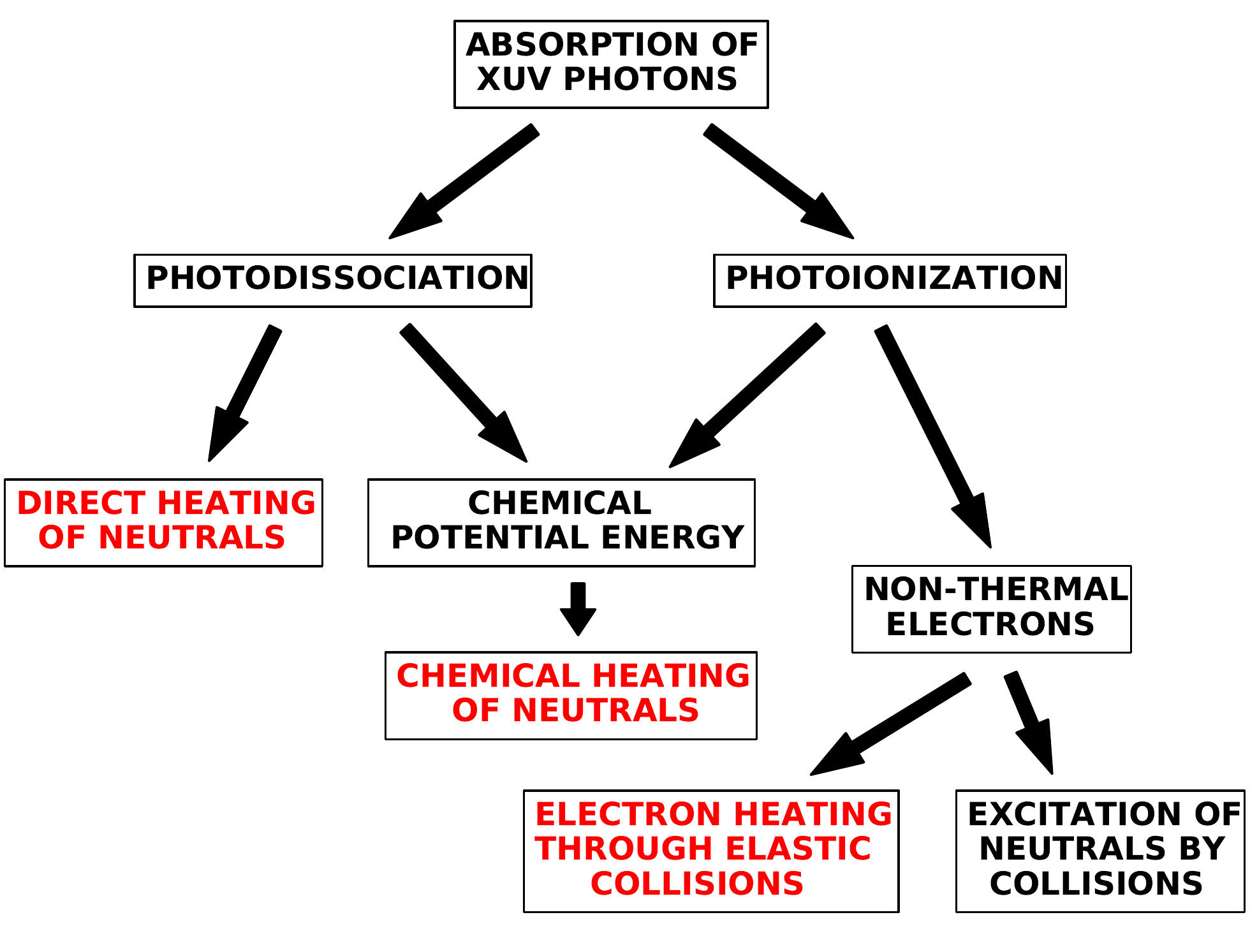}}
\caption{
Simplified cartoon illustrating the main pathways taken in our model by the energy that is removed from the XUV field by absorption due to photodissociation and photoionization reactions. 
In both cases, much of the energy used to cause the photoreaction is released due to exothermic chemical reactions. 
For photodissociation, the remaining energy from the absorbed photon is given to the thermal energy budget of the neutral gas directly.
For photoionization, the remaining energy is released as kinetic energy of the produced electron, and is then lost as the electron collides with the ambient gas; collisions with neutral species are inelastic and lead to excitation, dissociation, and ionization, whereas collisions with ambient thermal electrons are elastic and lead to heating of the electron gas.
}
\label{fig:xuvpathwayscartoon}
\end{figure}

In this model, the important energy sources are stellar XUV (10--4000\AA) and IR (1--20$\mu$m) radiation.
We also include a simple treatment of Joule heating. 
The total energy deposition rate from a radiation field travelling a distance $dx$ through an absorbing gas is \mbox{$-dF/dx$}, where $F$ is the energy flux (\mbox{$=\int I_\nu d\nu$}). 
In reality, the absorbed energy is not directly added to the thermal energy budget of the gas, and not all of the energy deposited is eventually converted to heat.
A common way to calculate the heating is to multiply the total energy deposition rate by a heating efficiency factor (e.g. \citealt{Erkaev13}; \citealt{Johnstone15}); this assumption is generally undesirable since it adds an unconstrained free parameter into the model.
We instead use a more complete heating model where the energy release from different processes are calculated individually.
The heating processes considered are 
direct heating by the stellar XUV field, 
electron heating by elastic collisions with non-thermal electrons, 
heating from exothermic chemical reactions, 
direct heating by the stellar IR field, 
and Joule heating.

When a XUV photon is absorbed, a large part of its energy is used to cause either a dissociation, an ionization, or both. 
For photodissociation reactions, the remaining photon energy is given to the products as kinetic energy, and ultimately dissipated as heat in the gas.
It is this heating that we consider the direct heating by the stellar XUV field.
To calculate this direct heating, we need to consider each photodissociation reaction and each energy bin in the XUV field separately.
The full equation for the heating rate is  
\begin{equation} \label{eqn:XUVheatingequation}
Q_\mathrm{xuv} = \sum\limits_k Q_\mathrm{xuv,k} ,
\end{equation}
where $Q_\mathrm{xuv,k}$ is the heating rate by the $k$th reaction and is given by
\begin{equation}
Q_\mathrm{xuv,k} = \int\limits_{E_{\mathrm{T},k}}^\infty \left( E - E_{\mathrm{T},k} \right) \frac{I_E}{E} \sigma_k (E) [R_k] dE ,
\end{equation}
where the sum is over all photodissociation reactions, $E_{\mathrm{T},k}$ is the energy required for the reaction to take place, \mbox{$\sigma_k (E)$} is the reaction cross-section at energy $E$, $[R_k]$ is the number density of the reactant, and $I_E$ is the irradience in units of energy flux per unit energy. 
The term \mbox{$I_E / E$} is the photon flux per unit photon energy at energy $E$ and the term \mbox{$(I_E / E) \sigma_k (E) [R_k]$} is the rate at which the $k$th reaction takes place per unit volume per unit photon energy.
The energy released per reaction is \mbox{$E - E_{\mathrm{T},k}$}, which when multipled by the reaction rate and integrated over all photon energies gives the heating rate for the $k$th reaction.

For photoionization reactions, we do not give the remaining photon energy to the gas as heat directly, but instead assume that all this energy is given to the resulting free electron and we calculate the non-thermal electron spectrum, as described in Section~\ref{sect:photoelectron}.
This energy is either given to the neutral gas by inelastic collisions, typically exciting atoms/molecules or causing secondary ionizations, or it is given to the thermal electrons by elastic collisions.
We assume that the energy given to the neutral gas is all lost by radiative relaxation and consider therefore only the heating of the electron gas.
Using the expression given by \citet{SchunkNagy78}, we calculate the electron heating rate as
\begin{equation} \label{eqn:electronheating}
\begin{aligned}
Q_\mathrm{e} 
= 
\int\limits_0^{E_\mathrm{t}} & \left( E_\mathrm{e} - \frac{3}{2} k_\mathrm{B} T_\mathrm{e} \right) P_\mathrm{e} (E_\mathrm{e})  dE_\mathrm{e}
+
\int\limits_{E_\mathrm{t}}^\infty n_\mathrm{e} L_\mathrm{e} (E_\mathrm{e}) \phi_\mathrm{e} (E_\mathrm{e}) dE_\mathrm{e}
\\ 
 & +
\left( E_\mathrm{t} - \frac{3}{2} k_\mathrm{B} T_\mathrm{e} \right) n_\mathrm{e} L_\mathrm{e} (E_\mathrm{t}) \phi_\mathrm{e} \left( E_\mathrm{t} \right) ,
\end{aligned}
\end{equation}
where $E_\mathrm{e}$ is the electron kinetic energy, \mbox{$P_\mathrm{e} (E_\mathrm{e})$} is the production spectrum of electrons (see Eqn.~\ref{eqn:PEproduction}), $E_\mathrm{t}$ is the energy above which the non-thermal flux is larger than the thermal flux, and \mbox{$L_\mathrm{e} (E_\mathrm{e})$} is the loss function given by
\begin{equation}
L_\mathrm{e} (E_\mathrm{e}) = \frac{3.37 \times 10^{-12}}{E_\mathrm{e}^{0.94} n_\mathrm{e}^{0.03}} \left( \frac{ E_\mathrm{e} - E_\mathrm{th} }{ E_\mathrm{e} - 0.53 E_\mathrm{th} } \right)^{2.36} ,
\end{equation}
where \mbox{$E_\mathrm{th} = 8.618 \times 10^{-5} T_\mathrm{e}$} (\citealt{swartz1971analytic}).
The three terms on the RHS of Eqn.~\ref{eqn:electronheating} give respectively the heating/cooling by the direct production of thermal electrons by photoionization reactions, heating of thermal electrons by elastic collisions with non-thermal electrons, and a surface term related to the crossover between the thermal and non-thermal electron spectra.
Although a more accurate version of the surface term was derived by \citet{Hoegy84}, we use the version given above because it is simpler and is sufficiently accurate for our purposes.

Much of the photon energy used to cause a photoreaction is not lost, but is instead converted into chemical potential energy that can then be released as heat in exothermic chemical reactions.
The heating rate at a given point by chemical reactions is given by
\begin{equation}
Q_\mathrm{chem} = \sum\limits_k R_k Q_{\mathrm{chem},k} ,
\end{equation}
where the sum is over all chemical reactions, and $R_k$ and $Q_{\mathrm{chem},k}$ are the reaction rate and energy released per reaction for the $k$th reaction.
The values of $Q_{\mathrm{chem},k}$ are given for each reaction in Table.~\ref{table:chemicalnetwork}.
These energies are mostly taken from \citet{Tian08a} or from the KIDA database, and when the energy for a given reaction is not available in either of these sources, we simply assume it does not contribute to the heating.
 
We consider also heating of the atmosphere by the abosrption of IR radiation.
We assume that all of the energy removed from the IR spectrum is input into the neutral gas as thermal energy, giving a heating rate of 
\begin{equation} \label{eqn:irheat}
Q_\mathrm{IR} = \sum\limits_j \int\limits_\nu \sigma_{\nu,j} [R_j] I_\nu d\nu,
\end{equation}
where the sum is over all absorbing species, the integral is over the entire IR spectrum that we consider, and \mbox{$\sigma_{\nu,j}$} and $[R_j]$ are the absorption cross-section and number density of the $j$th absorbing species.
In reality, this energy is first used to excite CO$_2$ and is then released as heat through collisional deexcitation, which we take into account with an additional excitation term in the equations for CO$_2$ cooling. 

Additionally, the upper atmospheres of magnetized planets are heated by two magnetospheric processes: these are energetic particle precipitation and Joule heating.
For the Earth, during quiet geomagnetic conditions these two processes are likely similar in magnitude, and Joule heating tends to dominate during geomagnetic storms (e.g. \citealt{chappell2016magnetosphere}).
This process could become important for planets that are exposed to extreme space weather (\citealt{Cohen14}).
Both processes are most significant at high latitudes, but tend not to influence the global heat budget significantly during quiet conditions.
In this paper, we model Joule heating using the simplified model described in \citet{Roble87} and \citet{SmithtroSojka05}.
The two input parameters are the ambient magnetic field strength, which we assume is 0.5~G everywhere, and the total global Joule heating rate, which we assume is \mbox{$1.4 \times 10^{18}$~erg~s$^{-1}$}.
This is double the value used in \citet{Roble87}, which is typical for quiet levels of geomagnetic activity (\citealt{Foster83}); we double the value to take into account also the energy input expected from particle precipitation.
We calculate the heating rate at each altitude using
\begin{equation}
Q_\mathrm{J} = \sigma_\mathrm{P} E^2 ,
\end{equation}
where $E$ is the electric field strength and $ \sigma_\mathrm{P}$ is the Pedersen conductivity (\citealt{Foster83}).
We do not calculate the electric field, but instead assume that it is a constant and use it as a free parameter that can be scaled in order to give us the desired total global Joule heating rate.
The Pedersen conductivity varies with altitude, and at a given point depends on the densities of individual ion and neutral species, the gas temperature, and the ambient magnetic field strength.
The $\sigma_\mathrm{P}$ profiles are calculated self-consistently within the model using the equations described in Section~5.11 of \citet{schunk2000ionospheres}.
The equation for $\sigma_\mathrm{P}$ is
\begin{equation}
\sigma_\mathrm{P} = \sum\limits_i \sigma_i \frac{\nu_i^2}{\nu_i^2 + \omega_i^2} ,
\end{equation}
where the sum is over all considered ion species, and $\sigma_i$, $\nu_i$, and $\omega_i$ are the ion conductivity, ion-neutral collision frequency, and angular gyrofrequency of the $i$th ion species.
The angular gyrofrequency is given by \mbox{$\omega_i = q_i B / m_i$}.
The ion conductivity is given by $\sigma_i = (n_i q_i^2)/(m_i \nu_i)$, where $n_i$, $q_i$, and $m_i$ are the ion number density, charge, and mass respectively.
The ion-neutral collision frequency, $\nu_i$, is calculated as the sum over the collision frequencies with individual neutral species, such that \mbox{$\nu_i = \sum\limits_n \nu_{in}$}, where $\nu_{in}$ is the frequency of collisions between the $i$th ion species and the $n$th neutral species.
For this, we use the same collisions and $\nu_{in}$ values described in Section~\ref{sect:heatexchange} for ion-neutral heat exchange.

\subsubsection{Cooling} \label{sect:cooling}

\begin{figure}
\includegraphics[trim = 0mm 0mm 0mm 0mm, clip=true,width=0.49\textwidth]{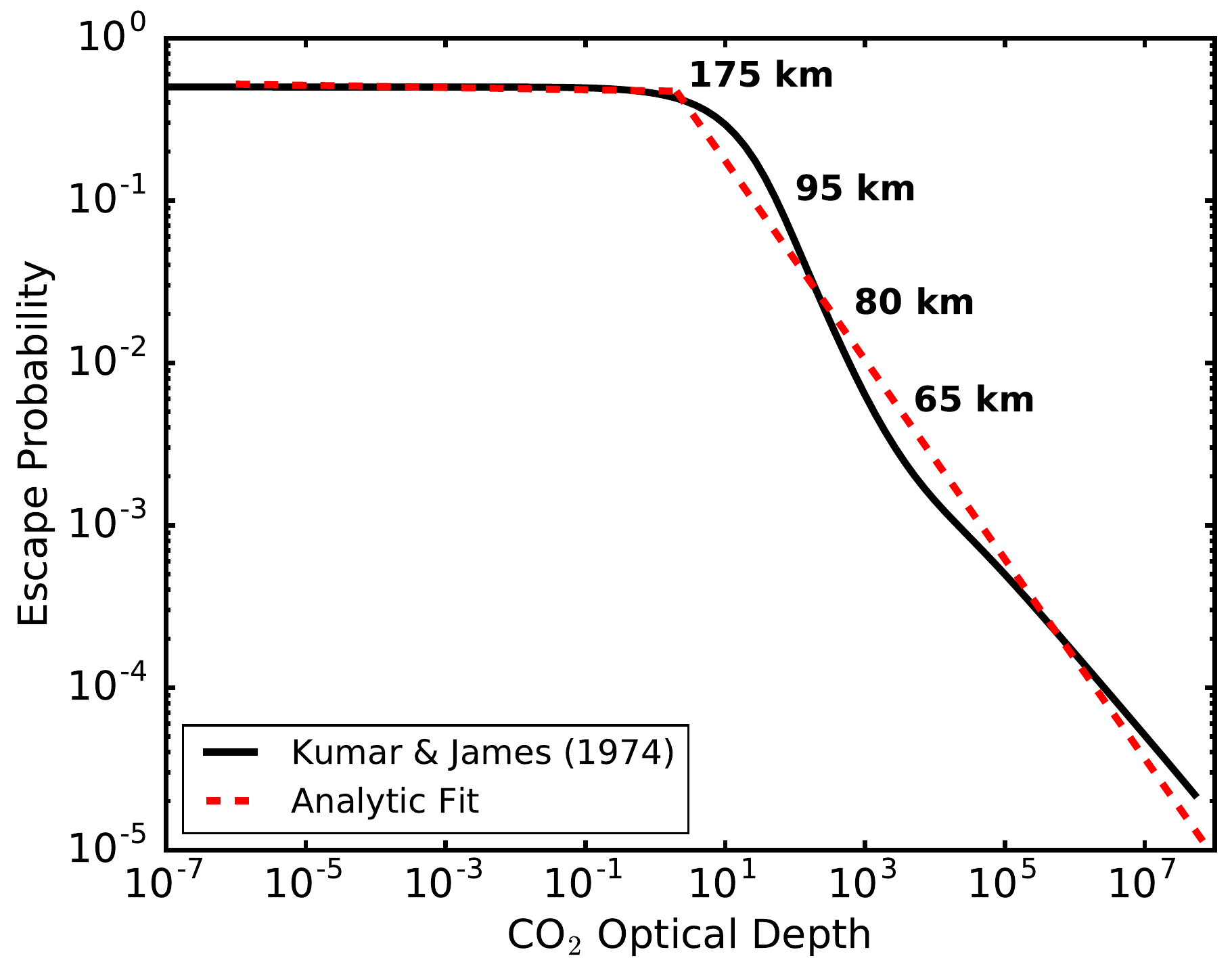}
\includegraphics[trim = 0mm 0mm 0mm 0mm, clip=true,width=0.49\textwidth]{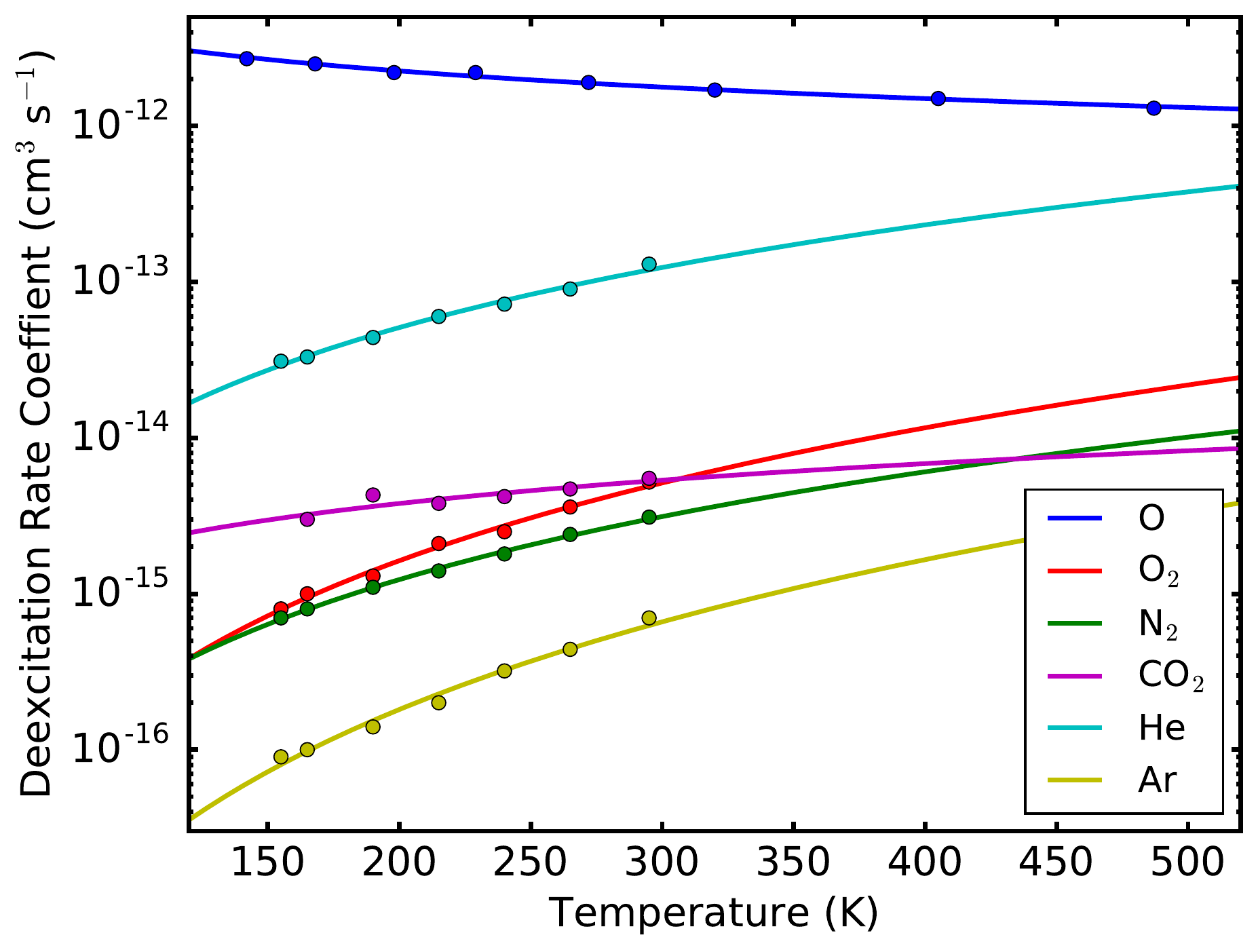}
\caption{
Figures showing the escape probability of a 15~$\mu$m photon as a function of CO$_2$ optical depth (\emph{upper-panel}) and the CO$_2$ deexcitation rate coefficients as a function of temperature (\emph{lower-panel}).
In the upper panel, the numbers show the approximate altitudes in our current Earth model where these points on the line are reached. 
In the lower panel, the data points are measurements from \citet{Siddles94} and \citet{Castle12}, and the solid lines are our power-law fits, given in Table.~\ref{table:CO2coefficients}.
}
\label{fig:CO2coolingstuff}
\end{figure}

\begin{table*}
\centering
\begin{tabular}{|c|cccccc|}
\hline
Species & O & O$_2$ & N$_2$ & CO$_2$ & He & Ar \\
\hline
A & $5.10 \times 10^{-11}$ & $4.97 \times 10^{-22}$ & $6.43 \times 10^{-21}$ & $4.21 \times 10^{-17}$ & $4.73 \times 10^{-19}$ & $8.13 \times 10^{-24}$ \\ 
B & -0.59 & 2.83 & 2.30 & 0.85 & 2.19 & 3.19 \\
\hline
\end{tabular}
\caption{
Table giving our best fit parameters for the CO$_2$ deexcitation rate coefficients, as shown in Fig~\ref{fig:CO2coolingstuff}.
The fit equation is \mbox{$k_{d,M} (T_\mathrm{n}) = A T_\mathrm{n}^{B}$}, where $k_{d,M}$ and $T_\mathrm{n}$ have units of cm$^3$~s$^{-1}$ and K respectively.
}
\label{table:CO2coefficients}
\end{table*}

We consider the effects of IR cooling by CO$_2$, H$_2$O, NO, and O.
In all cases, cooling happens when atoms/molecules are excited by collisions with other particles and then radiate the energy away before they are deexcited by further collisions.
Collisions cause there to be a continuous transfer of energy from the atmosphere's thermal energy reservoir to the various forms of energy within the individual atoms and molecules, and a corresponding transfer of energy back to the thermal energy reservoir. 
However, due to radiative relaxation (i.e. spontaneous/stimulated emission) and the loss of many of the emitted photons to space, the rate at which energy is transferred back to the thermal reservoir is reduced, and the resulting inbalance is the cooling that we are interested in. 
The calculation of the cooling rates is seldom trivial and ideally would involve calculating the full transport of the emitted IR spectrum through the atmosphere and tracking the populations of each of the various excited states in the relevant species (e.g. see \mbox{\citealt{Wintersteiner92}}).
In this paper, we take into account all of these processes in a simpler way and aim to implement more sophisticated treatments of cooling in future studies.

Cooling by CO$_2$ is dominated by emission at 15~$\mu$m.
We use the cool-to-space approximation (e.g. \citealt{Dickinson72}); the fundamental assumption is that the cooling at each point is caused entirely by emitted photons that escape directly to space. 
Ignoring stimulated emission, this means
\begin{equation} \label{eqn:}
Q_{\mathrm{CO}_2} = (h\nu)_{15\mu \mathrm{m} } A_{10} [ \mathrm{CO}_2^* ] \epsilon ,
\end{equation}
where  \mbox{$(h\nu)_{15\mu \mathrm{m} }$} is the energy of a single 15~$\mu$m photon (\mbox{$= 1.325 \times 10^{-13}$~erg}), $A_{10}$ is the Einstein coefficient for spontaneous emission, $[ \mathrm{CO}_2^* ]$ is the density of excited CO$_2$ molecules, and $\epsilon$ is the probability that a 15~$\mu$m photon emitted from a given point escapes to space. 
The definition of $\epsilon$ is such that it takes into account the fact that photons emitted downwards are not lost, and therefore approaches a maximum of 0.5 at high altitudes. 

In order to calculate \mbox{$[ \mathrm{CO}_2^* ]$}, we consider three excitation and two deexcitation mechanisms.
The excitation mechanisms are collisional excitation, the absorption of 15~$\mu$m photons previously emitted by excited CO$_2$ molecules, and the absorption of photons from the host star's IR spectrum. 
For the second process, the fundamental simplifying assumption is that all photons that are emitted and are not lost to space are reabsorbed locally where the emission took place. 
The excitation rate is given by
\begin{equation} \label{eqn:CO2excite}
\begin{aligned}
\frac{ d [\mathrm{CO}_2^*] }{ dt } 
= 
\left( \sum\limits_M k_{e,M} [M] \right) & \left( [\mathrm{CO}_2] - [\mathrm{CO}_2^*] \right) \\
 & +
A_{10} [ \mathrm{CO}_2^* ] \left( 1 - \epsilon \right)
+
S_\mathrm{IR} ,
\end{aligned}
\end{equation}
where the sum is over all species that collisionally excite CO$_2$, $k_{e,M}$ is the rate coefficient for collisional excitation, and $S_\mathrm{IR}$ is the additional excitation term due to the absorption of stellar IR radiation given by Eqn.~\ref{eqn:infraredCO2excitation}.
The term \mbox{$[\mathrm{CO}_2] - [\mathrm{CO}_2^*]$} is the density of non-excited CO$_2$ molecules. 
The deexcitation mechanisms are collisional deexcitation and radiative relaxation.
The deexcitation rate is given by
\begin{equation} \label{eqn:CO2deexcite}
- \frac{ d [\mathrm{CO}_2^*] }{ dt } 
= 
\left( \sum\limits_M k_{d,M} [M] \right) [\mathrm{CO}_2^*] 
+
A_{10} [ \mathrm{CO}_2^* ]  .
\end{equation}
Where the first term on the RHS dominates, the atmosphere is in the local thermodynamic equilibrium (LTE) regime, and where the two terms are similar, or the second term dominates, the atmosphere is in the non-local thermodynamic equilibrium (non-LTE) regime.
Assuming a steady state, Eqns.~\ref{eqn:CO2excite} and \ref{eqn:CO2deexcite} add up to zero, giving 
\begin{equation} \label{eqn:CO2exciteddensity}
[ \mathrm{CO}_2^* ]
=
\frac{
\sum\limits_M k_{e,M} [M] [\mathrm{CO}_2] + S_\mathrm{IR}
}{
\sum\limits_M \left( k_{e,M} + k_{d,M} \right) [M] + A_{10} \epsilon
} .
\end{equation}
The Einstein coefficient, $A_{10}$, is 0.46~s$^{-1}$ (\citealt{curtis1956thermal}), and the rate coefficients are related by \mbox{$k_{e,M} = 2 k_{d,M} \exp{\left(-667/T_\mathrm{n} \right)}$} (\citealt{Castle06}).
For the escape probabilities, we use the tabulated values given by \citet{KumerJames74} which depend entirely on the amount of CO$_2$ above the considered altitude, $z$, given by \mbox{$N_\mathrm{CO_2} = \int_z^\infty [\mathrm{CO}_2] dz$}.
We fit their tabulated values with 
\begin{equation} \label{eqn:15micronescape}
\epsilon 
=
\begin{cases}
0.7202 \left( \sigma N_\mathrm{CO_2} \right)^{-0.613} & \mathrm{if} \hspace{1mm}  \sigma N_\mathrm{CO_2} > 2 , \\
0.4732 \left( \sigma N_\mathrm{CO_2} \right)^{-0.0069} & \mathrm{if} \hspace{1mm}  \sigma N_\mathrm{CO_2} < 2 . \\
\end{cases}
\end{equation}
where \mbox{$\sigma =  6.43\times10^{-15}$~cm$^2$}.
The dependence of $\epsilon$ on $\sigma N_\mathrm{CO_2}$ is shown in Fig.~\ref{fig:CO2coolingstuff}.
For the collisional excitation/deexcitation rate, we consider the influences of O, O$_2$, N$_2$, CO$_2$, He, and Ar. 
For O, we use the experimentally measured values of $k_{d,M}$ given by \citet{Castle12}, and for the other species, we use the measured values given by \citet{Siddles94}.
In all cases, $k_{d,M}$ have temperature dependences that we fit using power-laws of the form \mbox{$k_{d,M} (T_\mathrm{n}) = A T_\mathrm{n}^{B}$}, where the values of $A$ and $B$ are given in Table~\ref{table:CO2coefficients}.
In Fig.~\ref{fig:CO2coolingstuff}, we show the measured deexcitation rates and our analytic fit formulae for each species. 
A signficant worry with our fit formulae for $k_{d,M}$ is that all of the measurements that we use are for low gas temperatures, and therefore our fit formulae might be inaccurate at high temperatures.
This problem is not likely to influence our results in this paper since CO$_2$ cooling is only significant in regions of the atmospheres that are within the experimental temperature ranges.

For NO cooling, we use the model given by \citet{Oberheide13}.
We consider emission in the vibrational band at 5.3~$\mu$m assuming two excitation mechanisms: these are collisional excitation by O atoms and radiative pumping by earthshine.
We assume that all photons emitted by NO molecules escape to space (i.e. \mbox{$\epsilon = 1$}), which is realistic for the Earth since NO cooling is only significant in the thermosphere (\citealt{Kockarts80}). 
The cooling rate is given by
\begin{equation} \label{eqn:}
Q_{\mathrm{NO}} = (h\nu)_{5.3\mu \mathrm{m} } A_{10} [ \mathrm{NO}^* ] ,
\end{equation}
where \mbox{$(h\nu)_{5.3\mu \mathrm{m} } = 3.75 \times 10^{-13}$~erg} and \mbox{$[ \mathrm{NO}^* ]$} is the density of excited NO molecules, given by
\begin{equation} \label{eqn:}
[ \mathrm{NO}^* ]
=
\frac{
k_{e,O} [O] + S_\mathrm{E}
}{
\left( k_{e,O} + k_{d,O} \right) [O] + S_\mathrm{E} + A 
} [\mathrm{NO}],
\end{equation}
where $S_\mathrm{E}$ is the excitation rate due to earthshine, $k_{e,O}$ and $k_{d,O}$ are the collisional excitation and deexcitation rate coefficients, and $A$ is the Einstein coefficient for spontaneous emission.
As in \citet{Oberheide13}, we use \mbox{$S_\mathrm{E} = 1.06\times10^{-4}$~s$^{-1}$}, \mbox{$k_{d,O} = 2.8\times10^{-11}$~cm$^{3}$~s$^{-1}$}, \mbox{$A = 12.54$~s$^{-1}$}, and \mbox{$k_{e,O} = k_{d,O} \exp{\left( -2700/T_\mathrm{n} \right)}$}.

For O cooling, we consider emission at 63~$\mu$m and 147~$\mu$m using the parameterization derived by \citet{bates1951temperature} (see Eqn.~14.57 and Eqn.~14.58 of \citealt{banks1973aeronomy}) given by
\begin{equation} \label{eqn:}
Q_\mathrm{O} = Q_{\mathrm{O},63\mu\mathrm{m}} + Q_{\mathrm{O},147\mu\mathrm{m}} ,
\end{equation}
where
\begin{equation} \label{eqn:}
Q_{\mathrm{O},63\mu\mathrm{m}} 
= 
\frac{
1.67 \times 10^{-18} \exp\left( -228/T_\mathrm{n} \right) [\mathrm{O}]
}{
1 + 0.6 \exp\left( -228/T_\mathrm{n} \right) + 0.2 \exp\left( -326/T_\mathrm{n} \right)
},
\end{equation}
\begin{equation} \label{eqn:}
Q_{\mathrm{O},147\mu\mathrm{m}} 
= 
\frac{
4.59 \times 10^{-20} \exp\left( -326/T_\mathrm{n} \right) [\mathrm{O}]
}{
1 + 0.6 \exp\left( -228/T_\mathrm{n} \right) + 0.2 \exp\left( -326/T_\mathrm{n} \right)
},
\end{equation}
where [O] is in cm$^{-3}$, $T_\mathrm{n}$ is in K, and the cooling rates are in erg~s$^{-1}$~cm$^{-3}$.
More sophisticated modelling  of O cooling will be used in future models. 

For cooling by H$_2$O, we use the parametitzation for emission in rotational bands by \citet{Hollenbach79} and summarized in \citet{Kasting83} (see their Eqns.~32--38).
In this model, H$_2$O is excited by collisions with H atoms only.
Given the length of the set of equations involved, we do not write them here.

\subsubsection{Conduction} 

In the Earth's upper thermosphere, cooling of the neutral gas is not strong enough to balance heating, and a steady state is only reached because conduction downwards into the cooler lower thermosphere removes this excess energy.
Since the temperatures of the neutrals, ions, and electrons are evolved separately, separate conductivities must be used for each of these components. 
For ion and electron conductivities, we ignore the effects of the magnetic field, which reduces the conduction in directions perpendicular to the magnetic field.
The conduction equations are solved using the implicit Crank-Nicolson method, as described in Appendix~\ref{appendix:conduction}.

For the neutral gas, we consider eddy conduction, which is dominant in the lower atmosphere, and molecular conduction, which is dominant in the upper atmosphere.
The neutral conduction equation is 
\begin{equation} \label{eqn:neutralconduction}
\frac{\partial e_\mathrm{n}}{\partial t} = \frac{1}{r^2} \frac{\partial}{\partial r} \left[ r^2 \kappa_\mathrm{mol} \frac{\partial T_\mathrm{n}}{\partial r} + r^2 \kappa_\mathrm{eddy} \left( \frac{\partial T_\mathrm{n}}{\partial r} + \frac{g}{c_\mathrm{P}} \right) \right] ,
\end{equation}
where $\kappa_\mathrm{mol}$ is the molecular conductivity, $\kappa_\mathrm{eddy}$ is the eddy conductivity, $g$ is the gravitational acceleration, and $c_\mathrm{P}$ is the specific heat at constant pressure. 
The term $g/c_\mathrm{P}$ for the eddy conduction is the adiabatic lapse rate.
The eddy conductivity is related to the eddy diffusion coefficient by \mbox{$\kappa_\mathrm{eddy} = \rho c_\mathrm{P} K_\mathrm{E}$} (\citealt{Hunten74}).
The molecular conductivity is dependent on the temperature and composition of the gas. 
We estimate $\kappa_\mathrm{mol}$ using the equations given in Section~14.3 of \citet{banks1973aeronomy} with some minor simplifications.
The molecular conductivity of the $k$th species is given by
\begin{equation} \label{eqn:molcondcomponent}
\kappa_k = A_k T_\mathrm{n}^{s_k} ,
\end{equation}
where $A_k$ and $s_k$ are coefficients that depend on the species.
We assume the total conductivity of the gas is given by
\begin{equation} \label{eqn:molcond}
\kappa_\mathrm{mol} = \sum\limits_k \frac{ n_k \kappa_k }{ \sum\limits_j n_j \phi_{kj} } ,
\end{equation}
where
\begin{equation} \label{eqn:molcondphi}
\phi_{kj} = \frac{ 
\left[ 1 + \left( \kappa_k / \kappa_j \right)^{\frac{1}{2}} \left( m_j / m_k \right)^{\frac{1}{4}} \right]^2
}{
2 \sqrt{2} \left[ 1 +  \left( m_j / m_k \right) \right]^{\frac{1}{2}}
} ,
\end{equation}
where $m_k$ is the molecular mass of the $k$th species.
The sums in Eqn.~\ref{eqn:molcond} should be over all neutral species, but in reality we only consider species for which we have $A_k$ and $s_k$ values.
The species we consider are N$_2$, O$_2$, CO$_2$, CO, O, He, H, and Ar, with values for $A_k$ and $s_k$ taken from Table~13 of \citet{BauerLammerBook} for Ar, and Table~10.1 of \citet{schunk2000ionospheres} for the others. 

For the ions, the conduction equation is
\begin{equation} \label{eqn:ionconduction}
\frac{\partial e_\mathrm{i}}{\partial t} = \frac{1}{r^2} \frac{\partial}{\partial r} \left[ r^2 \kappa_\mathrm{i} \frac{\partial T_\mathrm{i}}{\partial r}  \right],
\end{equation}
where $\kappa_\mathrm{i}$ is the ion conductivity.
For $\kappa_\mathrm{i}$, we use Eqn.~22.122 from \citet{banks1973aeronomy} which expresses the conductivity of ion gases made of a single ion species as \mbox{$4.6\times10^4 A^{-1/2} T_\mathrm{i}^{5/2}$} eV~cm$^{-1}$~s$^{-1}$~K$^{-1}$, where $A$ is the atomic mass of the species and $T_\mathrm{i}$ is the ion temperature. 
For a gas mixture, they recommend using a density weighted average thermal conductivity, so we adopt the form
\begin{equation} \label{eqn:ionconductivity}
\kappa_\mathrm{i} = 4.6\times10^4 \frac{ \sum_k n_k A_k^{-\frac{1}{2}} }{\sum_k n_k} T_\mathrm{i}^{\frac{5}{2}},
\end{equation}
where the sums are over all ion species and $n_k$ is in cm$^{-3}$, $T_\mathrm{i}$ is the ion temperature in K, and $\kappa_\mathrm{i}$ is in eV~cm$^{-1}$~s$^{-1}$~K$^{-1}$.

For the thermal electron gas, the conduction equation is the same as Eqn.~\ref{eqn:ionconduction} with the subscript i replaced by e. 
We calculate the electron conductivity using Eqn.~22.116 of \citet{banks1973aeronomy}:
\begin{equation} \label{eqn:electronconductivity}
\kappa_\mathrm{e} = \frac{ 7.7 \times 10^{5} T_\mathrm{e}^{5/2} }{1 + 3.22 \times 10^4 \left( T_\mathrm{e}^2 / n_\mathrm{e} \right) \sum\limits_k n_k \bar{Q}_{\mathrm{D},k}  } ,
\end{equation}
where $\bar{Q}_{\mathrm{D},k}$ is the average momentum transfer cross-section of the $k$th species. 
The sum in the denominator should technically be over all neutral species, but in reality only the main species contribute significantly.
In this sum, we take into account the effects of N$_2$, O$_2$, O, H, and He using the temperature dependent equations for $\bar{Q}_{\mathrm{D},k}$ given in Table~9.2 of \citet{banks1973aeronomy}.
The numerator in Eqn.~\ref{eqn:electronconductivity} is the electron conductivity of a fully ionized gas, and the denominator corrects for the reduction in conductivity caused by collisions with neutrals reducing the mean free paths of thermal electrons.

\subsubsection{Energy Exchange} \label{sect:heatexchange}

The neutral, ion, and electron gases exchange energy by collisions.
In our model, the electrons lose energy only by collisions with neutrals and ions.
The energy gained by the ions is then given to the neutrals by further collisions, which is the most important neutral heating mechanism  in the upper thermosphere. 
The energy exchange equations are solved using the implicit Crank-Nicolson method, as described in Appendix~\ref{appendix:heatexchange}.

For the electron-ion energy exchange, we take into account elastic Coulomb collisions only and the total energy exchange rate is calculated by summing over the rates for individual ion species. 
The basic equation is 
\begin{equation} \label{eqn:eiheatexchange}
Q_\mathrm{ei} = - 3 k_\mathrm{B} \left( T_\mathrm{e} - T_\mathrm{i} \right) \sum_k \frac{n_k m_k \nu_{\mathrm{e}k}}{m_\mathrm{e} + m_k}   ,
\end{equation}
where the sum is over all ion species and $\nu_{\mathrm{e}k}$ is the momentum transfer collision frequency between electrons and the $k$th ion species, 
This equation, derived by \mbox{\citet{Schunk75}}, requires several assumptions, including that the temperature difference between the electrons and ions are small. 
The definition of $Q_\mathrm{ei}$ is such that a positive value means energy is taken from the ions and given to the electrons. 
To calculate $\nu_{\mathrm{e}k}$ for a given ion, we use \mbox{$\nu_{\mathrm{e}k} = 54.5 n_k Z_k^2 / T_\mathrm{e}^{3/2}$}, where $Z_k$ is the charge of the ion (see Section~4.8 of \citealt{schunk2000ionospheres}). 

The total ion-neutral energy exchange rate is the sum of the exchange rates of individual species pairs.
The equation for the energy exchange rate is 
\begin{equation} \label{eqn:inheatexchange}
Q_\mathrm{in} = - 3 k_\mathrm{B} \left( T_\mathrm{i} - T_\mathrm{n} \right) \sum_n \sum_k \frac{n_i m_i \nu_{in}}{m_i + m_n}   ,
\end{equation}
where the subscripts $n$ and $i$ are for the $n$th neutral and $i$th ion species.
The definition of $Q_\mathrm{in}$ is such that a positive value means energy is taken from the neutrals and given to the ions. 
The neutrals that we consider in the sums are H, He, N, O, CO, N$_2$, O$_2$, and CO$_2$; the ions that we consider are H$^+$, He$^+$, C$^+$, N$^+$, O$^+$, CO$^+$, N$_2^+$, NO$^+$, O$_2^+$, and CO$_2^+$.
The ion-neutral heat exchange, described in detail in \citet{schunk2000ionospheres}, is dominated by two types of interactions: resonant and non-resonant interactions which dominate at high (>300~K) and low temperatures respectively. 
The resonant interactions happen when a neutral approaches its ion equivalent (e.g. O and O$^+$) and charge exchanges with it, with the changing identities of the particles representing a net energy exchange between the ion and neutral gases.
We use the temperature-dependent equations for the momentum transfer collision frequencies, $\nu_{in}$, for individual ion-neutral pairs given in Table~4.5 of \citet{schunk2000ionospheres}.
Non-resonant interactions involve neutral species and dissimilar ions.
For these interactions, $\nu_{in}$ can be described simply by \mbox{$\nu_{in} = C_{in} n_{n}$}, where we use the coefficients $C_{in}$ for individual ion-neutral pairs given in Table.~4.4 of \citet{schunk2000ionospheres}.

Several important mechanisms exist that cause thermal electrons to lose energy to the neutral gas. 
Electron-neutral heat exchange is very important for the electron temperature structure in the low thermosphere of the Earth, and normally proceeds through inelastic collisions that excite neutral atoms or molecules.
The most important of these processes, at least for the current Earth, is inelastic collisions that cause fine structure transitions in ground state atomic oxygen; for this process, we use the scaling laws derived by \citet{Hoegy76}. 
We also consider the excitation of ground state oxygen to the O($^1$D) excited state. 
In addition, we take into account energy exchange from electron collisions with neutral molecules that cause the exitation of rotational or vibrational modes in the molecule; for these, we consider collisions with N$_2$, O$_2$, H$_2$, CO$_2$, CO, and H$_2$O.
We use the scaling laws for these processes that are convieniently listed in Section~9.7 of \citet{schunk2000ionospheres}.

\begin{figure}
\includegraphics[trim = 0mm 0mm 0mm 0mm, clip=true,width=0.49\textwidth]{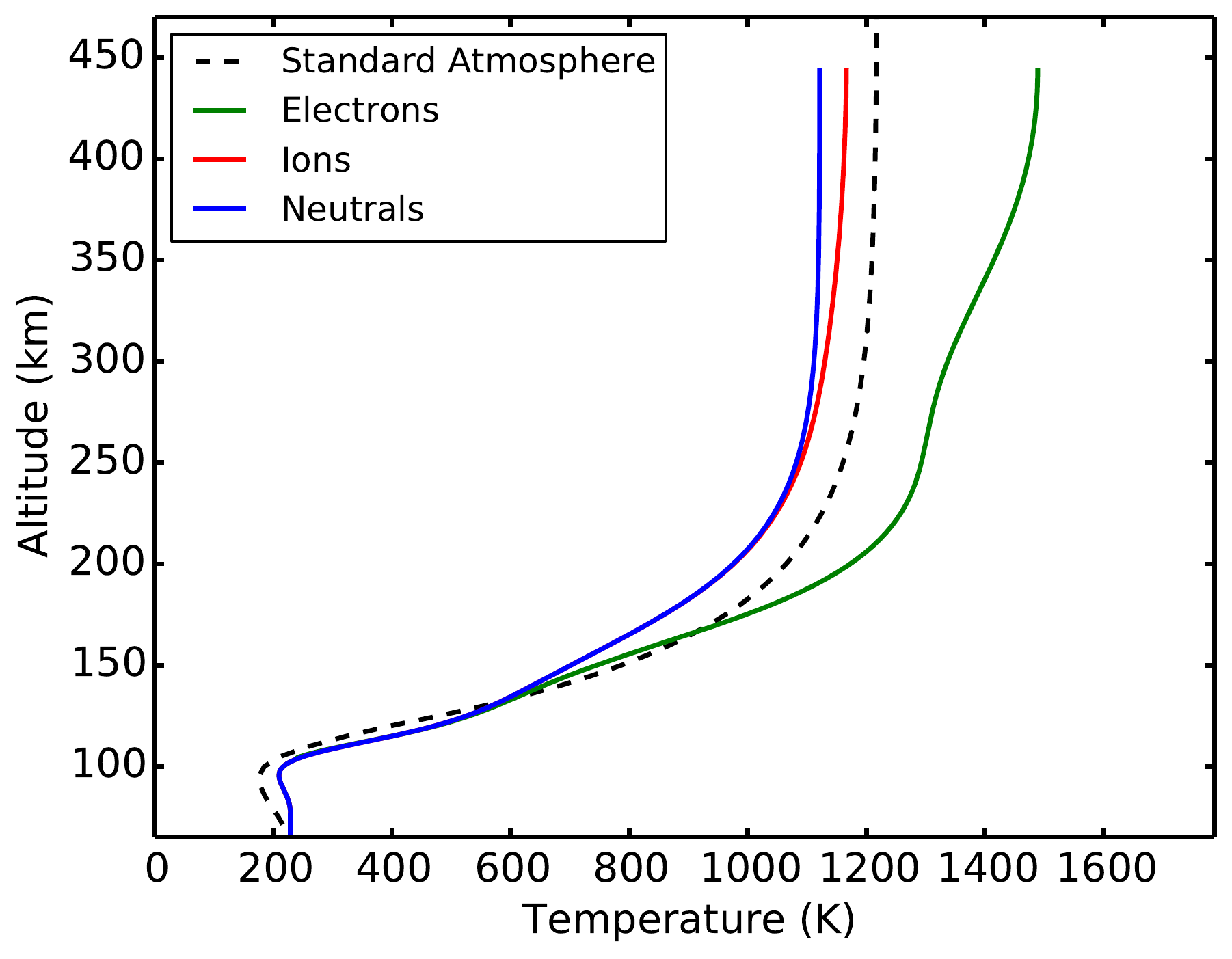}
\caption{
Figures showing the neutral, ion, and electron temperature structures of our current Earth model.
The dashed black line is for the empirical NRLMSISE-00 model. 
}
\label{fig:earthtemp}
\end{figure}

\begin{figure} [!h]
\includegraphics[trim = 0mm 0mm 0mm 0mm, clip=true,width=0.42\textwidth]{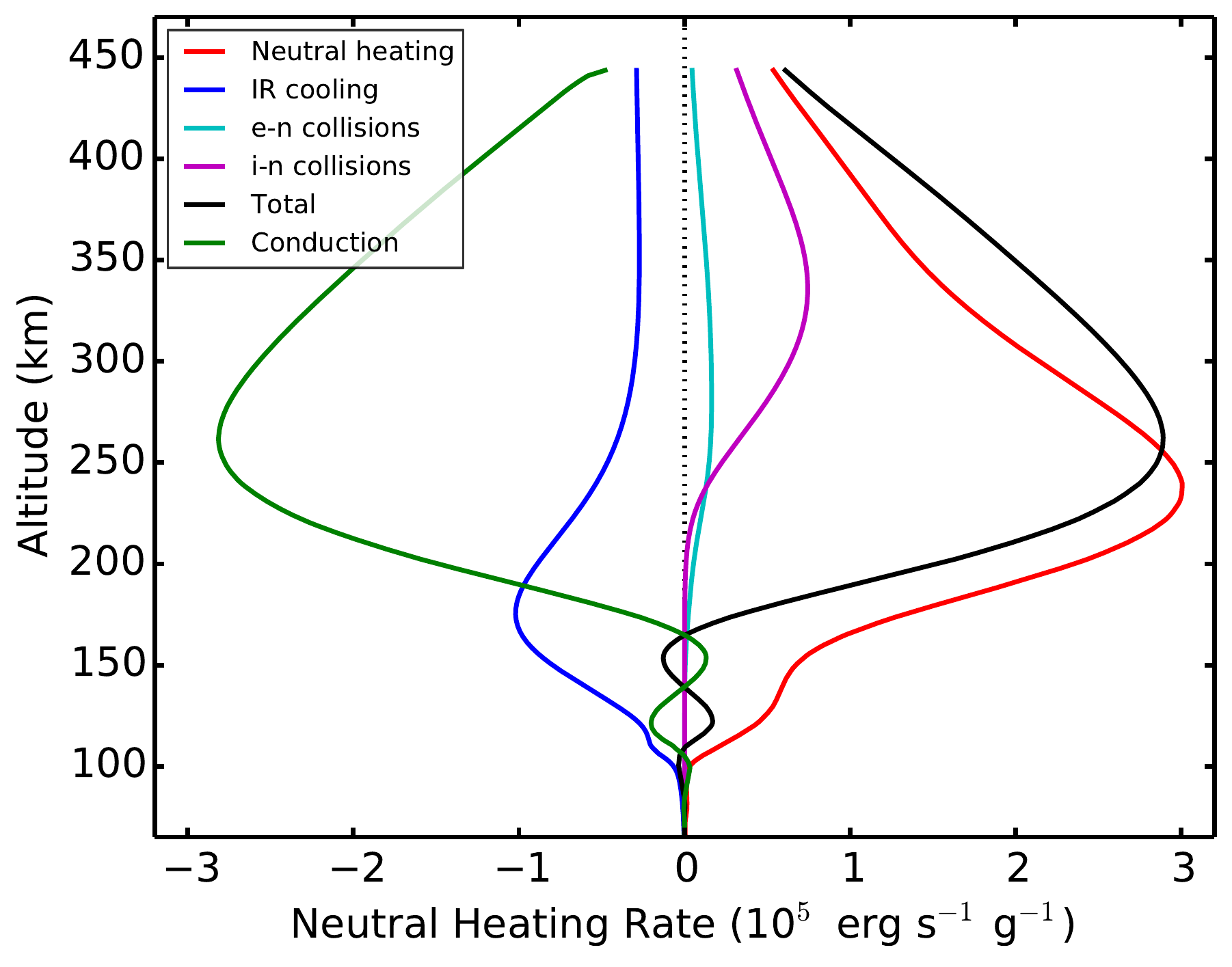}
\includegraphics[trim = 0mm 0mm 0mm 0mm, clip=true,width=0.42\textwidth]{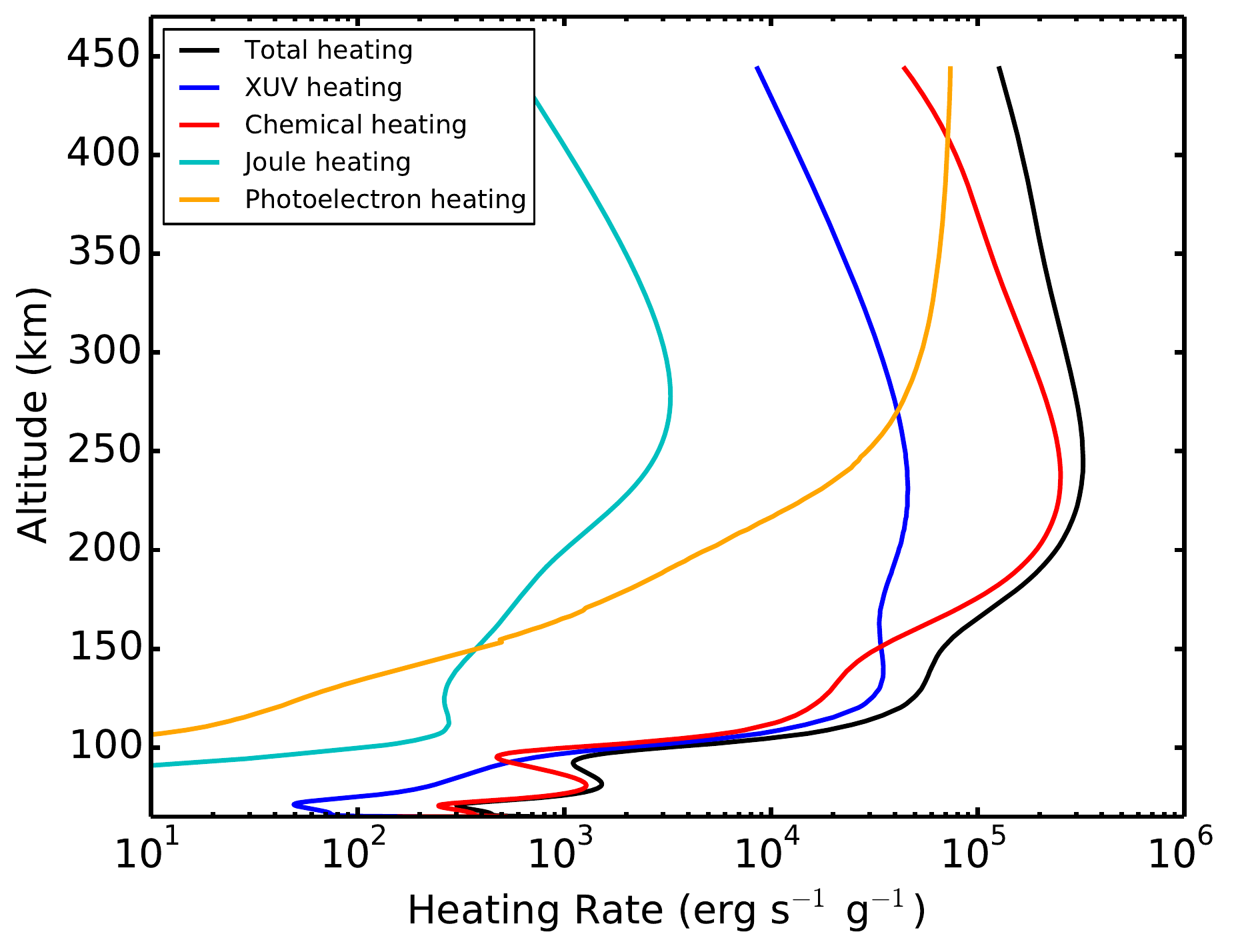}
\includegraphics[trim = 0mm 0mm 0mm 0mm, clip=true,width=0.42\textwidth]{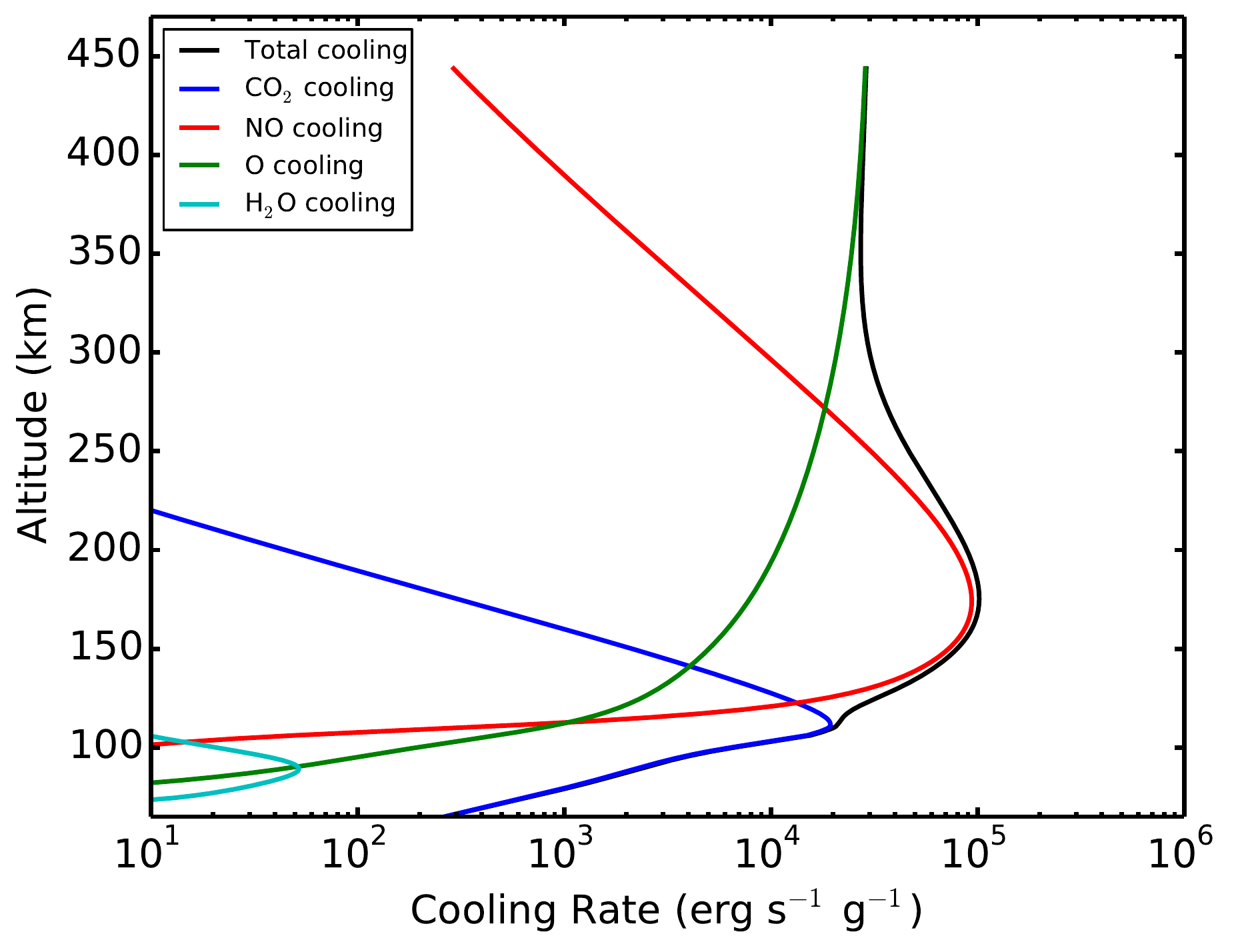}
\caption{
Figures showing the thermal processes for our Earth model.
In the upper panel, the red line shows the total XUV, chemical, and Joule heating, the blue line shows the total IR cooling, the cyan and magenta lines show the neutral heating by collisions with electron and ions, the black line shows the sum of all of these, and the green line shows the effects conduction. 
In the middle and lower panels, we show the contributions of individual mechanisms to the total heating and cooling.
}
\label{fig:earthheatcool}
\end{figure}

\begin{figure}
\includegraphics[trim = 0mm 0mm 0mm 0mm, clip=true,width=0.45\textwidth]{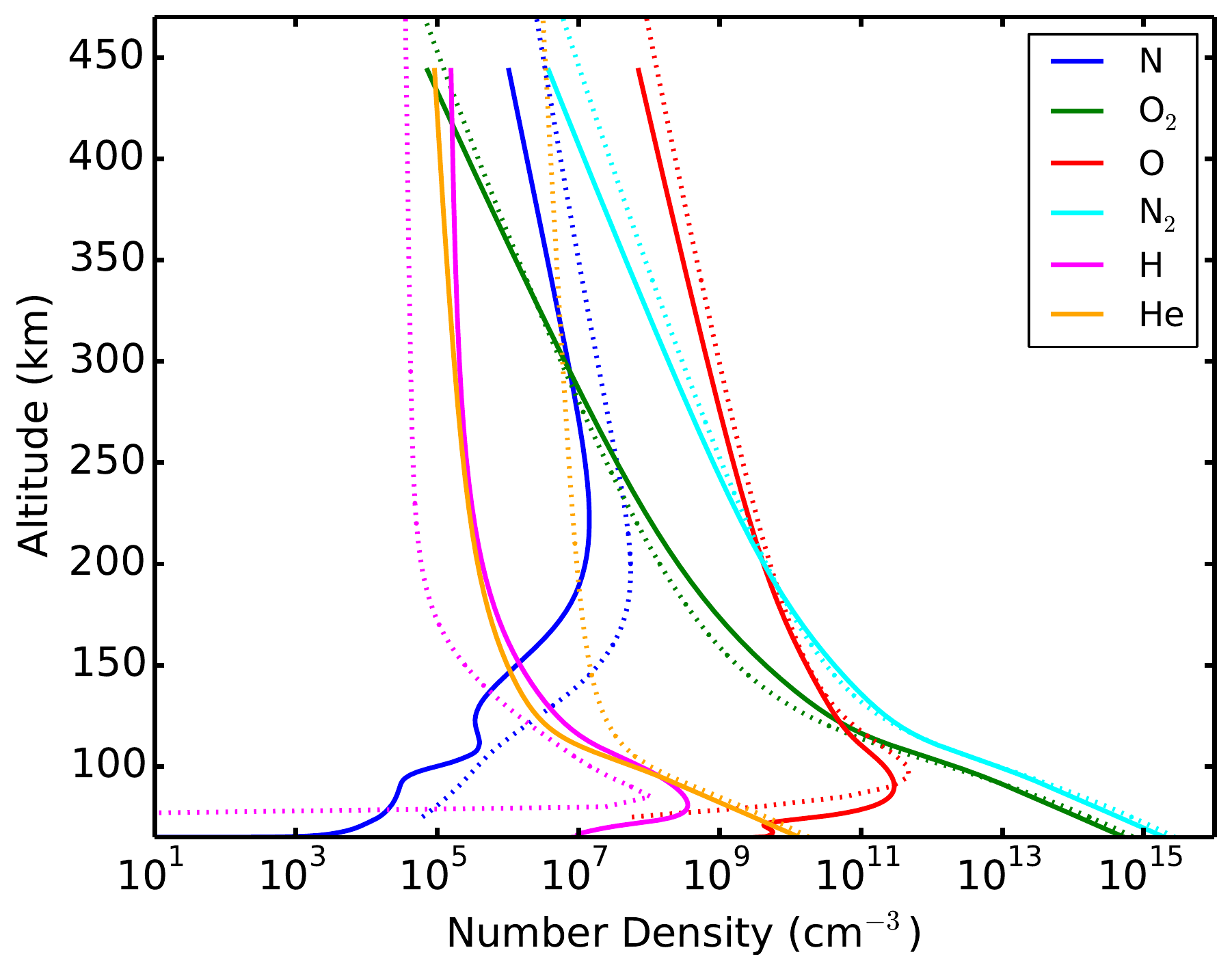}
\includegraphics[trim = 0mm 0mm 0mm 0mm, clip=true,width=0.45\textwidth]{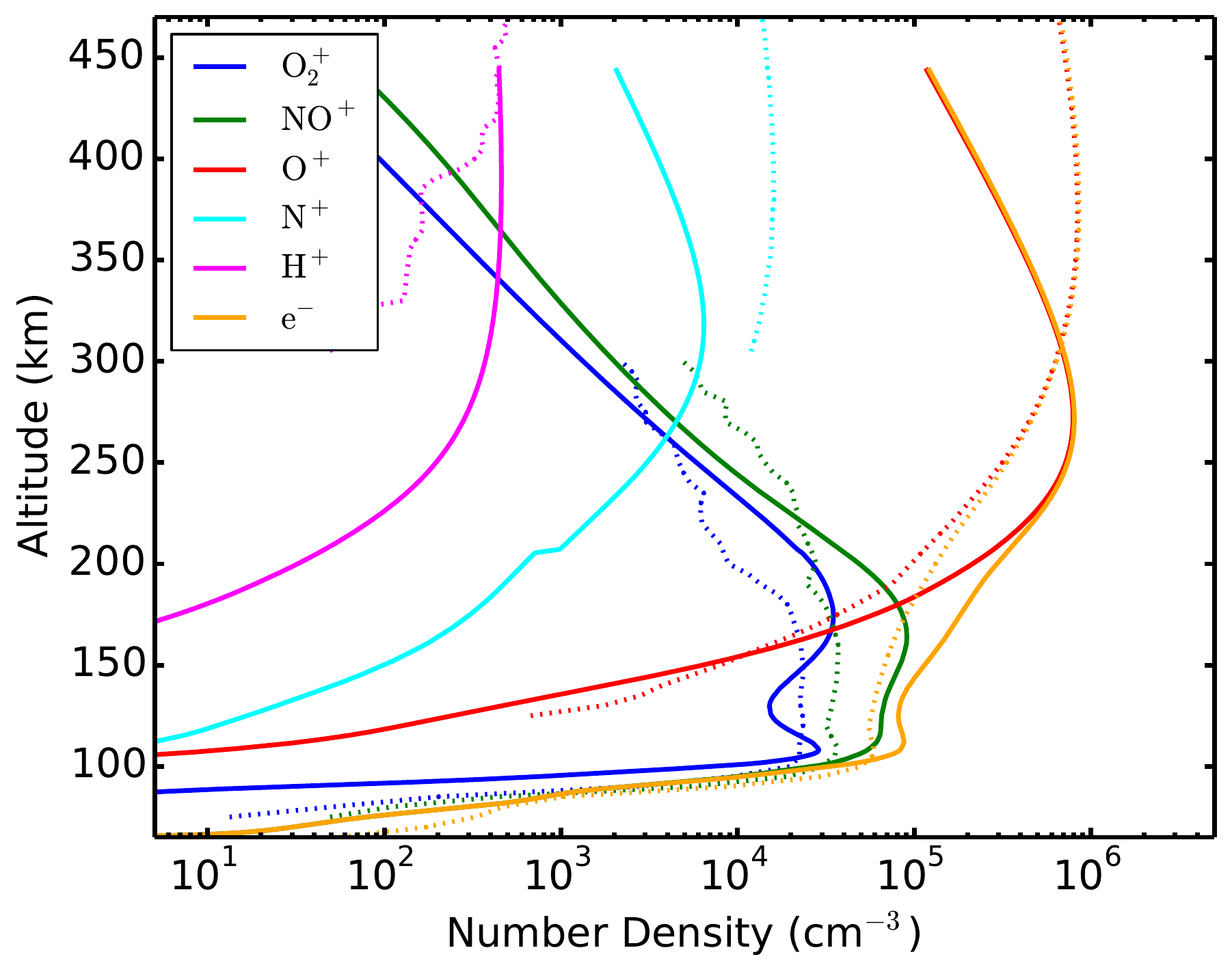}
\caption{
Figures showing the density structures of several important neutral (\emph{upper-panel}) and ion (\emph{lower-panel}) species.
The solid lines show the predictions of our model and the dotted lines are from the empirical NRLMSISE-00 model for the neutrals and the IRI-2007 model for the ions.
}
\label{fig:earthchem}
\end{figure}

\begin{figure*}
\includegraphics[trim = 0mm 0mm 0mm 0mm, clip=true,width=0.49\textwidth]{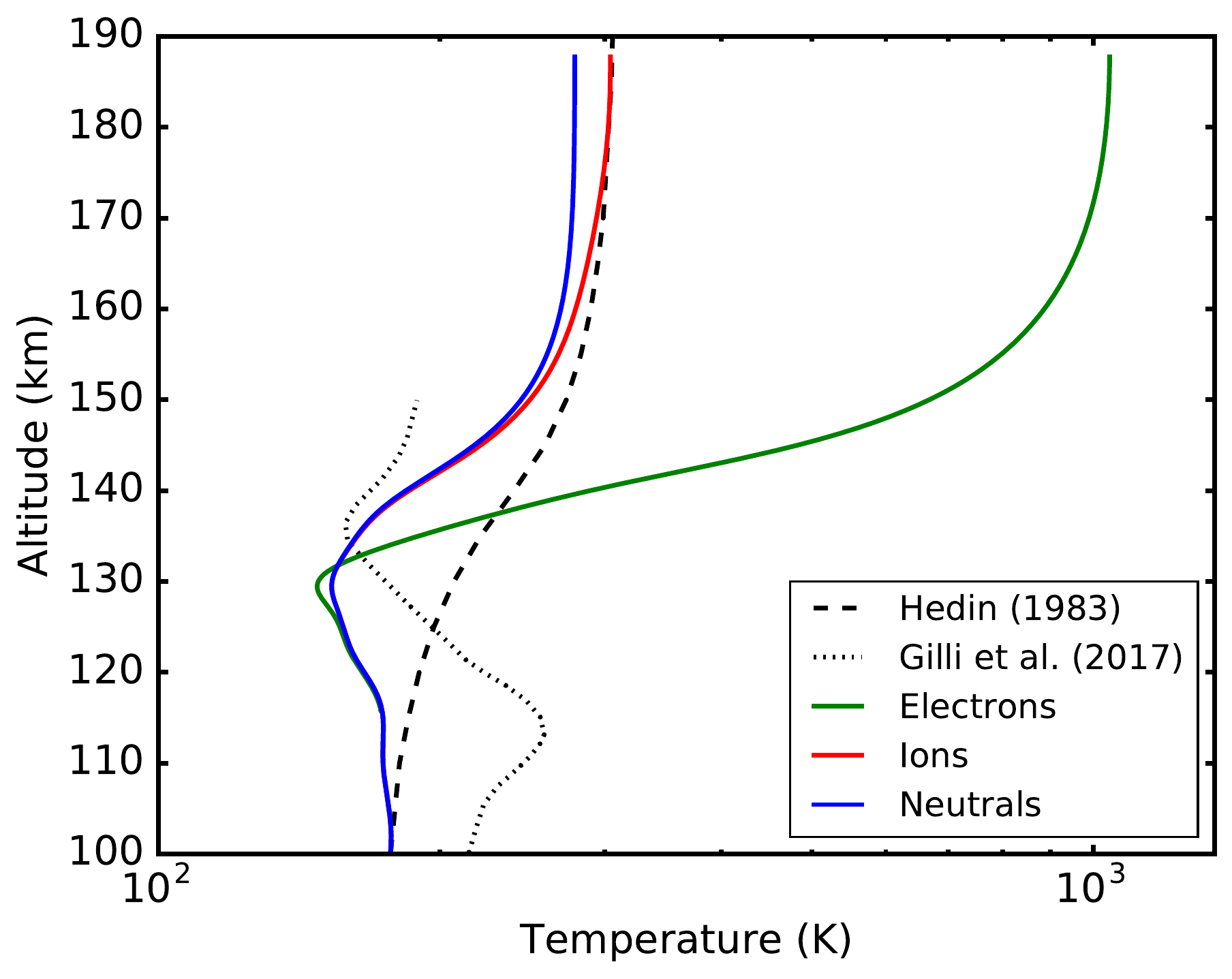}
\includegraphics[trim = 0mm 0mm 0mm 0mm, clip=true,width=0.49\textwidth]{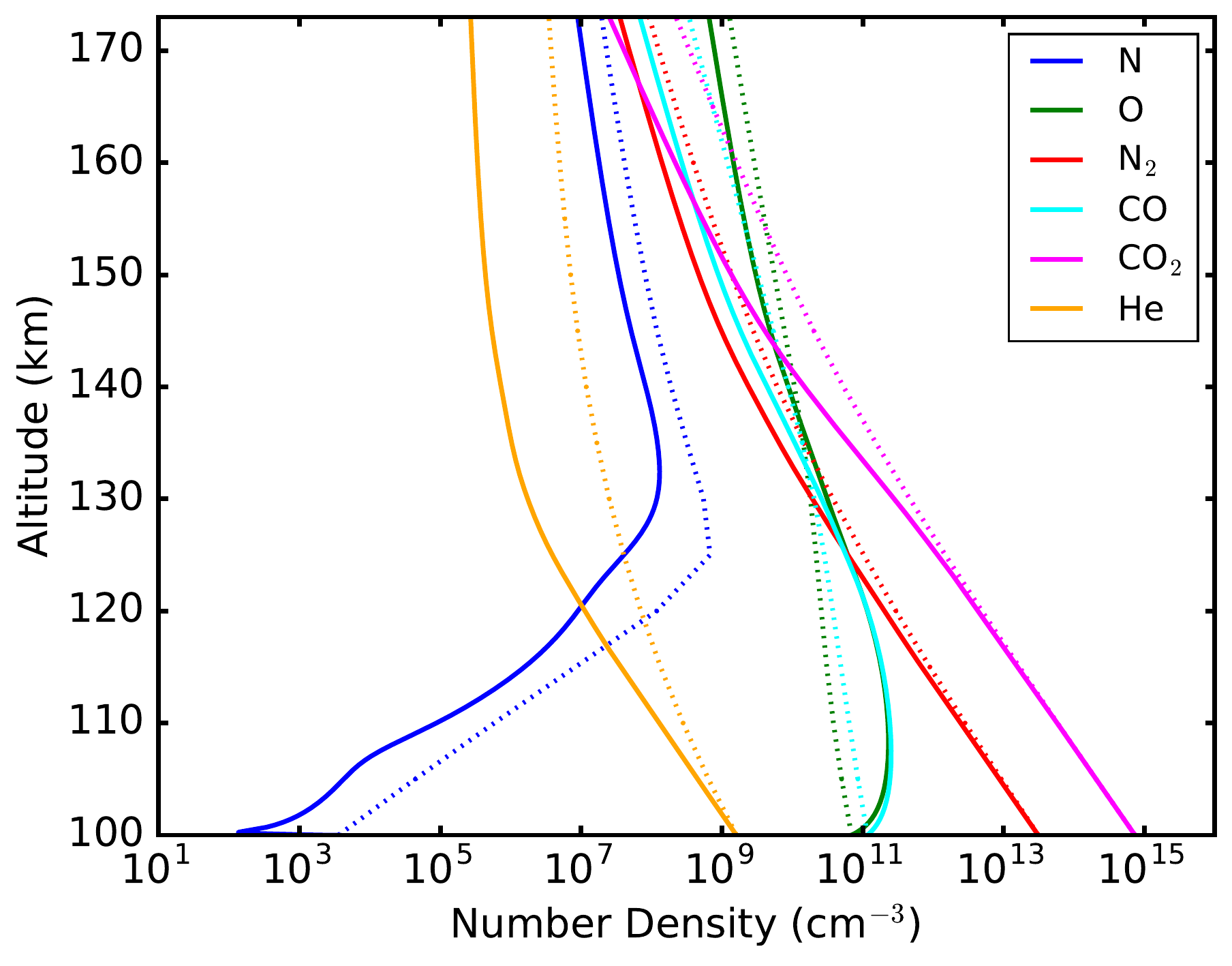}
\includegraphics[trim = 0mm 0mm 0mm 0mm, clip=true,width=0.49\textwidth]{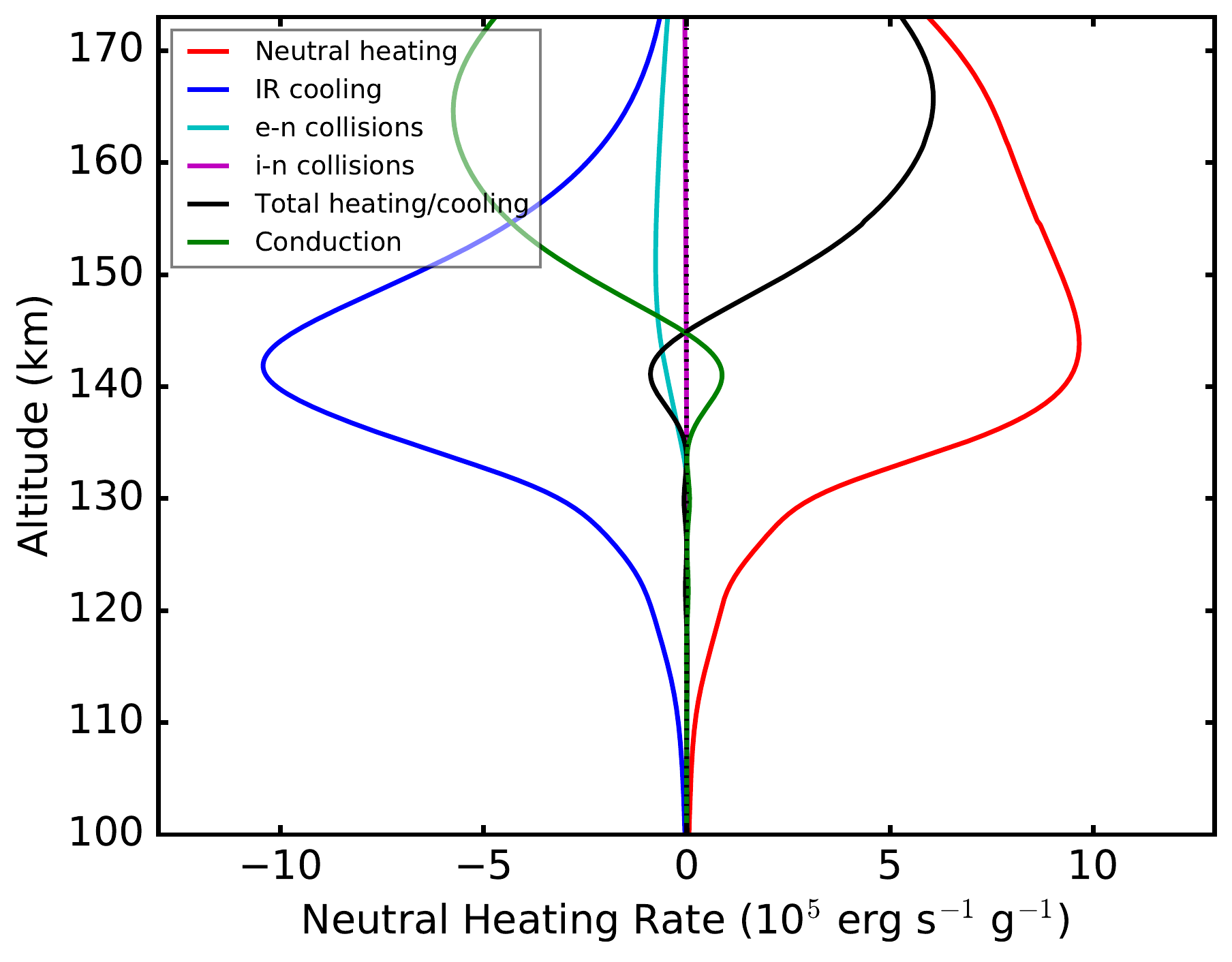}
\includegraphics[trim = 0mm 0mm 0mm 0mm, clip=true,width=0.49\textwidth]{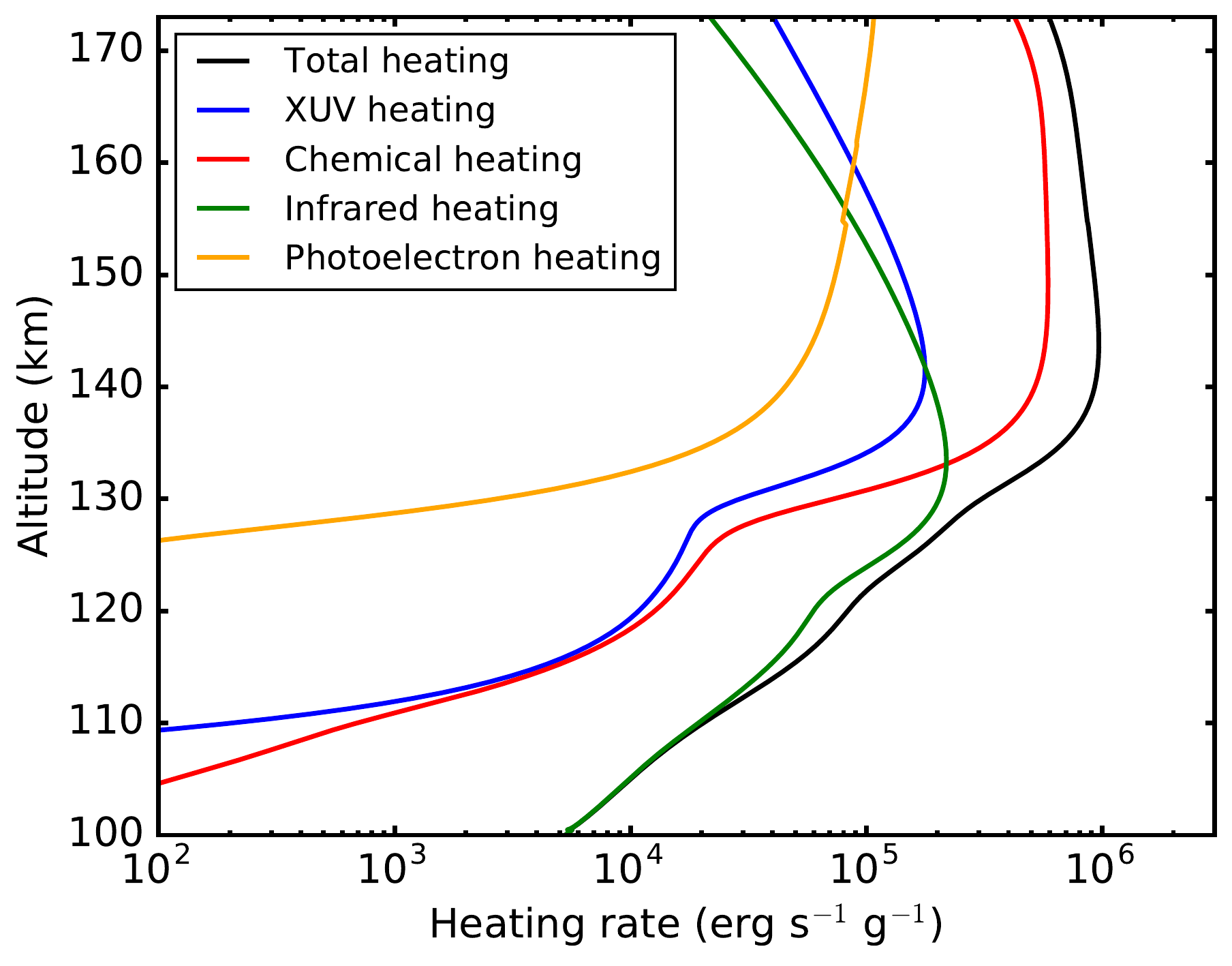}
\caption{
Figures showing the results of our Venus model.
In the upper-left panel, showing the temperature profiles, the dashed black line shows the empirical model by \citet{Hedin83} and the dotted black line shows the daytime profiles shown in Fig.~9 of \citet{Gilli17}. 
In the upper-right panel, showing the density profiles of several important species, the solid lines show the results of our model and the dotted lines show the profiles given by \citet{Hedin83}. 
The lower-left panel shows the heating and cooling mechanisms for the neutral gas, and the lower-right panel shows all of the heating mechanisms.
}
\label{fig:venusresults}
\end{figure*}


\section{Model Validation} \label{sect:validation}

To validate our model, we calculate the upper atmospheres of modern Earth and Venus in this section.
In both simulations, we use the modern solar XUV spectrum given by \mbox{\citet{Claire12}}, which represents the Sun approximately at the maximum of its activity cycle.

\subsection{Earth} \label{sect:currentEarth}

To validate our model for the Earth, we compare our results to those of two empirical models.
For the neutral gas, we use the atmospheric density and temperature profiles of the empirical NRLMSISE-00 model (\citealt{Picone02}).
This model produces vertical profiles for the Earth's atmosphere at arbitrary longitudes and latitudes and at arbitrary dates; the output profiles that we use are for temperature, and the densities of N$_2$, O$_2$, N, O, H, Ar and He. 
For the ion densities, we use International Reference Ionosphere 2007 (IRI-2007; \citealt{BilitzaReinisch08}).
This model produces vertical profiles for O$_2^+$, NO$^+$, O$^+$, N$^+$, H$^+$, and electrons.
We use these two standard models to obtain vertical atmospheric profiles for all longitudes and latitudes on the 1st January 1990, when the Sun was at approximately peak activity. 
We then calculate globally averaged profiles for this date.
For our own Earth simulation, we assume a zenith angle of 66$^\circ$, which we show below provides a good approximation for the globally averaged profiles.

We model the Earth's upper atmosphere between an altitude of 65~km and the exobase.
At the lower boundary, we use the values for temperature and density from this altitude in the NRLMSISE-00 model for comparison purposes, and additionally assume CO$_2$ and H$_2$O mixing ratios of \mbox{$4 \times 10^{-4}$} and \mbox{$6 \times 10^{-6}$} respectively, which are reasonable values for the Earth's middle atmosphere (\citealt{korner2001global}).
In Fig.~\ref{fig:earthtemp}, we show the thermal structure of our current Earth model.
The dashed line shows the standard atmosphere model that we use for comparison for the neutral gas temperature. 
The comparison between our results and the standard atmosphere model is very good, though an exact match between the models should not be expected especially since our input solar XUV spectrum will not match exactly the one used to produce the standard model.
In Fig.~\ref{fig:earthheatcool}, we show the strengths of the various heating and cooling mechanisms for this model.
The results resemble very closely those of other global upper atmosphere models (e.g. \mbox{\citealt{Roble95}}; \mbox{\citealt{Tian08a}}).

In Fig.~\ref{fig:earthchem}, we show the densities as a function of altitude of several important species in our simulation. 
The species are chosen to be those output by the NRLMSISE-00 and IRI-2007 models that we use for comparison.
Our predicted density structures are very similar to those of the comparison models, with the only exception being He, which we predict to be less abundant at high altitudes than expected.
Clearly our model is able to realistically predict the structure of the Earth's atmosphere.

\subsection{Venus} \label{sect:currentVenus}

In this section, we further validate our model by calculating the structure of the upper atmosphere of Venus.
This is especially useful since Venus' atmosphere is made up mostly of CO$_2$ and has therefore much stronger atmospheric cooling, allowing us to test that our model realistically responds to large changes in the CO$_2$ content.  
For a reference atmosphere, we use the empirical thermosphere model given in Table~3b of \citet{Hedin83}, which is for Venus' atmosphere at noon during approximately solar maximum conditions.
We therefore assume a zenith angle of 0$^\circ$ in our model and use the values given by \citet{Hedin83} for the temperature and species densities at the lower boundary.
We do not include Joule heating in our Venus model.

In Fig.~\ref{fig:venusresults}, we show a summary of the results of our Venus model.
We compare our calculated temperature structures to the standard model given by \mbox{\citet{Hedin83}} and to the recent 3D global models for the full atmosphere of Venus by \citet{Gilli17}.
The neutral temperature profile resembles that of \mbox{\citet{Hedin83}}, with very similar exobase temperatures, though we find that our model is colder in the lower thermosphere, with the largest difference being around 40~K around 130~km.
These differences in the temperature profiles do not suggest that there is a problem with our simulations; the \citet{Gilli17} model is also colder than the \mbox{\citet{Hedin83}} profiles at 140~km, though it is much warmer than both our model and the \mbox{\citet{Hedin83}} lower in the thermosphere.
In Fig.~\ref{fig:venusresults}, the density profiles for several species from our simulations can be compared to those from \mbox{\citet{Hedin83}}.
As with the case of the Earth's thermosphere, we underestimate the He abundances at high altitudes.
Our other density profiles are similar to those of \mbox{\citet{Hedin83}}, with the differences being mostly due to the different temperatures.


\section{Results} \label{sect:results}

We present two applications of our code. 
In Section~\ref{sect:CO2effect}, we explore the effects of enhancing the CO$_2$ abundance in the current Earth's atmosphere on the upper atmospheric structure. 
In Section~\ref{sect:Earthevo}, we explore the response of Earth's upper atmosphere to the evolving XUV spectrum of the Sun.

\subsection{The influence of enhanced CO$_2$ abundance on the atmospheric structure} \label{sect:CO2effect}

\begin{figure}[!h]
\includegraphics[trim = 0mm 0mm 0mm 0mm, clip=true,width=0.49\textwidth]{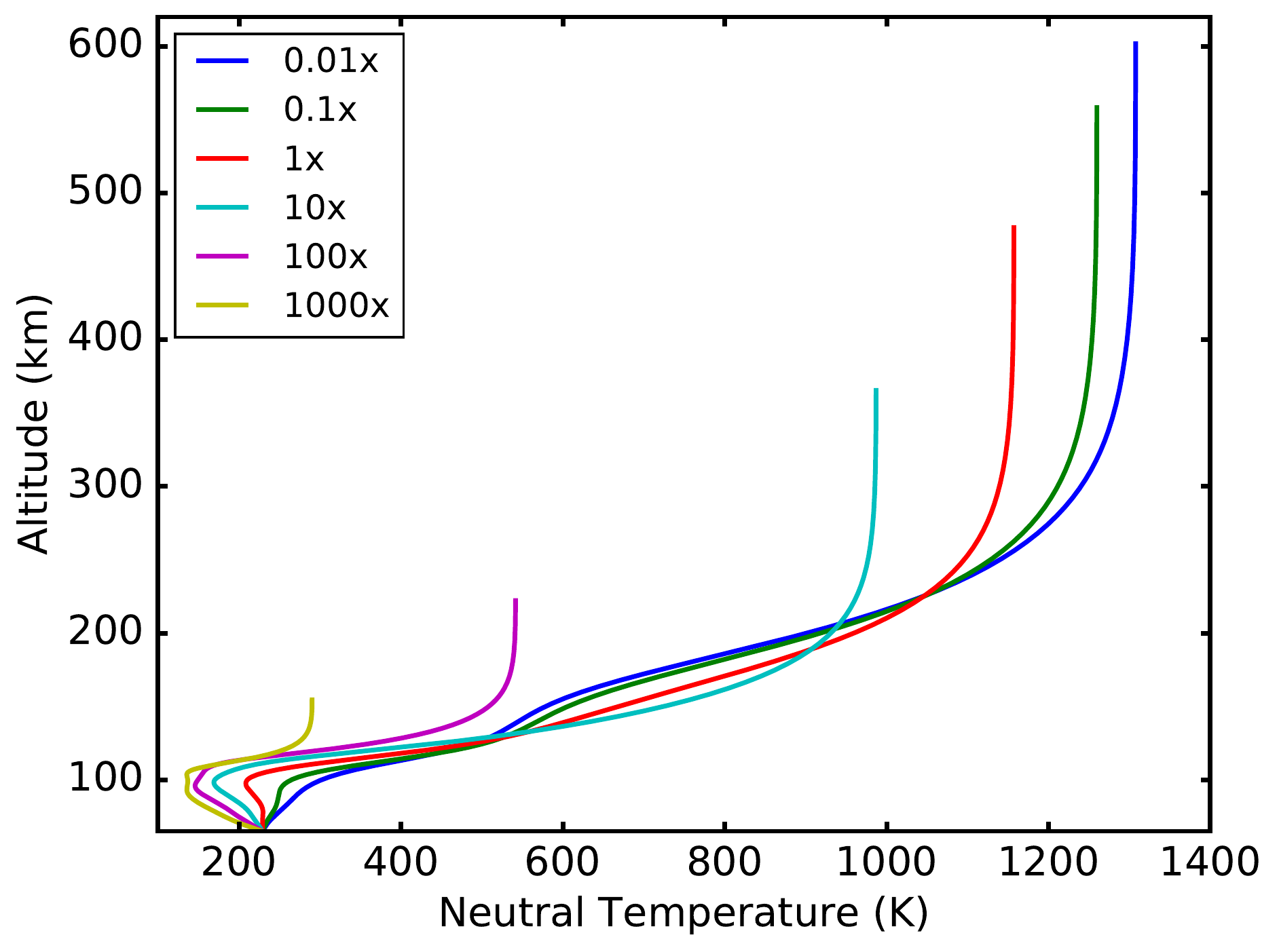}
\includegraphics[trim = 0mm 0mm 0mm 0mm, clip=true,width=0.49\textwidth]{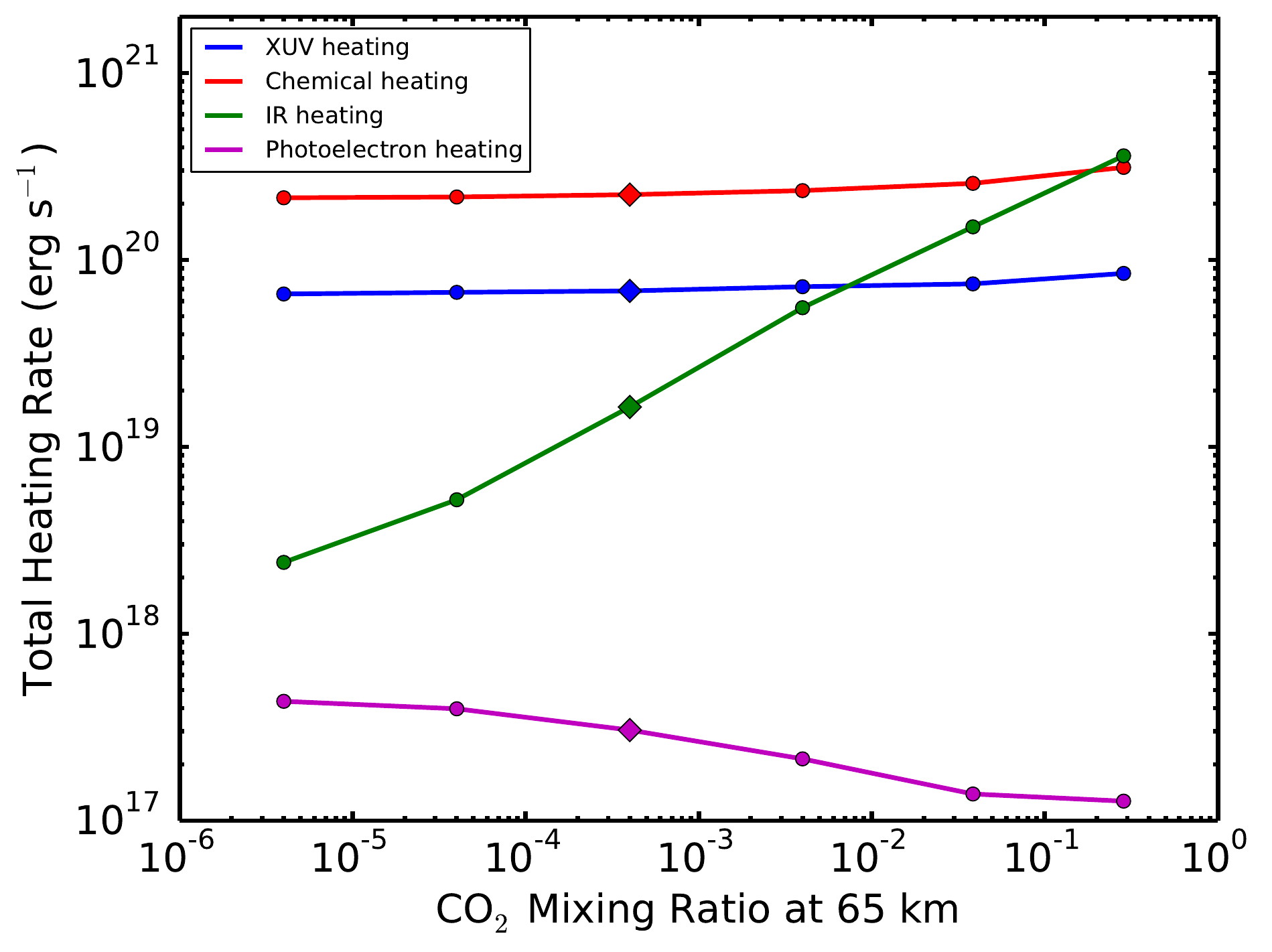}
\caption{
Figures showing the neutral temperature structures for our simulations with different base mixing ratios of CO$_2$(\emph{upper-panel}) and the total atmospheric heating rates due to the various heating processes as a function of CO$_2$ mixing ratio (\emph{lower-panel}).  
In the upper panel, the lines (which stop at the exobase) are simulations where the base CO$_2$ density is 0.01, 0.1, 1, 10, 100, and 1000 times the value for the current Earth, which has a mixing ratio at 65~km of \mbox{$4 \times 10^{-4}$}.
}
\label{fig:co2testtemp}
\end{figure}

\begin{figure}[!ht]
\includegraphics[trim = 0mm 0mm 0mm 0mm, clip=true,width=0.44\textwidth]{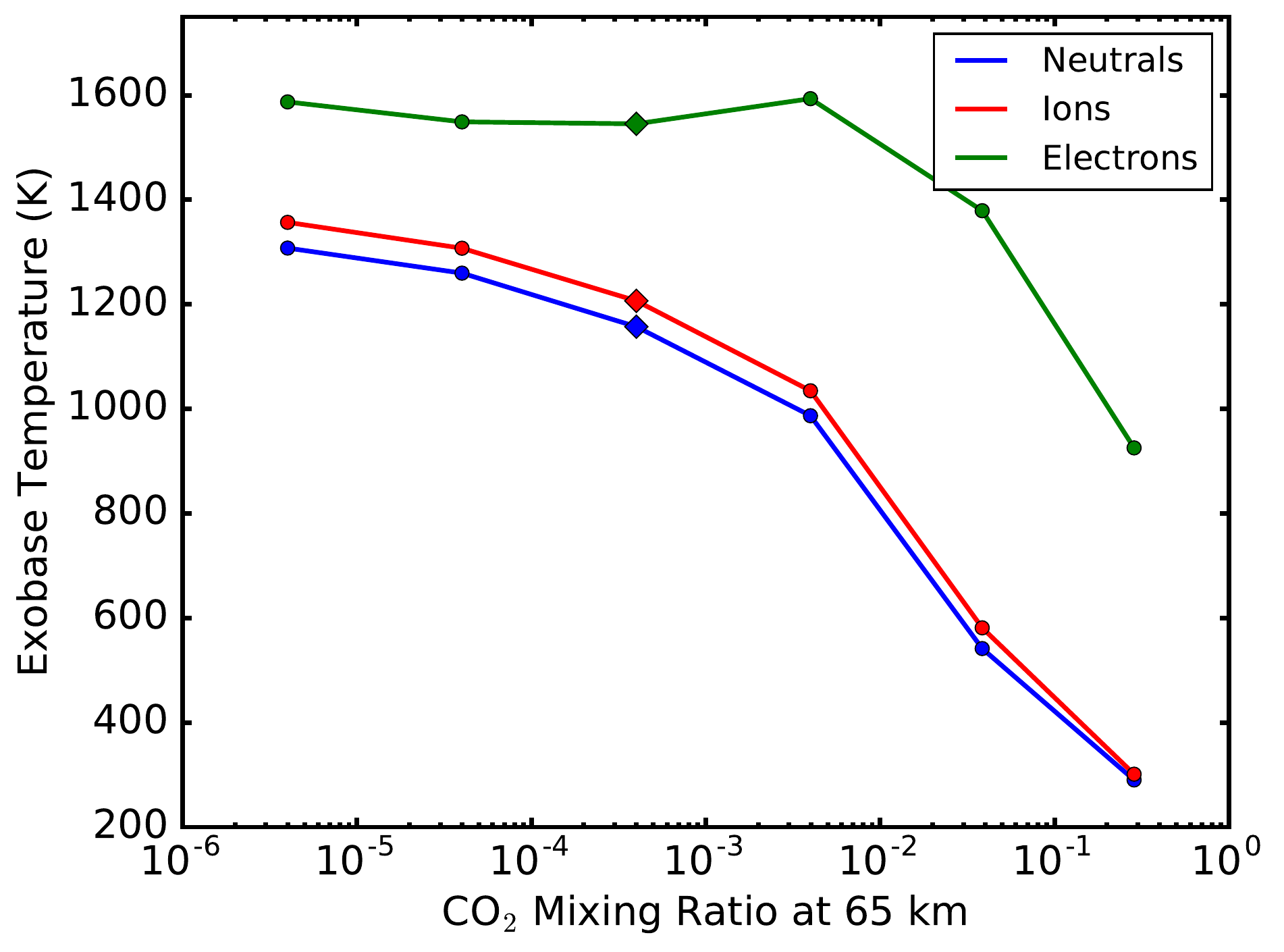}
\includegraphics[trim = 0mm 0mm 0mm 0mm, clip=true,width=0.44\textwidth]{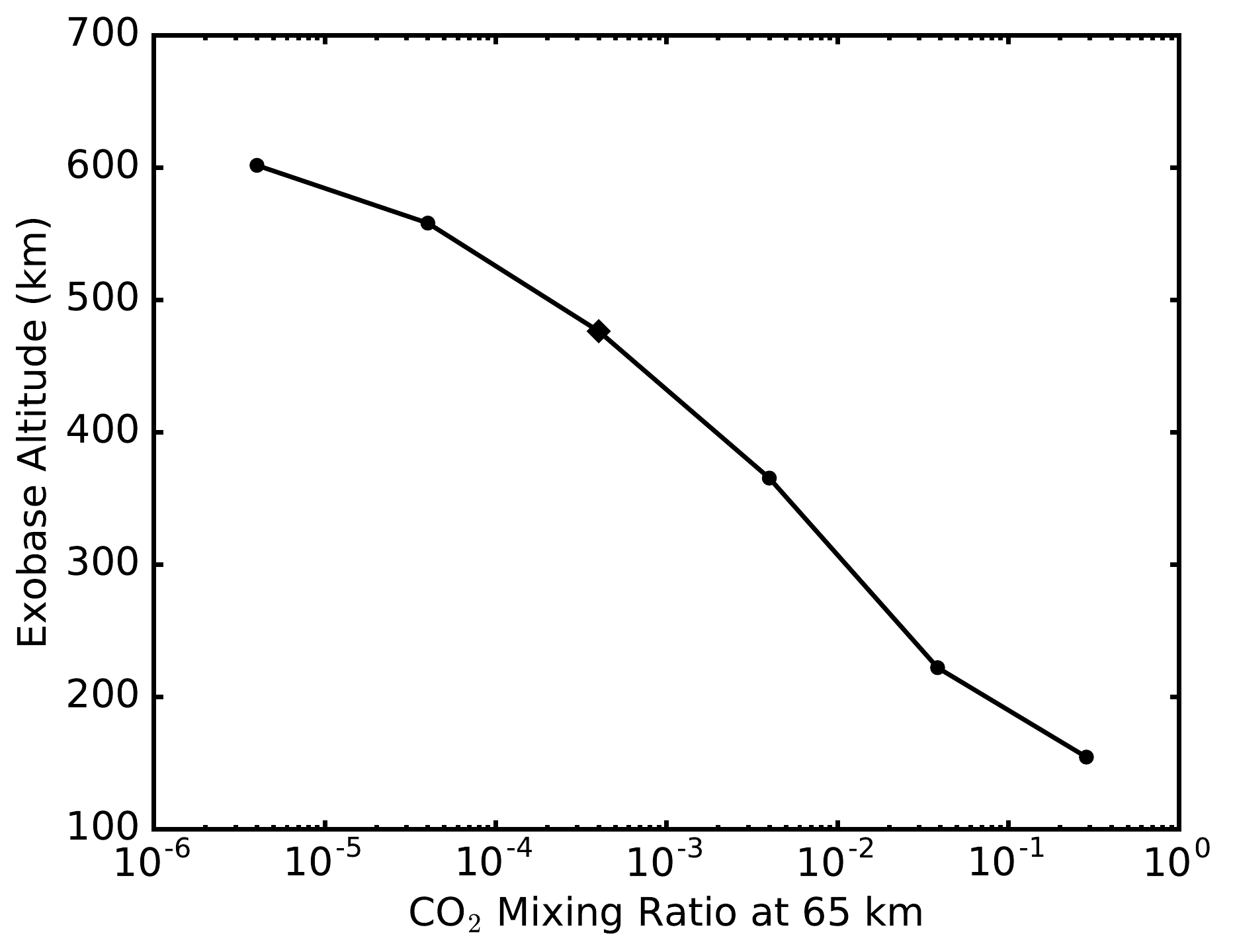}
\includegraphics[trim = 0mm 0mm 0mm 0mm, clip=true,width=0.44\textwidth]{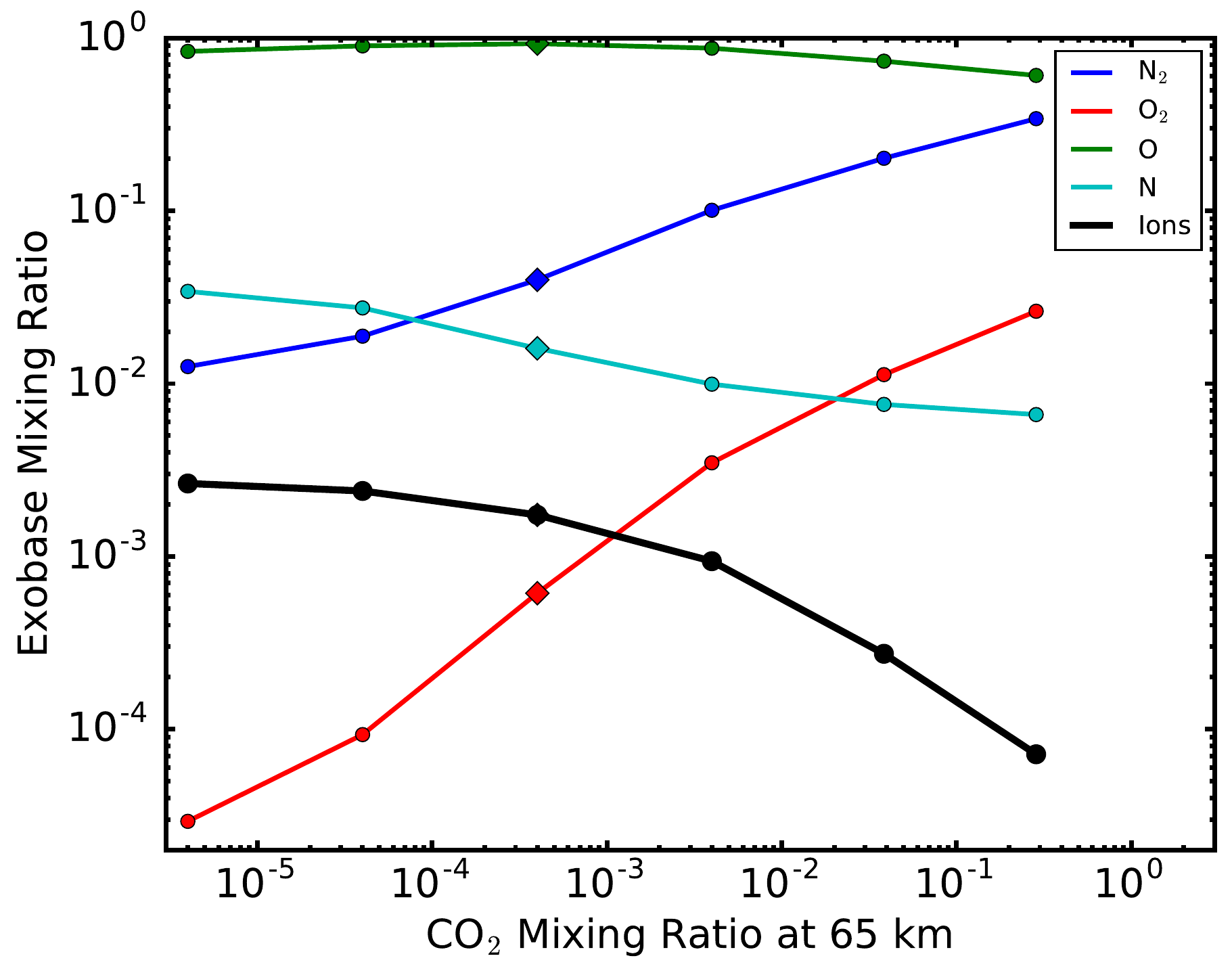}
\caption{
Figures showing the temperatures (\emph{upper-panel}), altitudes (\emph{middle-panel}), and mixing ratios of several important species of the exobase (\emph{lower-panel}) as a function of the CO$_2$ mixing ratio at 65~km.
In the lower panel, the black line gives the total ion mixing ratio (i.e. the ionization fraction).
}
\label{fig:co2testexo}
\end{figure}

To study the response of the Earth's upper atmosphere to large changes in the CO$_2$ abundance, we calculate several models for the current Earth, varying only the lower boundary density of CO$_2$. 
We calculate models where the CO$_2$ base density is 0.01, 0.1, 1, 10, 100, and 1000 times the value for the current Earth.
We refer to these models by this multiplicative factor, such that our model with 1000 times more CO$_2$ is the 1000$\times$ model.
In our 1000$\times$ model, CO$_2$ has a mixing ratio of approximately 0.3, which is a factor of a few lower than that of Venus.
As in previous sections, our input solar spectrum is for the current Sun at activity maximum.

It is important to note that the fact that we do not include the effects of changing the CO$_2$ abundances on the lower atmosphere.
In reality, large changes in the CO$_2$ abundances will cause also changes in these lower boundary properties. 
This has the effect that the total heating and cooling rates approximately balance only in our 1$\times$ model.
In models with lower CO$_2$ abundances, energy is removed from the computational domain at the lower boundary by downward conduction, causing the total cooling to be less than the total heating.
In models with higher CO$_2$ abundances, the opposite effect takes place. 

In Fig.~\ref{fig:co2testtemp}, we show the neutral temperature structures for each model.
As expected, increasing the CO$_2$ abundance leads to a significant decrease in the thermospheric temperature due to enhanced CO$_2$ cooling. 
Similarly, decreasing the amount of CO$_2$ leads to the thermosphere becoming hotter. 
In the 1000$\times$ model, the maximum temperature reached in the thermosphere is only $\sim$300~K, similar to that of Venus, and the mesosphere is cooled to a minimum temperature of $\sim$130~K, which is much cooler than the minimum temperature of $\sim$200~K that we find in our current Earth model. 
Fig.~\ref{fig:co2testtemp} also shows the total atmospheric heating rate\footnotemark~as a function of CO$_2$ mixing ratio for different heating processes.
Note however that these quantities are relatively crude measures of how important the various heating mechanisms are; for example, the total photoelectron heating for the current Earth is very small, but the effect is relatively large because it takes place high in the thermosphere where the gas densities are low and therefore less energy is required for the heating to be significant.

\footnotetext{
For example, to calculate the total XUV heating rate, we calculate \mbox{$4 \pi \int r^2 Q_\mathrm{xuv} dr$}, where the integral is over all radii, $r$, and $Q_\mathrm{xuv}$ is the volumetric heating rate given by Eqn.~\ref{eqn:XUVheatingequation}.
}

In the current Earth case, the two dominant heating mechanisms are chemical and XUV heating.
These two mechanisms depend primarily on the input XUV energy flux, and therefore remain approximately constant.
Due to the decrease in the ionization fraction of the gas, the heating of the electrons by collisions with non-thermal photoelectrons decreases significantly as the CO$_2$ mixing ratio is increased (see the linear dependence between electron heating and density in Eqn.~\ref{eqn:electronheating}).
The mechanism that changes the most is the heating due to the absorption of stellar IR photons.
Unlike the stellar XUV photons, which are all absorbed in the upper atmosphere in all cases, most of the IR photons pass through the upper atmosphere unhindered; therefore, adding efficient IR absorbing gases influences significantly how much IR energy is absorbed.  
In the 1000$\times$ model, we find that most of the heating is from IR absorption, especially at low altitudes. 
In the 0.01$\times$ model, the IR heating is dominated by H$_2$O absorption.

It is interesting to compare our simulations to those of \mbox{\citet{Kulikov07}} who also modelled the effects on enhanced CO$_2$ abundances on the Earth's thermosphere (see their Fig.~2).
Our results are broadly similar which provides additional validation of our model. 
We find in general cooler upper thermospheric temperatures than they do for the enhanced CO$_2$ models.
Our models differ in a few important ways: for example, they did not calculate separate neutral, ion, and electron temperatures and they did not consider the effects of  chemical reactions on the density structures of individual species.
The differences between our results likely have two sources. 
Firstly, the XUV heating in their model is based on an assumed heating efficiency parameter that in reality might vary with CO$_2$ abundance.
Secondly, they put the lower boundaries of their models at the mesopause at an altitude of 100~km with an approximately fixed temperature, whereas we calculate the upper mesosphere from 65~km. 
In the enhanced CO$_2$ models, we get significant additional mesospheric cooling, and therefore much lower mesopause temperatures, which leads to cooler temperatures also at higher altitudes.

The changes in the exobase are demonstrated in Fig.~\ref{fig:co2testexo}.
In the models with larger CO$_2$ abundances, the exobase altitudes and gas temperatures are much lower.
Interestingly, the electron temperature does not decrease as much as the neutral and ion temperatures. 
This is largely because the total photoionization rate is approximately the same in all of our atmosphere models, meaning that the amount of energy in the non-thermal electron spectrum is also approximately constant. 
The volumetric heating rate for electrons is lower in simulations with higher CO$_2$ abundances because the electron densities are lower, but the heating rate per electron is in fact higher. 

The lower panel in Fig.~\ref{fig:co2testexo} shows the mixing ratios of several important species at the exobase as a function of CO$_2$ abundance, including the total ion mixing ratio.
The exobase composition changes significantly, especially for N$_2$ and O$_2$; although O remains the most abundant species at the exobase in all simulations, in the 1000$\times$ simulation, the N$_2$ mixing ratio is similar to that of O.
The changes in the exobase N$_2$ and O$_2$ mixing ratios are primarily because of the change in the exobase altitude.
As the atmosphere cools and the exobase moves to lower altitudes, the molecular diffusion rates and the distance between the homopause and the exobase are reduced, giving molecular diffusion less of a chance to separate the heavier and lighter species.


\begin{figure*}
\includegraphics[trim = 0mm 0mm 0mm 0mm, clip=true,width=0.50\textwidth]{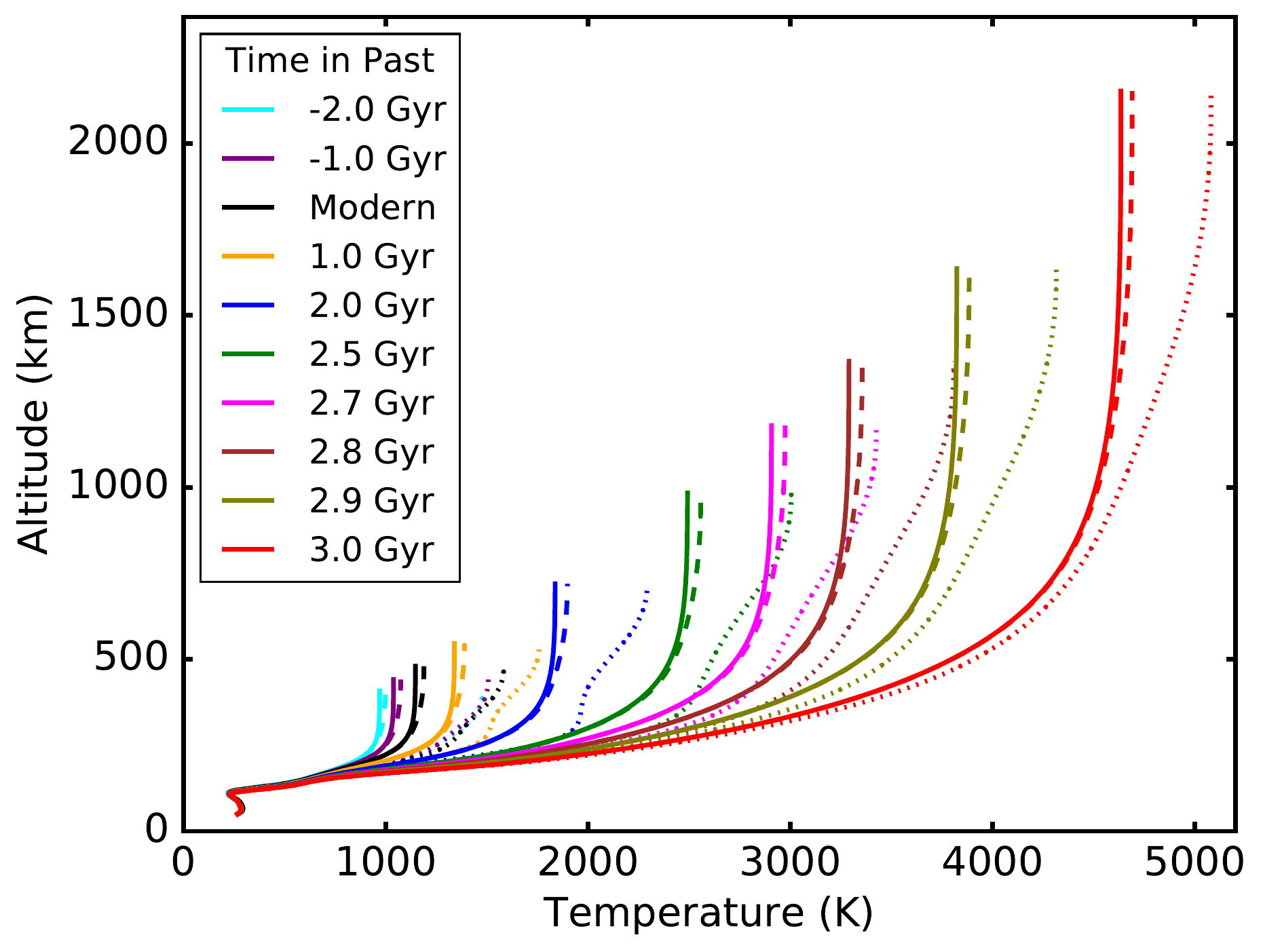}
\includegraphics[trim = 0mm 0mm 0mm 0mm, clip=true,width=0.49\textwidth]{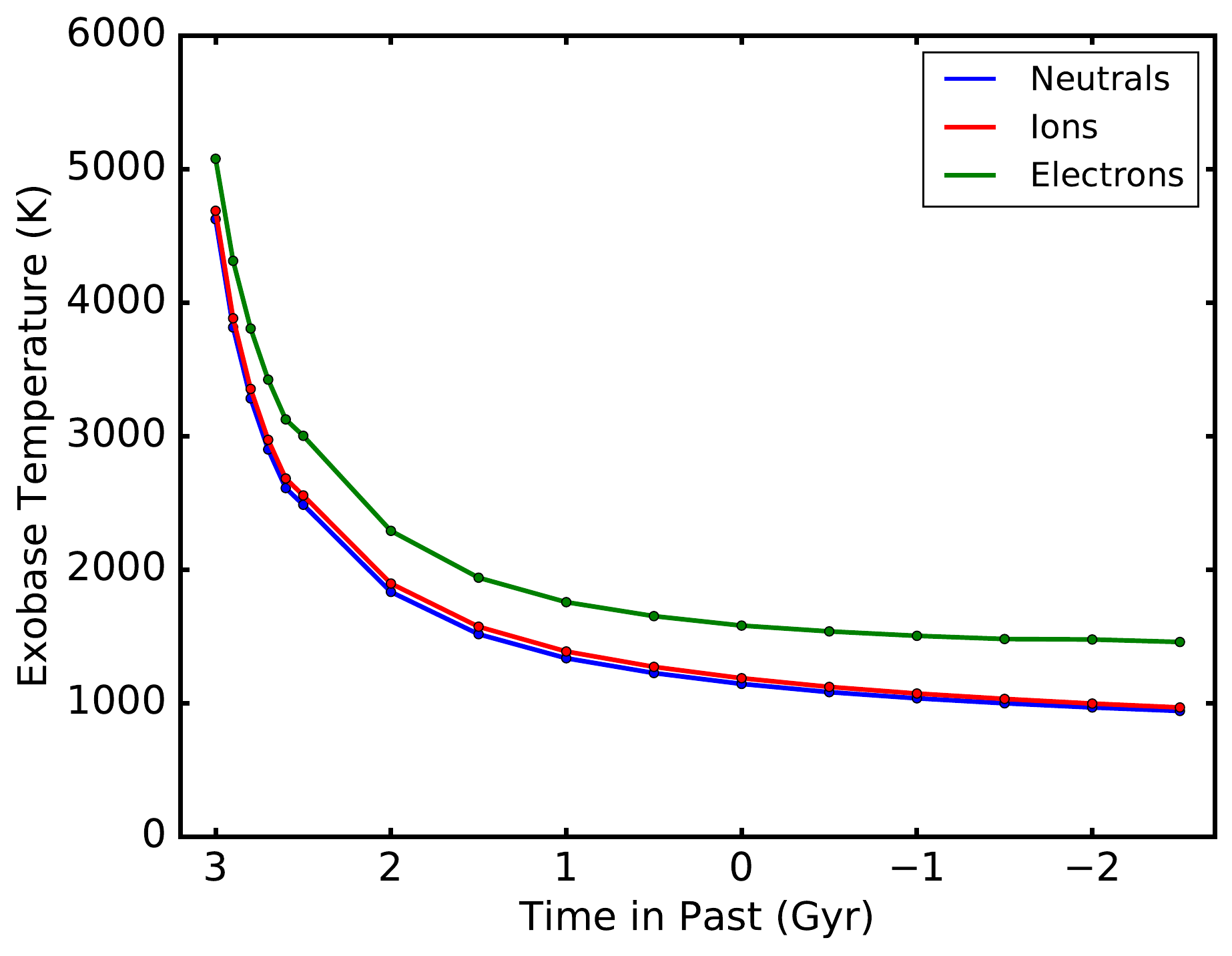}
\includegraphics[trim = 0mm 0mm 0mm 0mm, clip=true,width=0.49\textwidth]{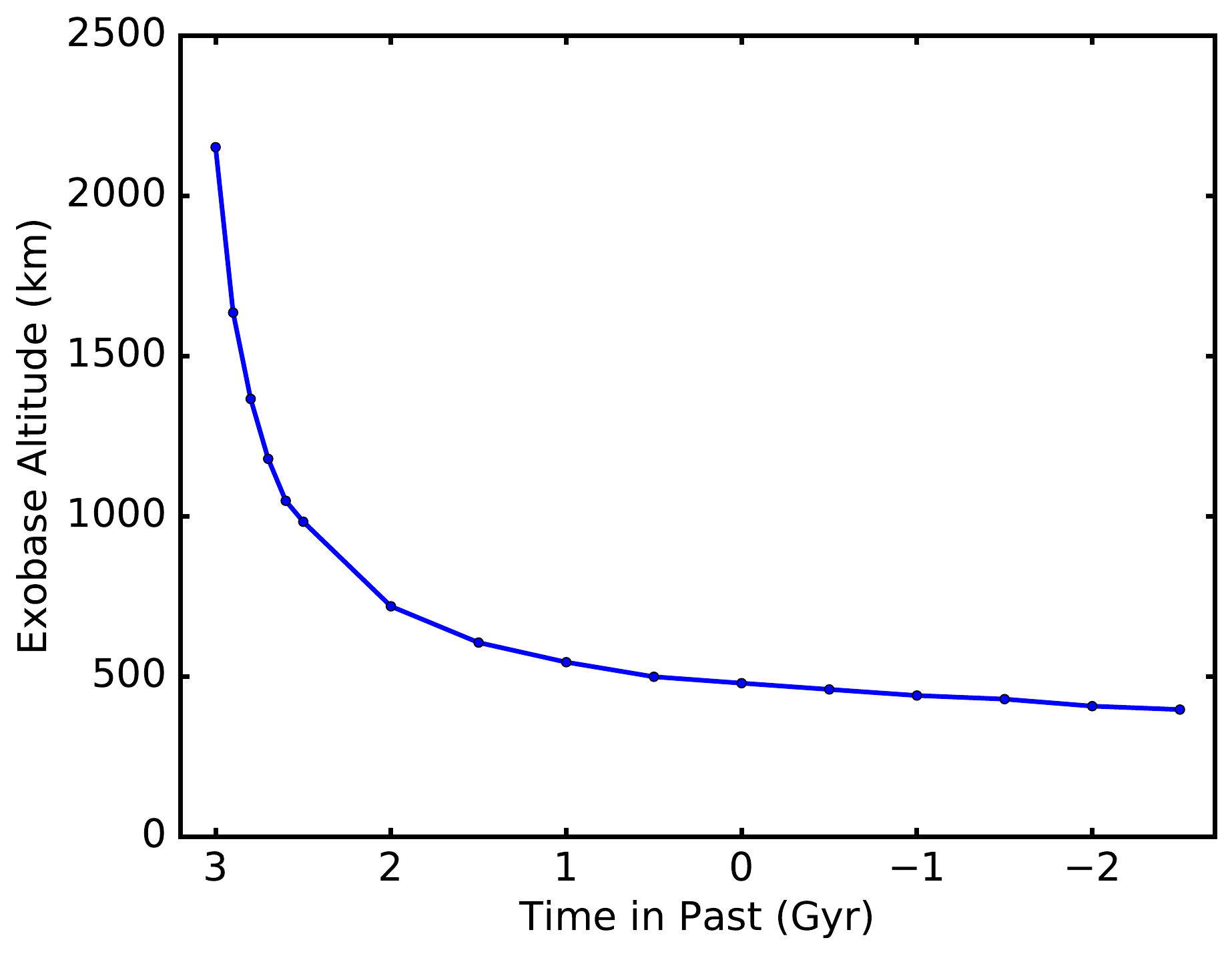}
\includegraphics[trim = 0mm 0mm 0mm 0mm, clip=true,width=0.49\textwidth]{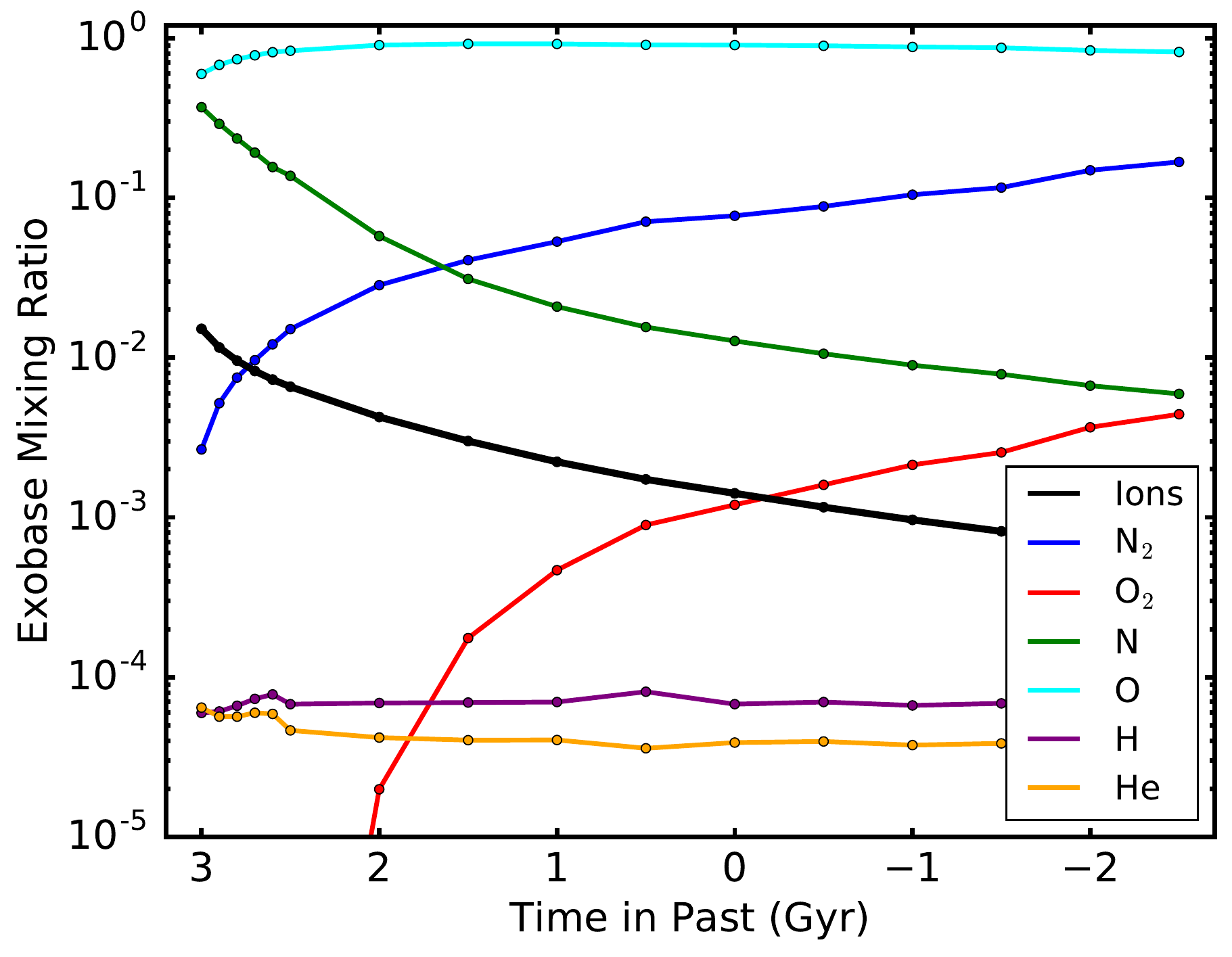}
\includegraphics[trim = 0mm 0mm 0mm 0mm, clip=true,width=0.50\textwidth]{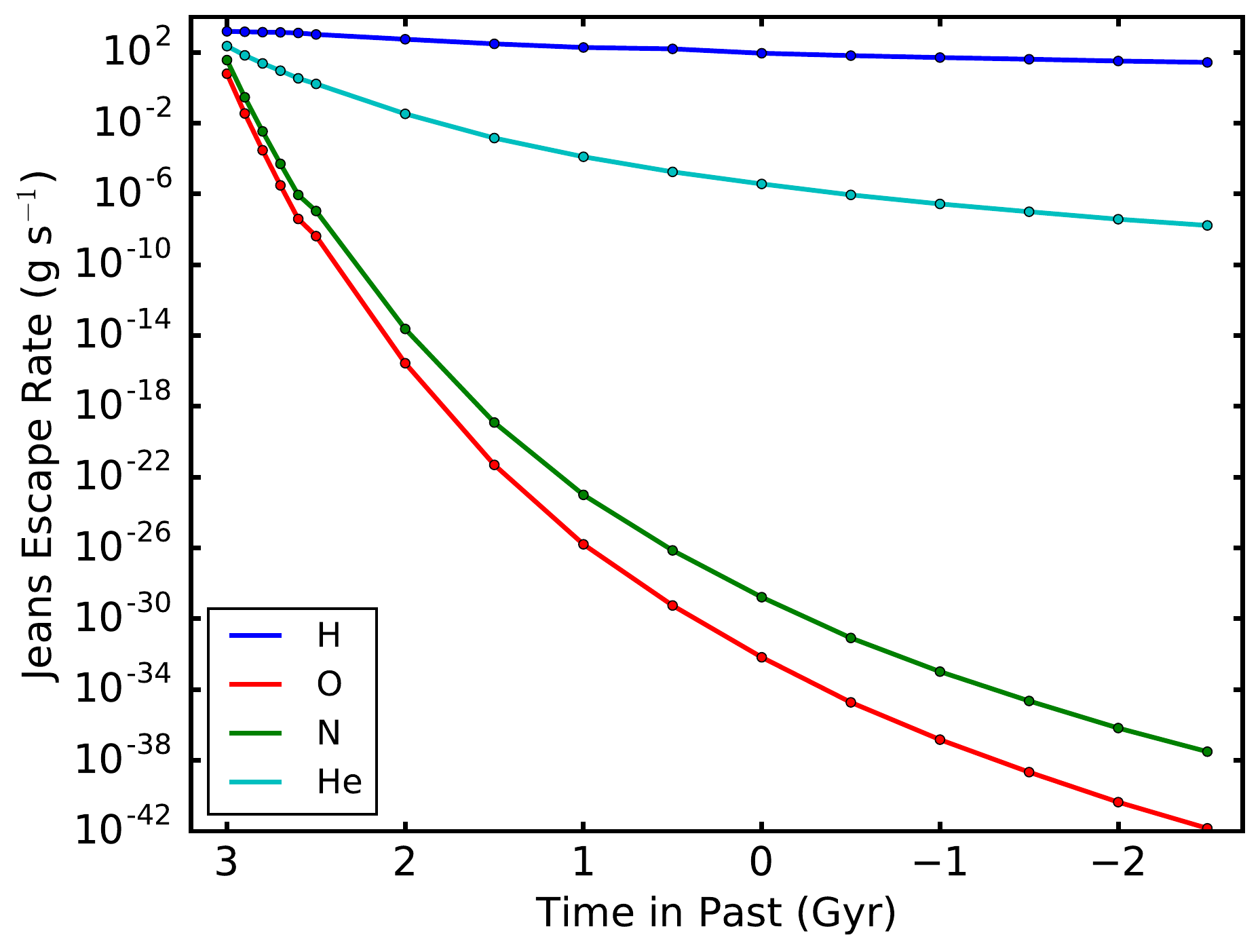}
\includegraphics[trim = 0mm 0mm 0mm 0mm, clip=true,width=0.49\textwidth]{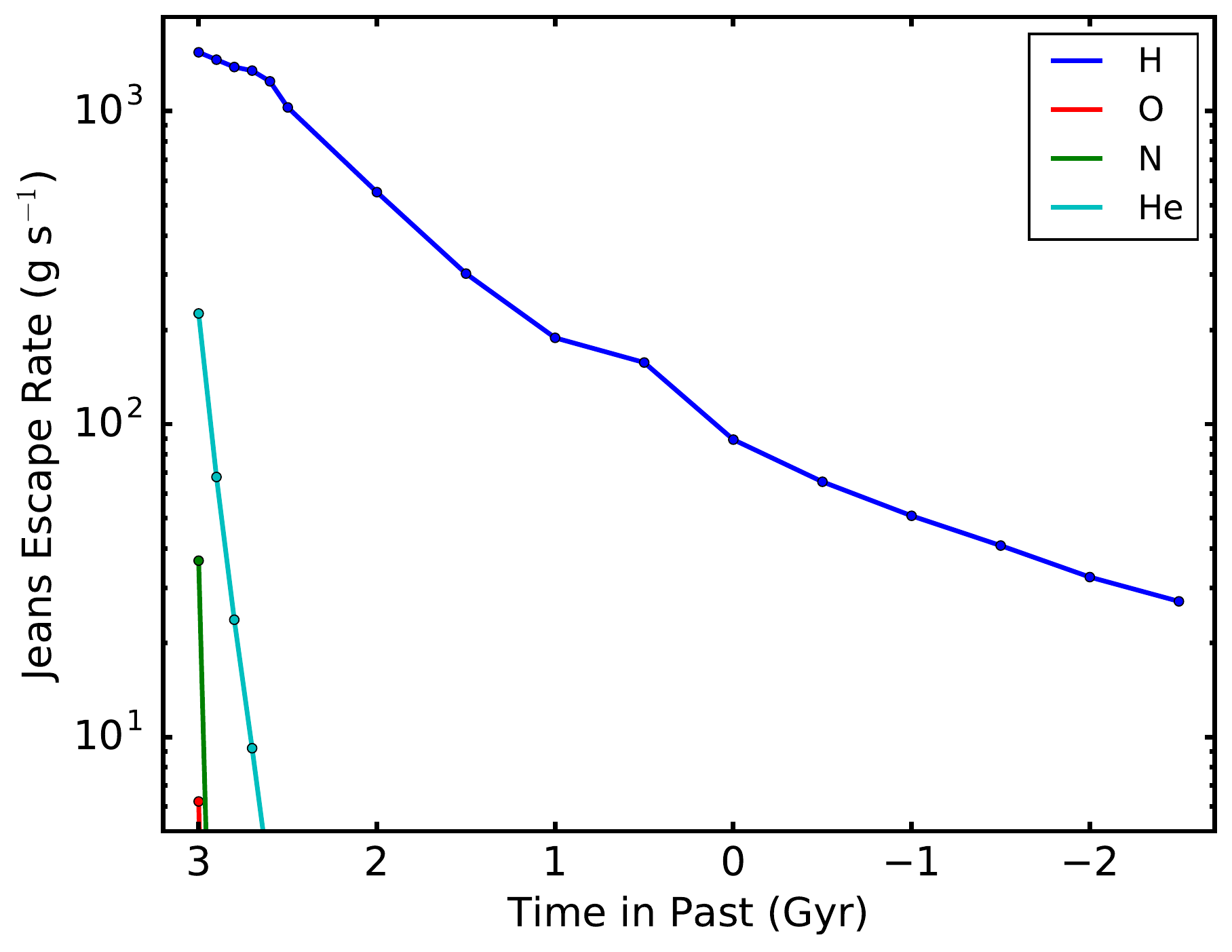}
\caption{
Figures summarising our simulations for the response of the Earth's atmosphere to the evolving XUV spectrum of the Sun. 
In the upper-left panel, we show the temperature structures for several of these models, where the different colors are for different ages and the solid, dashed, and dotted lines show the neutral, ion, and electron temperatures respectively.
In all cases, the lines end at the exobase.
In the upper-right, middle-left, and middle-right panels, we show the exobase temperatures, altitudes, and chemical compositions as functions of age.
In the lower-left and lower-right panels, we show the evolution of Jeans escape for H, O, N, and He, with the difference between the two plots being the range on the y-axis.
In each figure, the small circles show the exact locations of each simulation. 
}
\label{fig:EarthEvo}
\end{figure*}

\begin{figure}[!h]
\includegraphics[trim = 0mm 0mm 0mm 0mm, clip=true,width=0.46\textwidth]{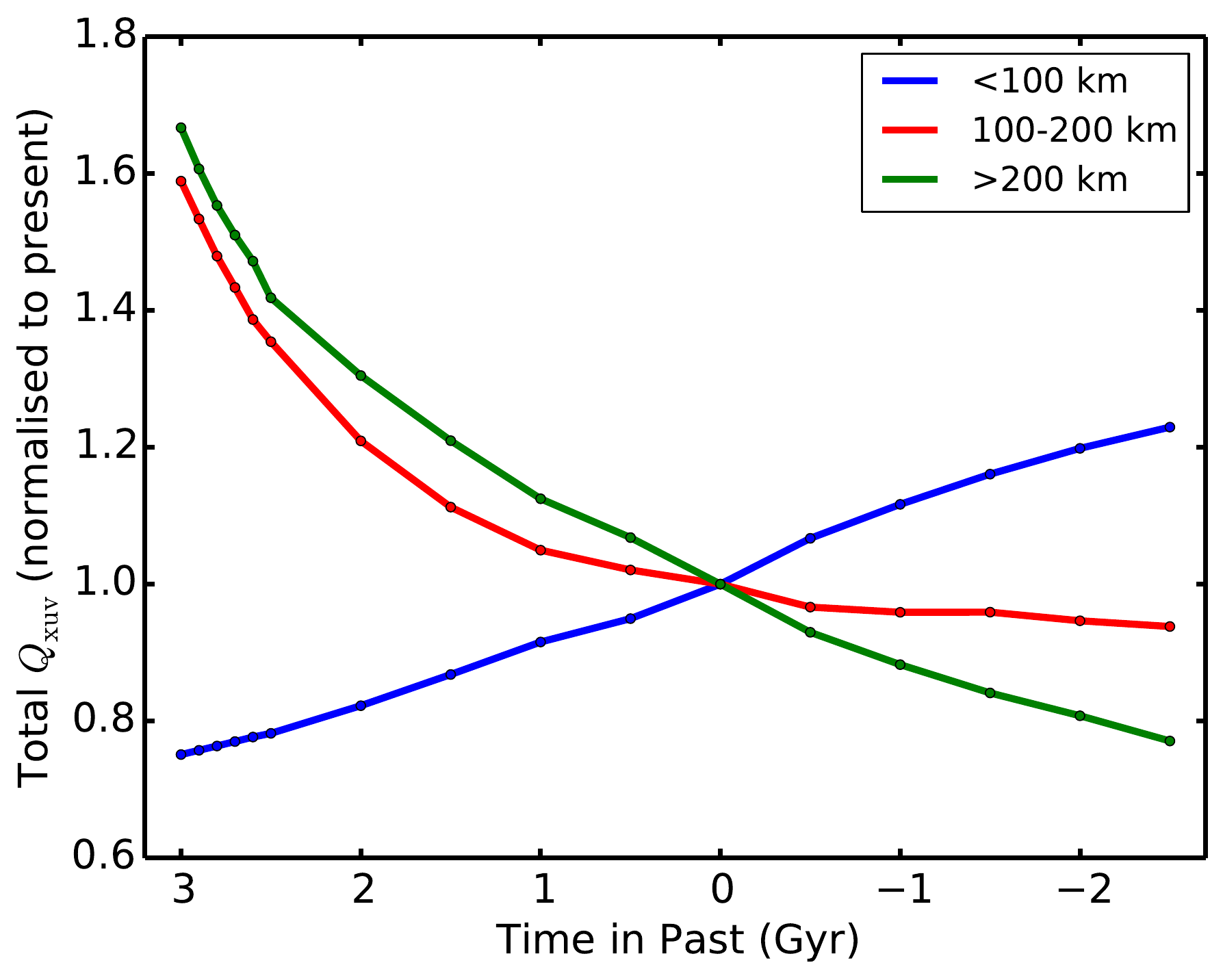}
\includegraphics[trim = 0mm 0mm 0mm 0mm, clip=true,width=0.46\textwidth]{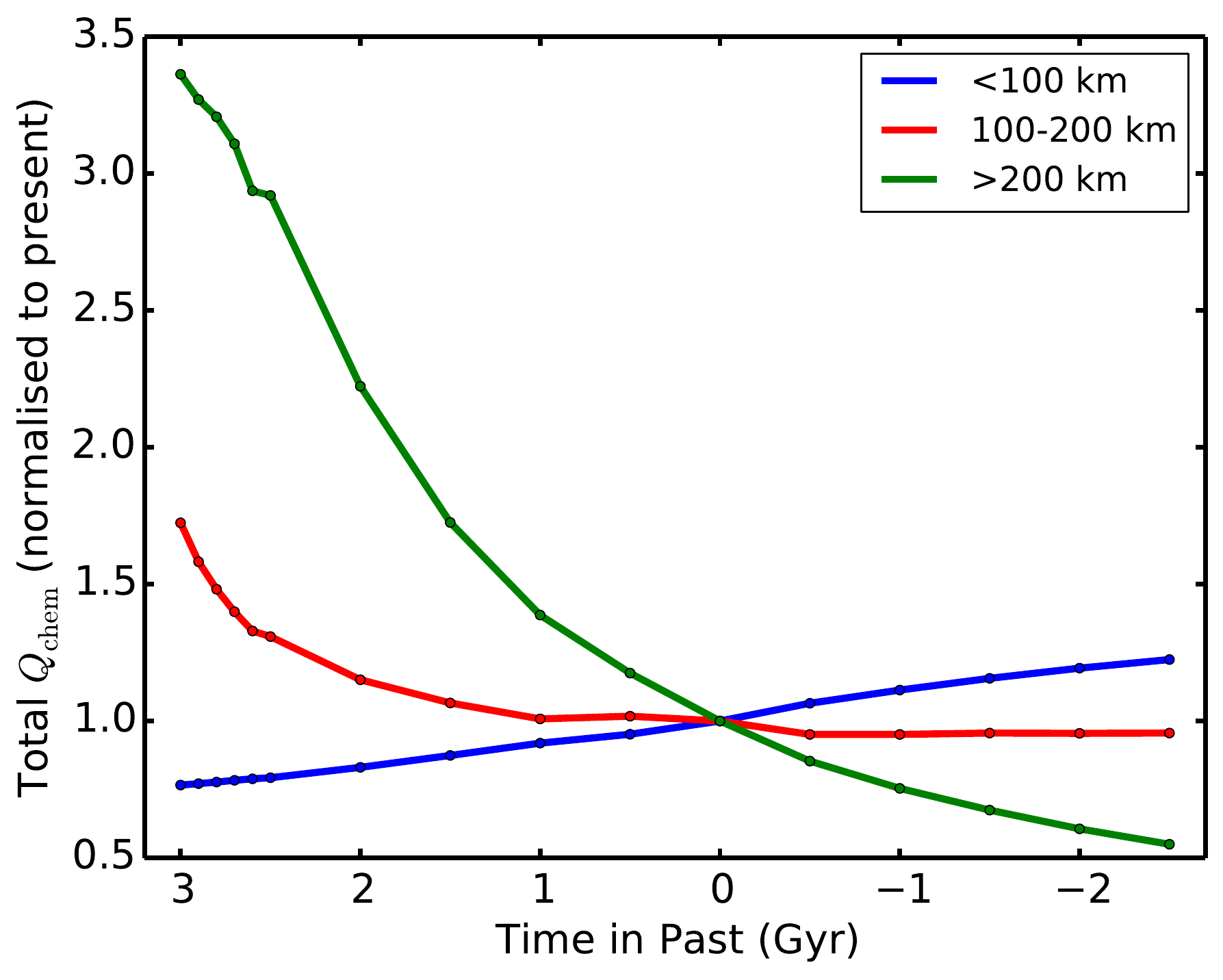}
\includegraphics[trim = 0mm 0mm 0mm 0mm, clip=true,width=0.46\textwidth]{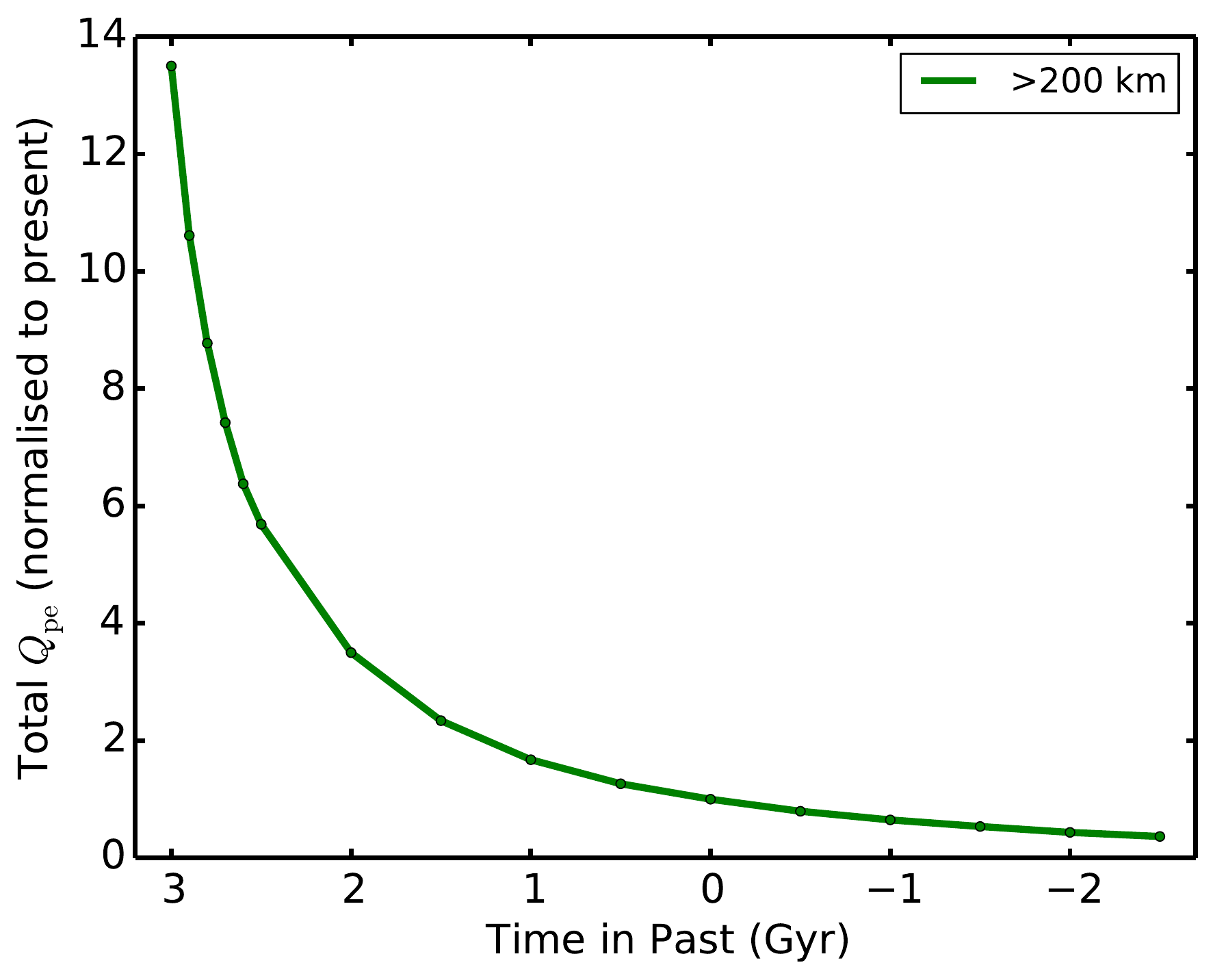}
\caption{
Figures showing the evolutions of direct XUV heating (\emph{upper-panel}), chemical heating (\emph{middle-panel}), and photoelectron heating (\emph{lower-panel}) within three altitude ranges in the Earth's atmosphere. 
The quantities plotted are the total heating rates integrated over the relevant volumes and normalised to the values for the modern Earth.
For photoelectron heating, only the values for altitudes between 200~km and the exobase are shown since at lower altitudes this process is negligible.
}
\label{fig:EarthEvoHeating}
\end{figure}

\subsection{The evolution of Earth's upper atmosphere} \label{sect:Earthevo}

In this section, we explore the responses of the upper atmospheres of Earth to the evolving XUV spectrum of the Sun between 3~Gyr in the past and 2.5~Gyr in the future.
We do not consider the effects of the evolving lower atmospheric composition, which we will study in future work.
We use the XUV spectra from \citet{Claire12}, who produced solar spectra as a function of age for all wavelengths that are of interest to us.
Note that at young ages, the XUV spectra of solar mass stars are not unique functions of age given that they follow different rotational and activity evolution tracks (\citealt{Johnstone15wind}), which can be very important for the evolution of a planet's atmosphere (\citealt{Johnstone15}).
Since we do not know how rapidly the Sun was rotating at young ages, we do not know its early XUV output (\citealt{Tu15}).
However, at the ages that we consider, the rotation rates of young solar mass stars have converged to unique age dependent values, and so the unique XUV evolution presented by \citet{Claire12} is valid. 

In Fig.~\ref{fig:EarthEvo}, we show our results for the Earth. 
In the upper-left panel, the temperature structures for our models at several ages are shown; the lines in this panel can be seen as an evolutionary sequence from right to left, with the thermospheres becoming cooler and less extended as the Sun's XUV spectrum decays.
In all cases, the neutral and ion temperatures are approximately the same at all altitudes and the electron temperatures are higher.
In the upper-right panel, the evolutions of the exobase temperatures are shown.
Interestingly, the electron temperatures at the exobase are higher by approximately the same amount (i.e. $\sim$500~K) at all ages. 
At 3~Gyr ago, our models suggest that the exobase neutral and electron temperatures were approximately 4600~K and 5100~K respectively. 
Our models also suggest that going into the future, we should not expect a large change in the upper atmosphere of the Earth simply due to the Sun's activity decay.
This is consistent with the power-law dependence of the solar activity on age, which means that the fastest changes take place at young ages; in our models, the atmosphere changes the most between 3~Gyr ago and 2~Gyr ago, and all changes after that are relatively slow in comparison.

Our results for the temperature structures are broadly consistent with those calculated by \citet{Tian08a}.
Since \citet{Claire12} used a more realistic method for estimating the spectra of the Sun at higher activity levels, we should not expect any exact agreement between our models and those of \citet{Tian08a} even for the same total input XUV flux.
Our model for 3~Gyr in the past corresponds approximately to their model with input fluxes of 4.9 times the current solar value, and we get similar results.
Although not studied in this paper, in models with an even more active Sun, we can also see the effects of adiabatic  cooling that were found by \citet{Tian08a}.

The middle row of Fig.~\ref{fig:EarthEvo} shows the exobase altitude and composition. 
The largest change in the altitude of the exobase takes place between 3~Gyr ago and 2~Gyr ago, dropping from 2000~km to approximately 600~km in that time. 
This is mostly due to the decrease in the thermospheric temperature, but is also partly due to the change in the chemical composition of the gas.
At the youngest age considered, we find that the chemical composition at the exobase contains much more N and much less N$_2$ than we find for the current Earth, though O is in all cases the dominant species. 
Interestingly, we do not find any change in the exobase mixing ratios of H and He, which stay at values of \mbox{$\sim 5 \times 10^{-5}$} at all ages.
The ionization fraction at the exobase also decreases with age by about an order of magnitude between 3~Gyr ago and now. 

Since the atmospheres are never fully hydrodynamic in our simulations, the dominant mass loss rates are likely to be non-thermal processes.
Calculating these requires the application of additional models, which we do not attempt in this paper.
We can calculate from our models the Jeans escape rates, which are shown for several species in the lower panels of Fig.~\ref{fig:EarthEvo}. 
For the current Earth, it is well known that the only species undergoing significant Jeans escape is H, due to its low molecular mass.
As we go to the past however, the Jeans escape rates of He, O, and N increase significantly, and in our earliest model are within $\sim$2 orders of magnitude of the escape rate of H.
If we were to go further into the past, the Jeans escape rates of these species would become comparable to that of H; it is around this time that we would find the atmosphere becoming hydrodynamic and the effects of adiabatic cooling on the temperature structures becoming important.

It is interesting to consider how the various heating mechanisms evolve with the evolving solar spectrum.
In Fig.~\ref{fig:EarthEvoHeating}, we show the evolutions of direct XUV heating, chemical heating, and photoelectron heating for three altitude ranges.
The quantities are the volumetric heating rates integrated over all the relevant altitudes and normalised to the modern Earth values.
For altitudes above 100~km, the total heating rates for all these processes decrease with age due to the decreasing solar magnetic activity.
However, at lower altitudes, the heating rates in fact increase with age.
This is because this heating comes almost entirely from O$_3$ absorption at wavelengths longer than 2000~\AA, which is generated in the solar photosphere.
Since the Sun's bolometric luminosity is increasing as it ages, the heating in the lower atmosphere of the Earth also increases.
Note that the heating in the lower regions of our model should in fact be lower at young ages due to the absence of O$_2$ in the atmosphere prior to the Great Oxidation Event.
The largest change is seen for heating of thermal electrons by non-thermal photoelectrons, which we find was around 14 times larger 3~Gyr ago than the modern value.
This is a result of the higher photoionization rates leading to there being more energy in the photoelectron spectrum and the greater ionization fractions of the gas leading to a larger fraction of that energy being transferred to the thermal electrons through elastic collisions, as opposed to being transferred to the neutrals through inelastic collisions.

\section{Discussion} \label{sect:discussion}


In this paper, we develop and validate a physical model for the thermal and chemical structures of planetary upper atmospheres.
We use this model to estimate how the Earth's upper atmosphere would look if the CO$_2$ abundance was varied by a large amount, and to explore the response of the upper atmosphere to the evolving XUV spectrum of the Sun. 
Increasing the CO$_2$ abundance causes the Earth's upper atmosphere to be much cooler and contracted, presumably causing a dramatic decrease in atmospheric losses.
This could also be important for the evolution of the Earth's atmosphere since without some form of protection, such as the enhanced CO$_2$ abundances expected during earlier epochs (\citealt{zahnle2010earth}), interactions with the solar wind could have stripped away the early Earth's atmosphere (\citealt{Lichtenegger10}). 
This is also important because, based on the examples of Venus, Mars, and likely the early Earth, CO$_2$ dominated atmospheres are likely common among terrestrial exoplanets.

Our physical model should be applicable to a range of situations with arbitrary atmospheric compositions and stellar input XUV and IR spectra.
For this purpose, we have attempted to develop the model in such a way that, as much as possible, it is based on first-principles physics.
This means that we have avoided adding free parameters in our model, and we rely as little as possible on parametrised scaling laws that have been developed for the solar system terrestrial planets.
For example, our model does not rely on parametarised heating efficiencies, but instead it calculates the heating rates for each of the contributing physical mechanisms explicitly.
However, our model still contains several weaknesses; for example, our treatment IR cooling should be improved with more detailed treatment of radiation transfer and the detailed interactions between the IR field and the various coolants.

The main reason that we are interested in the thermal and chemical structures of the atmospheres of terrestrial planets is their importance for atmospheric losses into space.
Even just for the case of the Earth, the atmosphere has radically changed many times, especially in the first billion years after the Earth's formation.
Models such as ours are essential tools for understanding this diverse set of atmospheric conditions. 
Estimating atmospheric loss rates atmospheres would require additional detailed modelling of the planetary exospheres (e.g. \mbox{\citealt{Kislyakova13}}) and of the interactions of the ionosphere with the planet's magnetic field (e.g. \mbox{\citealt{Glocer09}}), which we do not attempt in this paper.
In future work, we will combine the model developed in this paper with other such models to gain a more complete understanding of how losses to space influence atmospheric evolution. 


\section{Acknowledgments}

This study was carried out with the support by the FWF NFN project S11601-N16 ``Pathways to Habitability: From Disk to Active Stars, Planets and Life'' and the related subprojects S11604-N16, and S11607-N16.


\appendix

\section{The Kompot Code} \label{appendix:notes}

To solve the physical model described above, we have developed The Kompot Code, which calculates the 1D thermal and chemical structure of a planet's atmosphere.
The underlying equations for the physical model are described in Section~\ref{sect:model}, and the numerical methods used to solve these equations are described in the following appendices.
The code is designed to be very flexible, such that it can be applied to a wide range of atmospheres, and the underlying physics can be easily modified and improved.
The physical input parameters into the code are the temperature and species densities at the lower boundary, and the stellar XUV and IR spectra at the exobase. 
It is written in Fortran and Python and has been parallelized using OpenMP.
The Kompot Code will be made publicly available in the near future and will be obtainable by contacting the authors. 

Evolving the state of the atmosphere forward in time using the system of equations described in Section~\ref{sect:model} is not trivial since they contain many terms that are different in form.
For example, consider Eqn.~\ref{eqn:main_speciescontinuity} in the following form:-
\begin{equation} \label{eqn:}
\frac{\partial n_j}{ \partial t} =  
- \frac{1}{r^2} \frac{\partial \left( r^2  n_j v \right) }{\partial r} 
- \frac{1}{r^2} \frac{\partial \left( r^2 \Phi_{\mathrm{d},j} \right) }{\partial r} 
+ S_j .
\end{equation}
The first term on the RHS is for advection and contains a \mbox{$\partial n_j / \partial r$} term.
The second term is for diffusion and contains multiple \mbox{$\partial ^2 n_j / \partial r ^2$} terms.
The third term describes the chemical sources and is a stiff system of many ordinary differential equations.
No one numerical method is ideally suited to solve all of these problems simultaneously.
We instead use operator splitting to solve each one sequentially using different numerical methods.

The simulation takes place on a finite static spatial grid of cells and we calculate all quantities at cell centres.
For accuracy and efficiency we place our cells in such a way that the cell spacing increases linearly with altitude.
The code only considers the cells below the exobase, the location of which varies within the simulation.

The initial conditions of a simulation should be irrelevant for the final result, but this is only the case when they are not too unrealistic. 
We have three ways to calculate the initial conditions. 
Firstly, we can start the simulation assuming hydrostatic equilibrium, a uniform temperature, and uniform mixing ratios for all species.
For numerical reasons, it is often necessary to evolve the initial chemical structure of the atmosphere by an arbitrary amount of time before starting the simulation, mostly to avoid starting with no ions and electrons in the gas.
Secondly, we can start many simulations simply using the atmospheric structure from the results of previous simulations. 
Thirdly, in fully hydrodynamic simulations, we start the simulation assuming an isothermal Parker wind structure and uniform chemical composition.

We evolve the atmosphere forward in time by performing a large number of timesteps, with length \mbox{$\Delta t$}. 
We determine \mbox{$\Delta t$} by multiplying the minimum time taken for a sound wave to cross one grid cell by a fixed number, called the Courant number, $C$. 
Unless we are applying the full hydrodynamic method given in Appendix~\ref{appendix:hydro}, our simulations are not strictly constrained by the Courant–Friedrichs–Lewy condition that $C$ should be less than unity, though we often assume \mbox{$C \sim 1$} to ensure numerical stability.
At the beginning of a timestep, we first update the XUV, IR, and non-thermal electron spectra at each grid cell; these are then assumed to be constant for the entire timestep. 
We then update the atmospheric properties by applying each of the physical mechanisms successively in the following order: hydrodynamics, chemistry, diffusion, heating/cooling, energy exchange, conduction. 
At the end of each timestep, we then recalculate the exobase location. 

In order to make the simulations computationally more efficient, we generally do not do every part of the above set of steps in every timestep.
It is not necessary to recalculate the XUV and photoelectron spectra and the IR flux every timestep.
Instead, we update these quantities every 100 timesteps.
Similarly, we find that applying the chemical network to evolve the species densities is very time consuming and we get almost identical results by only making this step every 100 timesteps.
When we do evolve the chemistry, we do so for the entire 100 timesteps since it was last evolved.
In most applications of our model, we are interested in the steady-state atmospheric conditions that result in constant input parameters.
To calculate this, we perform successive timesteps starting from the initial conditions until the atmosphere reaches a steady state. 
Our model could also be used to simulate the time evolution of the atmosphere in response to changing input conditions (e.g. a stellar flare).

\section{Hydrodynamics solver} \label{appendix:hydro}

In this appendix, we give the algorithm used for solving the full set of time-dependent hydrodynamic equations.
The algorithm used is an explicit solver that is $2^\mathrm{nd}$ order accurate in time and $3^\mathrm{rd}$ order accurate in space. 
Although not used in this paper, we describe the solver here because it will be used in future studies with this model.

The hydrodynamic equations in spherical coordinates can be written
\begin{equation} \label{eqn:hydro_mass2}
\frac{\partial \rho}{ \partial t} 
+ \frac{\partial \left( \rho v \right) }{\partial r} 
= 
- \frac{2 \rho v}{r} ,
\end{equation}
\begin{equation} \label{eqn:hydro_momentum2}
\frac{\partial ( \rho v ) }{ \partial t} 
+ \frac{\partial \left( \rho v v \right) }{\partial r} 
=
- \frac{\partial p}{\partial r}
- \frac{2 \rho v^2 }{r} 
- \rho g ,
\end{equation}
\begin{equation} \label{eqn:hydro_neutralenergy2}
\frac{\partial e_\mathrm{n} }{ \partial t} 
+  \frac{\partial \left( e_\mathrm{n} v \right) }{\partial r} 
= 
- \frac{\partial \left( p_\mathrm{n} v \right) }{\partial r}
- \frac{2 v \left( e_\mathrm{n} + p_\mathrm{n} \right) }{r} 
- \rho_\mathrm{n} v g ,
\end{equation}
\begin{equation} \label{eqn:hydro_ionenergy2}
\frac{\partial e_\mathrm{i} }{ \partial t} 
+  \frac{\partial \left( e_\mathrm{i} v \right) }{\partial r} 
= 
- \frac{\partial \left( p_\mathrm{i} v \right) }{\partial r}
- \frac{2 v \left( e_\mathrm{i} + p_\mathrm{i} \right) }{r} 
- \rho_\mathrm{i} v g ,
\end{equation}
\begin{equation} \label{eqn:hydro_electronenergy2}
\frac{\partial e_\mathrm{e} }{ \partial t} 
+  \frac{\partial \left( e_\mathrm{e} v \right) }{\partial r} 
= 
- \frac{\partial \left( p_\mathrm{e} v \right) }{\partial r}
- \frac{2 v \left( e_\mathrm{e} + p_\mathrm{e} \right) }{r} 
- \rho_\mathrm{e} v g ,
\end{equation}
The LHSs have the forms of pure advection equations and the RHSs contain source terms for pressure (terms involving \mbox{$\partial /\partial r$}), the spherical geometry (terms involving \mbox{$1/r$}), and gravity (terms involving $g$).
Writing the equations in this way simplifies the problem significantly since we can then solve the advection and sources separately.

The above equations in vector form are
\begin{equation} \label{eqn:hydro_vector}
\frac{\partial \mathbf{U}}{\partial t} + \frac{\partial \mathbf{F}}{\partial r} = \mathbf{S} ,
\end{equation}
where 
\begin{equation} \label{eqn:hydro_vectorUFS}
\mathbf{U} 
= 
\begin{pmatrix}
\rho \\
\rho v \\ 
e_\mathrm{n} \\
e_\mathrm{i} \\
e_\mathrm{e} \\
\end{pmatrix}
, \hspace{0.5mm}
\mathbf{F}  = v \mathbf{U}
, \hspace{0.5mm}
\mathbf{S} 
= 
\begin{pmatrix}
- \frac{2 \rho v}{r}  \\
- \frac{\partial p}{\partial r} - \frac{2 \rho v^2 }{r} - \rho g \\ 
- \frac{\partial \left( p_\mathrm{n} v \right) }{\partial r} - \frac{2 v \left( e_\mathrm{n} + p_\mathrm{n} \right) }{r} - \rho_\mathrm{n} v g \\
- \frac{\partial \left( p_\mathrm{i} v \right) }{\partial r} - \frac{2 v \left( e_\mathrm{i} + p_\mathrm{i} \right) }{r} - \rho_\mathrm{i} v g \\
- \frac{\partial \left( p_\mathrm{e} v \right) }{\partial r} - \frac{2 v \left( e_\mathrm{e} + p_\mathrm{e} \right) }{r} - \rho_\mathrm{e} v g 
\end{pmatrix} .
\end{equation}
Here, $\mathbf{U}$ are the conserved quantities that we want to update, $\mathbf{F}$ are the advection fluxes of these quantities, and $\mathbf{S}$ are the sources. 
In the description below, we use the subscript $j$ to refer to the cell index and the superscript $n$ to refer to the timestep number, such that $\mathbf{U}_j^n$ is the value of $\mathbf{U}$ at the $j$th cell and the $n$th timestep.
The subscript \mbox{$j+1/2$} is used to refer to quantities at the boundary between the $j$th and \mbox{$(j+1)$th} cells at radius \mbox{$r_{j+1/2} = ( r_{j+1} - r_j ) / 2$}.

For the time integration, we split these equations into advection and source steps using Strang splitting.
This means that we first evolve $\mathbf{U}$ by half a timestep due to the sources, then evolve the updated values of $\mathbf{U}$ a full timestep due to advection, and finally evolve $\mathbf{U}$ again a half timestep due to the sources.\footnotemark
~For each one of these updates, we use the $2^\mathrm{nd}$ order Total Variation Diminishing (TVD) Runge-Kutte scheme given by \citet{gottlieb1998total}, given by
\begin{equation} \label{eqn:TVDRK1}
\mathbf{U}^{(1)} = \mathbf{U}^{n} + \Delta t f(\mathbf{U}^{n}) ,
\end{equation}
\begin{equation} \label{eqn:TVDRK2}
\mathbf{U}^{n+1} = \frac{1}{2} \left( \mathbf{U}^{n} + \mathbf{U}^{(1)} \right) + \frac{1}{2} \Delta t f(\mathbf{U}^{(1)}) ,
\end{equation}
where \mbox{$f(\mathbf{U}^{n})$} is \mbox{$\partial \mathbf{U} / \partial t$} calculated from $\mathbf{U}^{n}$.
The first step is the standard Forward Euler method, and the second step improves the approximation.
When doing the two source term updates, \mbox{$\Delta t$} above should be replaced with \mbox{$\Delta t / 2$}.

\footnotetext{
This can be written \mbox{$\mathbf{U}^{n+1} = L_{S,\Delta t / 2} L_{A,\Delta t} L_{S,\Delta t / 2}  \mathbf{U}^n$}, where $L_A$ and $L_S$ are the operators for updating $\mathbf{U}$ by advection and sources respectively.
Similarly, the simpler first-order accurate Gudonov type splitting is written \mbox{$\mathbf{U}^{n+1} = L_{S,\Delta t} L_{A,\Delta t}  \mathbf{U}^n$}.
}

For the advection part (i.e. \mbox{$\mathbf{S}=0$}), Eqn.~\ref{eqn:hydro_vector} can be expressed in terms of the boundary fluxes as
\begin{equation} \label{eqn:advection_flux}
f(\mathbf{U}_j^{n}) 
= 
-\left(\frac{\partial \mathbf{F}}{\partial r} \right)_j^n 
= 
- \frac{ \mathbf{F}_{j+\frac{1}{2}}^n - \mathbf{F}_{j-\frac{1}{2}}^n }{\Delta r_j},
\end{equation}
where \mbox{$\Delta r_j = r_{j+1/2} - r_{j-1/2}$} is the cell width.
To calculate the cell boundary fluxes, we use the MUSCL approach using the high-resolution TVD Lax-Friedrichs numerical flux and the minmod slope limiter (see \mbox{Sections~3.5.1 and 3.5.4} of \citealt{yee1989class}).
In the following discussion, we do not write the superscript $n$ and all quantities should be assumed to be from the $n$th timestep. 
The basic idea of the MUSCL approach is instead of calculating the flux $\mathbf{F}_{j+\frac{1}{2}}$ from the adjacent cell center values of $\mathbf{U}$ (i.e. $\mathbf{U}_{j}$ and $\mathbf{U}_{j+1}$), we calculate it from left and right values of $\mathbf{U}_{j+\frac{1}{2}}$, given by 
\begin{equation} \label{eqn:muscl1}
\mathbf{U}_{j+\frac{1}{2}}^\mathrm{L} 
= 
\mathbf{U}_{j} + \frac{1}{4} \left[ 
\left( 1 - \eta \right) \widetilde{\widetilde{\boldsymbol{\Delta}}}_{j-\frac{1}{2}} 
+ 
\left( 1 + \eta \right) \widetilde{\boldsymbol{\Delta}}_{j+\frac{1}{2}} 
\right] ,
\end{equation}
\begin{equation}
\mathbf{U}_{j+\frac{1}{2}}^\mathrm{R} 
= 
\mathbf{U}_{j+1} - \frac{1}{4} \left[ 
\left( 1 - \eta \right) \widetilde{\boldsymbol{\Delta}}_{j+\frac{3}{2}} 
+ 
\left( 1 + \eta \right) \widetilde{\widetilde{\boldsymbol{\Delta}}}_{j+\frac{1}{2}} 
\right] ,
\end{equation}
where
\begin{equation}
\widetilde{\boldsymbol{\Delta}}_{j+\frac{1}{2}} = \mathrm{minmod}( \boldsymbol{\Delta}_{j+\frac{1}{2}} , \omega \boldsymbol{\Delta}_{j-\frac{1}{2}} ),
\end{equation}
\begin{equation}
\widetilde{\widetilde{\boldsymbol{\Delta}}}_{j+\frac{1}{2}} = \mathrm{minmod}( \boldsymbol{\Delta}_{j+\frac{1}{2}} , \omega \boldsymbol{\Delta}_{j+\frac{3}{2}} ),
\end{equation}
and \mbox{$\boldsymbol{\Delta}_{j+1/2} = \mathbf{U}_{j+1} - \mathbf{U}_j$}.
The quantities $\eta$ and $\omega$ are discussed below.
The minmod slope limiter is given by
\begin{equation}
\mathrm{minmod}( x , \omega y ) = \frac{|x|}{x} \max \left[ 0 , \min \left( |x| ,  \frac{\omega y |x|}{x}  \right) \right].
\end{equation}
In the MUSCL scheme, the TVD Lax-Friedrichs numerical flux is 
\begin{equation}
\begin{aligned}
\mathbf{F}_{j+\frac{1}{2}} 
=
\frac{1}{2}  & \left[ 
\mathbf{F} \left( \mathbf{U}_{j+\frac{1}{2}}^\mathrm{L} \right) 
+ 
\mathbf{F} \left( \mathbf{U}_{j+\frac{1}{2}}^\mathrm{R} \right) 
\right] \\
 & \hspace{10mm} -
\frac{\Delta t}{2 \Delta r_{j+\frac{1}{2}}} \left( \mathbf{U}_{j+\frac{1}{2}}^\mathrm{R} - \mathbf{U}_{j+\frac{1}{2}}^\mathrm{L} \right) ,
\end{aligned}
\end{equation}
where \mbox{$\Delta r_{j+\frac{1}{2}} = r_{j+1} - r_j$}.
To calculate the fluxes on the RHS of this equation, we use
\begin{equation}
\mathbf{F} \left( \mathbf{U}_{j+\frac{1}{2}}^\mathrm{L} \right) 
=
\frac{\left( \rho v \right)_{j+\frac{1}{2}}^\mathrm{L}}{\rho_{j+\frac{1}{2}}^\mathrm{L}} \mathbf{U}_{j+\frac{1}{2}}^\mathrm{L} ,
\end{equation}
\begin{equation} \label{eqn:muscl2}
\mathbf{F} \left( \mathbf{U}_{j+\frac{1}{2}}^\mathrm{R} \right) 
=
\frac{\left( \rho v \right)_{j+\frac{1}{2}}^\mathrm{R}}{\rho_{j+\frac{1}{2}}^\mathrm{R}} \mathbf{U}_{j+\frac{1}{2}}^\mathrm{R}.
\end{equation}
Here, \mbox{$\left( \rho v \right)_{j+\frac{1}{2}}^\mathrm{L}$} and \mbox{$\rho_{j+\frac{1}{2}}^\mathrm{L}$} are elements of \mbox{$\mathbf{U}_{j+\frac{1}{2}}^\mathrm{L}$} and \mbox{$\left( \rho v \right)_{j+\frac{1}{2}}^\mathrm{R}$} and \mbox{$\rho_{j+\frac{1}{2}}^\mathrm{R}$} are elements of \mbox{$\mathbf{U}_{j+\frac{1}{2}}^\mathrm{R}$}. 
The value of $\eta$ chosen determines the spatial order of accuracy of the scheme; we take \mbox{$\eta = 1/3$}, making the scheme third-order accurate in space with an upwind bias. 
Given that value of $\eta$, the value of $\omega$ is an adjustable parameter with a maximum of 4; we assume \mbox{$\omega = 3.5$}.

The equation for the source term updates is
\begin{equation}
f(\mathbf{U}_j^{n}) 
= 
\mathbf{S}_j^n ,
\end{equation}
where $\mathbf{S}$ is given by Eqn.~\ref{eqn:hydro_vectorUFS}.
The only assumption that needs to be made to discreetize $\mathbf{S}$ is for the pressure source terms since they contain spatial derivatives. 
For these terms, we simply use central differencing.

We have performed several tests of our implementation of this algorithm, including using it to calculate the structure of a 1D isothermal stellar wind with a known analytic solution (\citealt{Parker58}).
In this test, our calculations match the analytic solutions exactly.
We have tested our hydrodynamics solver using the Versatile Advection Code (VAC; \citealt{Toth96}) and find that it performs well in the standard Sod shock tube test, though VAC is less diffusive at low spatial resolution.
Finally, we are able to reproduce well the hydrodynamic atmosphere simulations of \citet{Johnstone15}, also performed using VAC.

We must also evolve the densities of individual species due to advection. 
To do this, we convert the cell boundary fluxes for mass into cell boundary fluxes for individual species using a simple upwind approximation for the mixing ratios of individual species at the cell boundaries.
Specifically, the mixing ratios for each species are assumed to be equal to those at the center for the cell that the mass is flowing from, meaning that if the flux is positive at the $j+1/2$ boundary, then the mixing ratio at the $j$th cell is taken.
For the $k$th species, the flux is
\begin{equation} \label{eqn:}
F_{k,j+1/2} 
=
\begin{cases}
n_{k,j} F_{\rho,j+1/2} / \rho_{k,j+1/2}, & \mathrm{if} \hspace{1mm} F_{\rho,j+1/2} > 0, \\
0, & \mathrm{if} \hspace{1mm} F_{\rho,j+1/2} = 0, \\
n_{k,j+1} F_{\rho,j+1/2} / \rho_{k,j+1/2}, & \mathrm{if} \hspace{1mm} F_{\rho,j+1/2} < 0 .
\end{cases}
\end{equation}
With these fluxes, we then calculate the update for the densities in each cell using a simple forward Euler method for the time integration, such that
\begin{equation} \label{eqn:}
n_{k,j}^{n+1}
= 
n_{k,j}^n 
- \frac{\Delta t}{\Delta r_j} \left( F_{k,j+\frac{1}{2}} - F_{k,j-\frac{1}{2}} \right).
\end{equation}
After this update, small inconsistencies between the values of $\rho$ and the values of $n_k$ at each cell are corrected for by scaling the $n_k$ values such that \mbox{$\sum_k m_k n_k = \rho$}.

\section{Tridiagonal matrix algorithm} \label{appendix:diffusion}

The tridiagonal matrix algorithm is a commonly used algorithm and detailed descriptions of the method can be found in many textbooks. 
Since we apply this solver for several different parts of the model, it is necessary to repeat the main steps in the algorithm here for the explanations in the following appendices to make sense. 
The form of the algorithm presented here is adapted from \citet{Bodenheimer07}.

The aim of the algorithm is to solve the tridiagonal systems of equations, which can be written as a system of simultaneous equations, each given by
\begin{equation} \label{eqn:tridiagonal_sim}
a_j x_{j-1} + b_j x_j + c_j x_{j+1} = - d_j.
\end{equation}
In our solvers, we have one of these equations for each grid cell.
In the tridiagonal matrix algorithm, the coefficients $a_j$, $b_j$, $c_j$, and $d_j$ are known, and the aim is to calculate $x_j$, which represents the physical quantity of interest.
The aim in the next three appendices is to derive expressions for these coefficients, and $H_{j}$ and $Y_{j}$ discussed below, for the different physical mechanisms.

To solve this system of equations, consider the equation
\begin{equation} \label{eqn:tridiagonal_lin}
x_j = H_{j-1} x_{j-1} + Y_{j-1}.
\end{equation}
Firstly, the values of $H_{j}$ and $Y_{j}$ need to be calculated for each value of $j$ except for \mbox{$j=J$}.
Assuming we know $H_{J-1}$ and $Y_{J-1}$ which are derived separately for each problem, the other values are calculated by iterating downwards through the grid using
\begin{equation} \label{eqn:}
H_{j-1} = - \frac{a_j}{b_j + c_j H_j}
\end{equation}
and
\begin{equation} \label{eqn:}
Y_{j-1} = - \frac{d_j+c_i Y_i}{b_j + c_j H_j}.
\end{equation}
With $H_{j}$ and $Y_{j}$ known, the values of $x_j$ is calculated at each cell by iterating upwards using Eqn.~\ref{eqn:tridiagonal_lin} and assuming that the value of $x_1$ is known in advance (this is always the case since we use fixed lower boundary values for all quantities).

\section{Solver for semi-static hydrodynamic equations} \label{appendix:energyequationimplicit}

We discuss in this appendix first the method used to solve the energy equations and then the method used to solve the equations for $\rho$ and $v$ when solving the hydrodynamics using the semi-static method presented in Section~\ref{sect:hydro}.
It is convenient to update the energy each timestep using an implicit method since this avoids the restrictively short timestep sizes needed in the explicit schemes.
The hydrodynamic part of the energy equation including gravity is
\begin{equation} \label{eqn:}
\frac{\partial e }{ \partial t} =
 - \frac{1}{r^2} \frac{\partial \left[ r^2 \Phi_e \right] }{\partial r} 
 - \rho v g  ,
\end{equation}
where \mbox{$\Phi_e = v \left( e + p \right)$}.
For this section, we do not write the subscripts n, i, and e for the neutral, ion, and electron components; the method presented here is used to solve the energy equations for each component separately.
Using the Crank-Nicolson method for time discreetization, and the cell boundary values to discreetize cell centered spatial derivatives, we get
\begin{equation} \label{eqn:}
\begin{aligned}
\frac{ e_j^{n+2} - e_j^n }{ \Delta t} 
= &
 - \frac{1}{2 r_j^2} \frac{\left( r_{j+\frac{1}{2}}^2 \Phi_{e,j+\frac{1}{2}}^{n+1} - r_{j-\frac{1}{2}}^2 \Phi_{e,j-\frac{1}{2}}^{n+1} \right)}{\Delta r_j} \\
 & - \frac{1}{2 r_j^2} \frac{ \left( r_{j+\frac{1}{2}}^2 \Phi_{e,j+\frac{1}{2}}^{n} - r_{j-\frac{1}{2}}^2 \Phi_{e,j-\frac{1}{2}}^{n} \right) }{\Delta r_j} \\
& - \rho_j v_j g_j  .
\end{aligned}
\end{equation}
This can be rewritten 
\begin{equation} \label{eqn:energyimplicit1}
\begin{aligned}
 e_j^{n+2} & + k_j \left( r_{j+\frac{1}{2}}^2 \Phi_{e,j+\frac{1}{2}}^{n+1} - r_{j-\frac{1}{2}}^2 \Phi_{e,j-\frac{1}{2}}^{n+1} \right)
= \\
 & e_j^n - k_j \left( r_{j+\frac{1}{2}}^2 \Phi_{e,j+\frac{1}{2}}^{n} - r_{j-\frac{1}{2}}^2 \Phi_{e,j-\frac{1}{2}}^{n} \right)
- \rho_j v_j g_j \Delta t ,
\end{aligned}
\end{equation}
where
\begin{equation}
k_j = \frac{\Delta t}{2 r_j^2 \Delta r_j}.
\end{equation}
To get $\Phi_e$ at cell boundaries, we assume that the cell boundary values of spatially variable quantities are the average of their ell center values, e.g. \mbox{$e_{j+\frac{1}{2}} = (e_{j+1} + e_j)/2$}.
Therefore, the cell boundary fluxes is given by
\begin{equation} \label{eqn:energyfluximplicit}
\begin{aligned}
\Phi_{e,j+\frac{1}{2}}^{n} = & \frac{1}{2} \gamma_{j+\frac{1}{2}} v_{j+\frac{1}{2}} e_j^{n} + \frac{1}{2} \gamma_{j+\frac{1}{2}} v_{j+\frac{1}{2}} e_{j+1}^{n} \\ 
 & - \frac{1}{2} (\gamma_{j+\frac{1}{2}} - 1) \rho_{j+\frac{1}{2}} v_{j+\frac{1}{2}}^3.
\end{aligned}
\end{equation}
where we have used \mbox{$\Phi_e = v (e+p) = \gamma e v - \frac{1}{2} (\gamma - 1) \rho v^3$}.
The equations for $\Phi_{e,j-\frac{1}{2}}^{n}$, $\Phi_{e,j+\frac{1}{2}}^{n+1}$, and $\Phi_{e,j-\frac{1}{2}}^{n+1}$ have the same form and can be obtained with the appropriate substitutions of the subscripts and superscripts.

After substituting the equations for the cell boundary fluxes into Eqn.~\ref{eqn:energyimplicit1}, the resulting equation can be expressed in the form of Eqn.~\ref{eqn:tridiagonal_sim} with
\begin{equation}
a_j = - \frac{1}{2} k_j \gamma_{j-\frac{1}{2}} v_{j-\frac{1}{2}} r_{j-\frac{1}{2}}^2 ,
\end{equation}
\begin{equation}
b_j = 1 + \frac{1}{2} k_j \left( 
\gamma_{j+\frac{1}{2}} v_{j+\frac{1}{2}} r_{j+\frac{1}{2}}^2
- 
\gamma_{j-\frac{1}{2}} v_{j-\frac{1}{2}} r_{j-\frac{1}{2}}^2 
\right) ,
\end{equation}
\begin{equation}
c_j = \frac{1}{2} k_j \gamma_{j+\frac{1}{2}} v_{j+\frac{1}{2}} r_{j+\frac{1}{2}}^2 ,
\end{equation}
\begin{equation}
d_j = a_j e_{j-1}^n + ( 2 - b_j ) e_j^n + c_j e_{j+1}^n - \rho_j v_j g_j \Delta t .
\end{equation}

At the lower boundary, the energies are all assumed to be fixed, such that \mbox{$e_1^{n+1} = e_1^n$}.
At the outer boundary, we apply Eqn.~\ref{eqn:tridiagonal_lin} giving 
\begin{equation}
e_J^{n+1} = H_{J-1} e_{J-1}^{n+1} + Y_{J-1} .
\end{equation}
What we need are the values of $H_{J-1}$ and $Y_{J-1}$ ($H_{J}$ and $Y_{J}$ are not needed).
If we assume that the outer advective energy flux, $\Phi_{\mathrm{e,out}}$, is known and is a constant over the timestep, such that \mbox{$\Phi_{\mathrm{e,out}} = \Phi_{e,J+1/2}^n = \Phi_{e,J+1/2}^{n+1}$}, then Eqn.~\ref{eqn:energyimplicit1} can be written at the outer boundary as
\begin{equation} \label{eqn:energyimplicitoter}
\begin{aligned}
e_J^{n+2} & - r_{J-\frac{1}{2}}^2 \Phi_{e,J-\frac{1}{2}}^{n+1} 
= \\
& e_J^n - 2 k_J r_{J+\frac{1}{2}}^2 \Phi_{\mathrm{e,out}} - k_J r_{J-\frac{1}{2}}^2 \Phi_{e,J-\frac{1}{2}}^{n}
- \rho_J v_J g_J \Delta t .
\end{aligned}
\end{equation}
Inserting Eqn.~\ref{eqn:energyfluximplicit} for $\Phi_{e,J-\frac{1}{2}}^{n+1}$ into the above equation gives
\begin{equation}
H_{J-1}  \left( 1 - \frac{1}{2} k_J r_{J-\frac{1}{2}}^2 \gamma_{J-\frac{1}{2}} v_{J-\frac{1}{2}} \right) 
=
\frac{1}{2} k_J r_{J-\frac{1}{2}}^2 \gamma_{J-\frac{1}{2}} v_{J-\frac{1}{2}}  ,
\end{equation}
\begin{equation}
\begin{aligned}
Y_{J-1} & \left( 1 - \frac{1}{2} k_J r_{J-\frac{1}{2}}^2 \gamma_{J-\frac{1}{2}} v_{J-\frac{1}{2}} \right) 
= \\
 & e_J^n - 2 k_J r_{J+\frac{1}{2}}^2 \Phi_{\mathrm{e,out}} - k_J r_{J-\frac{1}{2}}^2 \Phi_{e,J-\frac{1}{2}}^{n} \\
 & - \rho_J v_J g_J \Delta t - \frac{1}{2} k_J r_{J-\frac{1}{2}}^2 \left( \gamma_{J-\frac{1}{2}} - 1 \right) \rho_{J-\frac{1}{2}} v_{J-\frac{1}{2}}^3 .
\end{aligned}
\end{equation}
These are used to calculate $H_{J-1}$ and $Y_{J-1}$.
In this paper, we set $\Phi_{\mathrm{e,out}}$ to zero. 

To calculate $v$ at all grid cells, we assume that $v$ at the exobase is already known and integrate downwards from the exobase to the lower boundary of our simulation domain using Eqn.~\ref{eqn:semistatic1}.
We use the Crank-Nicolson discretisation of the \mbox{$dv/dr$} term, which gives
\begin{equation}
v_j = v_{j+1} - \frac{1}{2} ( r_{j+1} - r_j ) ( F_{j+1} + F_j ),
\end{equation}
where \mbox{$F_j = (dv/dr)_j$} is calculated from Eqn.~\ref{eqn:semistatic1}.
When integrating from the $(j+1)$th to the $j$th cell, we assume \mbox{$dT/dr =  ( T_{j+1} - T_j ) / ( r_{j+1} - r_j )$} and make a similar assuming for $d\bar{m}/dr$.
To get a first estimate of $v_j$, we assume \mbox{$F_j = F_{j+1}$}, and the above equation becomes the Forward Euler method.
We then iteratively improve this estimate using Newton iteration, given by
\mbox{$v_j^{(m+1)} = v_j^{(m)} -  G_j^{(m)} / G_j'^{(m)} $}, 
where the superscript $(m)$ indicates the quantity is for the $m$th iteration.
The functions $G_j^{(m)}$ and $G_j'^{(m)}$ are given by
\begin{equation}
G_j^{(m)} =  v_j^{(m)} - v_{j+1} + \frac{1}{2} ( r_{j+1} - r_j ) ( F_{j+1} + F_j^{(m)} ),
\end{equation}
\begin{equation}
G_j'^{(m)} = \frac{ dG_j }{ dv_j } = 1 + \frac{1}{2} ( r_{j+1} - r_j ) F_j'^{(m)},
\end{equation}
where \mbox{$F_j' = dF_j / dv_j$}.
Differentiating Eqn.~\ref{eqn:semistatic1} with respect to $v$ gives
\begin{equation}
F_j' = \left( \frac{ v_j^{-2} + v_{0j}^{-2} }{ v_j^{-1} - v_{0j}^{-2} v_j } \right) F_j .
\end{equation}
We iteratively improve our estimate of $v_j$ until \mbox{$|1-v_j^{(m+1)}/v_j^{(m)} | < 10^{-5}$}, which indicates that the solution has converged.
To calculate $\rho$, we integrate upwards through the simulation domain using Eqn.~\ref{eqn:semistatic2} and the method is essentially the same as the method for $v$.

\section{Solver for diffusion equations} \label{appendix:diffusion}

In spherical coordinates, the equation for the rate of change of the density of a species at a certain point in space is 
\begin{equation} \label{eqn:diffusionpdf}
\frac{\partial n}{\partial t} = - \frac{1}{r^2} \frac{\partial (r^2 \Phi)}{\partial r} ,
\end{equation}
where $n$ is the species density and $\Phi$ is the diffusion flux given by Eqns.~\ref{eqn:diffusionflux} and \ref{eqn:diffusionvel}.
For this section, we remove the subscript d from the flux, such that \mbox{$\Phi = \Phi_\mathrm{d}$}.
We also remove the subscript that indicates which species the quantities refer to, and apply this method to each species separately.
As before, the subscript $j$ refers to the $j$th radial cell and the superscript $n$ refers to the $n$th timestep.

Consider the $j$th radial cell in the grid and assume that the subscripts \mbox{$j-1/2$} and \mbox{$j+1/2$} refer to the boundaries of this cell, meaning that the cell boundary diffusion fluxes are given by $\Phi_{j-1/2}$ and $\Phi_{j+1/2}$. 
For the spatial discreetisation, the spatial derivative in Eqn.~\ref{eqn:diffusionpdf} can be written
\begin{equation} \label{eqn:diffusion_spacediscreet}
\frac{\partial (r^2 \Phi)}{\partial r} = \frac{ r_{j+\frac{1}{2}}^2 \Phi_{j+\frac{1}{2}} - r_{j-\frac{1}{2}}^2 \Phi_{j-\frac{1}{2}} }{ \Delta r_j } ,
\end{equation}
where \mbox{$\Delta r_j = r_{j+1/2}-r_{j-1/2}$}.
For the time discreetisation, we use the Crank-Nicolson method, such that the time derivative in Eqn.~\ref{eqn:diffusionpdf} is written
\begin{equation} \label{eqn:diffusion_timediscreet}
\frac{ n_{j}^{n+1} - n_{j}^n }{ \Delta t }  = \frac{1}{2} \left[ \left( \frac{\partial n}{\partial t} \right)^{n+1} + \left( \frac{\partial n}{\partial t} \right)^n \right] ,
\end{equation}
where \mbox{$\Delta t = t^{n+1} - t^n$}.
The fact that the rate of change at the end of the update, which depends on the result of the update, is used to do the update itself is the reason that this method is implicit. 
These equations can be combined to give
\begin{equation} \label{eqn:diffusion_cranknicolson}
\begin{aligned}
n_{j}^{n+1} + & k_j \left[ 
r_{j+\frac{1}{2}}^2 \Phi_{j+\frac{1}{2}}^{n+1} - r_{j-\frac{1}{2}}^2 \Phi_{j-\frac{1}{2}}^{n+1} \right]
= \\
 & n_{j}^n - k_j \left[ 
r_{j+\frac{1}{2}}^2 \Phi_{j+\frac{1}{2}}^n - r_{j-\frac{1}{2}}^2 \Phi_{j-\frac{1}{2}}^n \right] ,
\end{aligned}
\end{equation}
where
\begin{equation}
k_j = \frac{ \Delta t }{ 2 r_j^2 \Delta r_j } .
\end{equation}

To simplify the evaluation of Eqn.~\ref{eqn:diffusion_cranknicolson}, the diffusion flux (Eqn.~\ref{eqn:diffusionflux} and Eqn.~\ref{eqn:diffusionvel}) can be rewritten as
\begin{equation} \label{eqn:diffusionflux_rewritten}
\begin{aligned}
\Phi 
= &
- (D + K_\mathrm{E}) \frac{\partial n}{\partial r} 
+ \frac{1}{N} \frac{\partial N}{\partial r} \left( \frac{m}{\bar{m}} D + K_\mathrm{E} \right) n\\
 & - D \left( 1 + \alpha_\mathrm{T} - \frac{m}{\bar{m}} \right) \frac{1}{T} \frac{\partial T}{\partial r} n,
\end{aligned}
\end{equation}
where the quantities are defined under Eqn.~\ref{eqn:diffusionvel}.
For simplicity, we assume that the cell boundary values are the averages of the cell center values, such that \mbox{$n_{j+\frac{1}{2}} = \frac{1}{2} ( n_{j} + n_{j+1} )$}.
For the derivatives, we use a simple central difference, such that \mbox{$\left( \partial n / \partial r \right)_{j+1/2} = ( n_{j+1} - n_{j} ) / ( r_{j+1} - r_j )$}.
We make the same assumptions for all other quantities when necessary. 
Since the cell width is not constant, the term \mbox{$r_{j+1} - r_j$} is different from the widths of the cells, given by $\Delta r_j$.
Inserting these into Eqn.~\ref{eqn:diffusionflux_rewritten} and rearranging gives
\begin{equation} \label{eqn:diffflux_upper}
\begin{aligned}
\Phi_{j+\frac{1}{2}}
= &
\left[ f_{j+\frac{1}{2}} + \frac{ D_{j+\frac{1}{2}} + K_{E,j+\frac{1}{2}} }{ r_{j+1} - r_j } \right] n_{j} \\
 & +
\left[ f_{j+\frac{1}{2}} - \frac{ D_{j+\frac{1}{2}} + K_{E,j+\frac{1}{2}} }{ r_{j+1} - r_j } \right] n_{j+1} ,
\end{aligned}
\end{equation}
where
\begin{equation} \label{eqn:diffflux_fi}
\begin{aligned}
f_{j+\frac{1}{2}}
= &
\frac{1}{2} \left( \frac{m}{\bar{m}_{j+\frac{1}{2}}} D_{j+\frac{1}{2}} + K_{E,j+\frac{1}{2}} \right) \frac{1}{N_{j+\frac{1}{2}}} \left( \frac{\partial N}{\partial r} \right)_{j+\frac{1}{2}} \\
 & -
\frac{1}{2} D_{j+\frac{1}{2}} \left( 1 + \alpha_\mathrm{T} - \frac{m}{\bar{m}_{j+\frac{1}{2}}} \right) \frac{1}{T_{j+\frac{1}{2}}} \left( \frac{\partial T}{\partial r} \right)_{j+\frac{1}{2}} .
\end{aligned}
\end{equation}
The equation for the flux at the other boundary is
\begin{equation} \label{eqn:diffflux_lower}
\begin{aligned}
\Phi_{j-\frac{1}{2}}
= &
\left[ f_{j-\frac{1}{2}} + \frac{ D_{j-\frac{1}{2}} + K_{E,j-\frac{1}{2}} }{ r_j - r_{j-1} } \right] n_{j-1} \\
 & +
\left[ f_{j-\frac{1}{2}} - \frac{ D_{j-\frac{1}{2}} + K_{E,j-\frac{1}{2}} }{ r_j - r_{j-1} } \right] n_{j} ,
\end{aligned}
\end{equation}
where the equation for $f_{j-\frac{1}{2}}$ can be obtained simply by substituting the subscripts \mbox{$j+1/2$} with \mbox{$j-1/2$} in Eqn.~\ref{eqn:diffflux_fi}.

A fundamental assumption that is made here is that all quantities except the species densities, $n$, in Eqn.~\ref{eqn:diffflux_upper} and Eqn.~\ref{eqn:diffflux_lower} are constant over the diffusion timestep and can be calculated using the state of the simulation at the beginning of the timestep. 
This is not strictly true since several of the quantities, most obviously $\bar{m}$, themselves evolve due to the changing values of $n$. 
This means that the $\Phi_{j+\frac{1}{2}}^n$ term can be obtained by replacing the $n_{j}$ and $n_{j+1}$ terms in Eqn.~\ref{eqn:diffflux_upper} with $n_{j}^n$ and $n_{j+1}^n$; similar substitutions can be made for $\Phi_{j+\frac{1}{2}}^{n+1}$, $\Phi_{j-\frac{1}{2}}^n$, and $\Phi_{j-\frac{1}{2}}^{n+1}$.

Using Eqns.~\ref{eqn:diffflux_upper}--\ref{eqn:diffflux_lower}, it is possible to rewrite Eqn.~\ref{eqn:diffusion_cranknicolson} in the form of Eqn.~\ref{eqn:tridiagonal_sim}, with \mbox{$x_j=n_j^{n+1}$} and the coefficients being given by
\begin{equation} \label{eqn:}
\begin{aligned}
a_j
= 
- k_j r_{j-\frac{1}{2}}^2
\left(
f_{j-\frac{1}{2}}
+
\frac{ D_{j-\frac{1}{2}} + K_{E,j-\frac{1}{2}} }{ r_j - r_{j-1} } 
\right)  ,
\end{aligned}
\end{equation}
\begin{equation} \label{eqn:}
\begin{aligned}
b_j
= 
1 + & k_j 
r_{j+\frac{1}{2}}^2 
\left( 
f_{j+\frac{1}{2}}
+
\frac{ D_{j+\frac{1}{2}} + K_{E,j+\frac{1}{2}} }{ r_{j+1} - r_j } 
\right) \\
& -
 k_j r_{j-\frac{1}{2}}^2 
\left( 
f_{j-\frac{1}{2}}
-
\frac{ D_{j-\frac{1}{2}} + K_{E,j-\frac{1}{2}} }{ r_j - r_{j-1} } 
\right) ,
\end{aligned}
\end{equation}
\begin{equation} \label{eqn:}
\begin{aligned}
c_j
= 
k_j r_{j+\frac{1}{2}}^2
\left(
f_{j+\frac{1}{2}}
-
\frac{ D_{j+\frac{1}{2}} + K_{E,j+\frac{1}{2}} }{ r_{j+1} - r_j } 
\right) ,
\end{aligned}
\end{equation}
\begin{equation} \label{eqn:}
\begin{aligned}
d_j = a_j n_{j-1}^n - \left( 2 - b_j \right) n_j^n + c_j n_{j+1}^n .
\end{aligned}
\end{equation}
At the lower boundary, the densities are all assumed to be fixed, such that \mbox{$n_1^{n+1} = n_1^n$}.
At the outer boundary, the values of $H_{J-1}$ and $Y_{J-1}$ are needed.
Assume the outward diffusion flux at the outer boundary is known and given by $\Phi_{\mathrm{out}}$.
Assuming also that the outward flux is a constant over the diffusion timestep, such that \mbox{$\Phi_\mathrm{out} = \Phi_{J+\frac{1}{2}}^n = \Phi_{J+\frac{1}{2}}^{n+1}$}, Eqn.~\ref{eqn:diffusion_cranknicolson} can be rewritten as
\begin{equation} \label{eqn:diffusion_upperboundary}
\begin{aligned}
n_{J}^{n+1} 
= & 
n_{J}^n 
- 2 k_J r_{J+\frac{1}{2}}^2 \Phi_\mathrm{out}
- k_J r_{J-\frac{1}{2}}^2 \left( \Phi_{J-\frac{1}{2}}^{n+1} + \Phi_{J-\frac{1}{2}}^n \right).
\end{aligned}
\end{equation}
Insertng Eqn.~\ref{eqn:diffflux_lower} for $\Phi_{J-\frac{1}{2}}^{n+1}$, this can be rewritten as
\begin{equation} \label{eqn:}
n_{J}^{n+1} = H_{J-1} n_{J-1}^{n+1} + Y_{J-1} ,
\end{equation}
where
\begin{equation} \label{eqn:}
H_{J-1} = \frac{ 
k_J r_{J-\frac{1}{2}}^2 \left( f_{J-\frac{1}{2}}
+
\frac{ D_{J-\frac{1}{2}} + K_{E,J-\frac{1}{2}} }{ r_J - r_{J-1} }  \right) 
}{
1
-
k_J r_{J-\frac{1}{2}}^2 \left( f_{J-\frac{1}{2}}
-
\frac{ D_{J-\frac{1}{2}} + K_{E,J-\frac{1}{2}} }{ r_J - r_{J-1} }  \right) 
} ,
\end{equation}
\begin{equation} \label{eqn:}
Y_{J-1} = \frac{ 
n_J^n - 2 k_J r_{J+\frac{1}{2}}^2 \Phi_{\mathrm{out}} + k_J r_{J-\frac{1}{2}}^2 \Phi_{J-\frac{1}{2}}^n 
}{
1
-
k_J r_{J-\frac{1}{2}}^2 \left( f_{J-\frac{1}{2}}
-
\frac{ D_{J-\frac{1}{2}} + K_{E,J-\frac{1}{2}} }{ r_J - r_{J-1} }  \right) 
} ,
\end{equation}
where $\Phi_{J-\frac{1}{2}}^n$ can be obtained by Eqn.~\ref{eqn:diffflux_lower}.
If outflow conditions are desired at the upper boundary instead of imposing an outward flux, then \mbox{$n_J^{n+1} = n_{J-1}^{n+1}$} and these expressions should be replaced with \mbox{$H_{J-1} = 1$} and \mbox{$Y_{J-1} = 0$}.

\section{Solver for conduction equations} \label{appendix:conduction}

In this appendix, we give the method for solving a general conduction equation with the same form as Eqn.~\ref{eqn:ionconduction}.
This can easily be applied therefore to solving the conduction equations for the ion and electron gases; for the conduction equation for the neutral gas, which includes extra terms related to eddy conduction, the necessary modifications are given at the end of this section. 
The conduction equation to solve is 
\begin{equation} \label{eqn:conductionflux}
\frac{\partial e}{\partial t} = - \frac{1}{r^2} \frac{\partial \left( r^2 \Phi_\mathrm{c} \right)}{\partial r} ,
\end{equation}
where the energy flux, $\Phi_\mathrm{c}$, is given by
\begin{equation} \label{eqn:}
\Phi_\mathrm{c} = - \kappa \frac{\partial T}{\partial r} .
\end{equation}
We assume here that the conductivity is a constant over the timestep, such that \mbox{$\kappa=\kappa^n=\kappa^{n+1}$} and can be calculated using the state of the system at the beginning of the conduction timestep.
This is not strictly true, since the conductivity is itself temperature dependent. 

Since over the conduction timestep, $e$ evolves only due to changes in the temperature, we can write
\begin{equation} \label{eqn:e_evolve_therm}
\frac{\partial e}{\partial t} = \frac{n k_\mathrm{B}}{\left(\gamma-1 \right) } \frac{\partial T}{\partial t} .
\end{equation}
Using cell boundary values to discreetize cell centered spatial derivatives, the Crank-Nicolson discreetization of Eqn.~\ref{eqn:conductionflux} is 
\begin{equation} \label{eqn:}
\begin{aligned}
T_j^{n+1} & - T_j^n = - \frac{\left( \gamma_j - 1 \right) \Delta t}{2 r_j^2 k_\mathrm{B} \Delta r_j } \times \\
 & \left( r_{j+\frac{1}{2}}^2 \Phi_{j+\frac{1}{2}}^{n+1} - r_{j-\frac{1}{2}}^2 \Phi_{j-\frac{1}{2}}^{n+1} + r_{j+\frac{1}{2}}^2 \Phi_{j+\frac{1}{2}}^{n} - r_{j-\frac{1}{2}}^2 \Phi_{j-\frac{1}{2}}^{n} \right) ,
\end{aligned}
\end{equation}
which can be rewritten as
\begin{equation} \label{eqn:conduction_cranknicolson}
\begin{aligned}
T_j^{n+1} + k_j & \left( r_{j+\frac{1}{2}}^2 \Phi_{j+\frac{1}{2}}^{n+1} - r_{j-\frac{1}{2}}^2 \Phi_{j-\frac{1}{2}}^{n+1} \right) \\ 
 & = 
T_j^n - k_j \left( r_{j+\frac{1}{2}}^2 \Phi_{j+\frac{1}{2}}^{n} - r_{j-\frac{1}{2}}^2 \Phi_{j-\frac{1}{2}}^{n} \right) ,
\end{aligned}
\end{equation}
where
\begin{equation} \label{eqn:}
k_j = \frac{\left( \gamma_j - 1 \right) \Delta t}{2 r_j^2 k_\mathrm{B} \Delta r_j } .
\end{equation}

For the cell boundary fluxes, we assume
\begin{equation} \label{eqn:condfluxJph}
\Phi_{j+\frac{1}{2}}^n = - \kappa_{j+\frac{1}{2}} \frac{ T_{j+1}^n - T_{j}^n }{ r_{j+1} - r_{j} },
\end{equation}
\begin{equation} \label{eqn:condfluxJmh}
\Phi_{j-\frac{1}{2}}^n = - \kappa_{j-\frac{1}{2}} \frac{ T_{j}^n - T_{j-1}^n }{ r_{j} - r_{j-1} },
\end{equation}
where \mbox{$\kappa_{j+\frac{1}{2}} = \left( \kappa_{j+1} + \kappa_{j} \right) / 2$} and \mbox{$\kappa_{j-\frac{1}{2}} = \left( \kappa_{j} + \kappa_{j-1} \right) / 2$}.
We can combine these two relations to get
\begin{equation} \label{eqn:}
\begin{aligned}
r_{j+\frac{1}{2}}^2 & \Phi_{j+\frac{1}{2}}^{n} - r_{j-\frac{1}{2}}^2 \Phi_{j-\frac{1}{2}}^{n}  \\
 & =
-f_{j+\frac{1}{2}} T_{j+1}^{n} + \left( f_{j+\frac{1}{2}} + f_{j-\frac{1}{2}} \right) T_{j}^{n} - f_{j-\frac{1}{2}} T_{j-1}^{n} ,
\end{aligned}
\end{equation}
where
\begin{equation} \label{eqn:conductionfluxdifference}
f_{j+\frac{1}{2}} = \frac{ r_{j+\frac{1}{2}}^2 \kappa_{j+\frac{1}{2}} }{ r_{j+1} - r_j }.
\end{equation}
Similarly for $\Phi^{n+1}$, we get 
\begin{equation} \label{eqn:}
\begin{aligned}
r_{j+\frac{1}{2}}^2 & \Phi_{j+\frac{1}{2}}^{n+1} - r_{j-\frac{1}{2}}^2 \Phi_{j-\frac{1}{2}}^{n+1}  \\
 & =
-f_{j+\frac{1}{2}} T_{j+1}^{n+1} + \left( f_{j+\frac{1}{2}} + f_{j-\frac{1}{2}} \right) T_{j}^{n+1} - f_{j-\frac{1}{2}} T_{j-1}^{n+1} .
\end{aligned}
\end{equation}

By inserting Eqn.~\ref{eqn:conductionfluxdifference} into Eqn.~\ref{eqn:conduction_cranknicolson}, an equation in the form of Eqn.~\ref{eqn:tridiagonal_sim} can be derived where \mbox{$x_j=T_j^n$} and 
\begin{equation} \label{eqn:}
a_j = - k_j f_{j-\frac{1}{2}} ,
\end{equation}
\begin{equation} \label{eqn:}
b_j = 1 +  k_j \left( f_{j+\frac{1}{2}} + f_{j-\frac{1}{2}} \right) ,
\end{equation}
\begin{equation} \label{eqn:}
c_j = - k_j f_{j+\frac{1}{2}} ,
\end{equation}
\begin{equation} \label{eqn:cond_e}
\begin{aligned}
e_j = a_j T_{j-1}^n - \left( 2 - b_j \right) T_j^n + c_j T_{j+1}^n .
\end{aligned}
\end{equation}
For the lower boundary, we assume \mbox{$T_1^{n+1} = T_1^n$}.
For the upper boundary, it is necessary to know \mbox{$H_{J-1}$} and \mbox{$Y_{J-1}$}.
If we assume that the outward conductive energy flux, $\Phi_\mathrm{c,out}$, is known in advance, and is constant over the timestep, we can rewrite Eqn.~\ref{eqn:conduction_cranknicolson} for the final cell as
\begin{equation} \label{eqn:}
\begin{aligned}
T_J^{n+1} 
=
T_J^n - 2 k_J r_{J+\frac{1}{2}}^2 \Phi_\mathrm{c,out} + k_J r_{J-\frac{1}{2}}^2 \left( \Phi_{J-\frac{1}{2}}^{n+1} + \Phi_{J-\frac{1}{2}}^{n} \right).
\end{aligned}
\end{equation} 
By inserting Eqn.~\ref{eqn:condfluxJmh} for $\Phi_{J-\frac{1}{2}}^{n+1}$, we get
\begin{equation} \label{eqn:}
H_{J-1} = \frac{k_J f_{ J - \frac{1}{2}} }{ 1 + k_J f_{J-\frac{1}{2}} } ,
\end{equation} 
\begin{equation} \label{eqn:cond_Y}
Y_{J-1} = \frac{
T_J^n - 2 k_J r_{J+\frac{1}{2}}^2 \Phi_\mathrm{c,out} + k_J r_{J-\frac{1}{2}}^2 \Phi_{J-\frac{1}{2}}^{n}
}{ 1 + k_J f_{J-\frac{1}{2}} },
\end{equation} 
where $\Phi_{J-\frac{1}{2}}^{n}$ should be calculated using Eqn.~\ref{eqn:condfluxJmh}.

The conduction equation for the neutral gas is slightly more complicated than the simple conduction equation solved here due to the additional eddy conduction process. 
If the conductivity is taken to be the sum of the molecular and eddy conductivities, \mbox{$\kappa = \kappa_\mathrm{mol} + \kappa_\mathrm{eddy}$}, then the full neutral conduction equation is
\begin{equation} \label{eqn:conductionflux_neutral}
\frac{\partial e}{\partial t} = - \frac{1}{r^2} \frac{\partial \left( r^2 \Phi_\mathrm{c} \right)}{\partial r} + \frac{1}{r^2} \frac{\partial \left( r^2 \rho g K_\mathrm{E} \right)}{\partial r},
\end{equation}
where $\rho$ is the mass density, $g$ is the gravitational acceleration, and $K_\mathrm{E}$ is the eddy diffusion coefficient.
Since the additional term on the RHS can be assumed to be constant over the conduction timestep, it only leads to additional terms in Eqn.~\ref{eqn:cond_e} for $e_j$ and in Eqn.~\ref{eqn:cond_Y} for $Y_{J-1}$.
These two equations should be replaced with
\begin{equation} \label{eqn:cond_e2}
\begin{aligned}
e_j = & a_j T_{j-1}^n - \left( 2 - b_j \right) T_j^n + c_j T_{j+1}^n \\
 & - \frac{\left( \gamma_j - 1 \right) \Delta t}{2 k_\mathrm{B} } \left[ \rho_j \frac{K_{\mathrm{E},j+1}-K_{\mathrm{E},j-1}}{r_{j+1}-r_{j-1}} + K_{\mathrm{E},j} \frac{\rho_{j+1}-\rho_{j-1}}{r_{j+1}-r_{j-1}} \right]
\end{aligned}
\end{equation}
and
\begin{equation} \label{eqn:}
\begin{aligned}
Y_{J-1} = & \frac{
T_J^n - 2 k_J r_{J+\frac{1}{2}}^2 \Phi_\mathrm{c,out} + k_J r_{J-\frac{1}{2}}^2 \Phi_{J-\frac{1}{2}}^{n}
}{ 1 + k_J f_{J-\frac{1}{2}} } \\
 & + \frac{
\frac{\left( \gamma_j - 1 \right) \Delta t}{2 k_\mathrm{B} } \left[ \rho_j \frac{K_{\mathrm{E},j+1}-K_{\mathrm{E},j-1}}{r_{j+1}-r_{j-1}} + K_{\mathrm{E},j} \frac{\rho_{j+1}-\rho_{j-1}}{r_{j+1}-r_{j-1}} \right]
}{ 1 + k_J f_{J-\frac{1}{2}} }.
\end{aligned}
\end{equation} 
The extra terms have been derived by first expanding the derivative in the final term in Eqn.~\ref{eqn:conductionflux_neutral} using the product rule and then using central differencing on the resulting derivatives.

\section{Solver for energy exchange} \label{appendix:heatexchange}

Evolving the energies of the neutral, ion, and electron gases due to energy exchange is not completely trivial.
Low in the atmosphere where the gas densities are high, the exchange rates can be large, meaning that restrictively small timesteps would be needed if an explicit integration scheme was adopted. 
To avoid this problem, we implement the energy exchange using the implicit Crank-Nicolson method.
Unlike in the previous three appendices, we do not use the tridiagonal matrix algorithm here.

The energy exchange equations involve a large number of terms representing different forms of exchange between many different pairs of species.
Most equations are not in a form that allows them to easily be solved implicitly.
To simplify the problem, we assume that the exchange rates vary proportionally to the temperature differences between the components. 
This gives
\begin{equation} \label{eqn:implicitQeiassumption}
Q_\mathrm{ei} = k_\mathrm{ei} \left( T_\mathrm{e} - T_\mathrm{i} \right) ,
\end{equation}
\begin{equation} \label{eqn:implicitQinassumption}
Q_\mathrm{in} = k_\mathrm{in} \left( T_\mathrm{i} - T_\mathrm{n} \right) ,
\end{equation}
\begin{equation} \label{eqn:implicitQenassumption}
Q_\mathrm{en} = k_\mathrm{en} \left( T_\mathrm{e} - T_\mathrm{n} \right) ,
\end{equation}
where $k_\mathrm{ei}$, $k_\mathrm{in}$, and $k_\mathrm{en}$ are assumed to be constants. 
At the beginning of the energy exchange timestep, we calculate the energy exchange rates between the different components using the full sets of equations discussed in Section~\ref{sect:heatexchange} and use these values, combined with the component temperatures, to calculate $k_\mathrm{ei}$, $k_\mathrm{in}$, and $k_\mathrm{en}$.

The evolution equations for the energy densities are
\begin{equation} 
\frac{\partial e_\mathrm{n}}{\partial t} = - Q_\mathrm{in} - Q_\mathrm{en},
\end{equation}
\begin{equation}
\frac{\partial e_\mathrm{i}}{\partial t} = - Q_\mathrm{ei} + Q_\mathrm{in},
\end{equation}
\begin{equation}
\frac{\partial e_\mathrm{e}}{\partial t} = Q_\mathrm{ei} + Q_\mathrm{en}.
\end{equation}
The time-discreetization of these equations using the Crank-Nicolson method gives
\begin{equation}
e_\mathrm{n}^{n+1} = e_\mathrm{n}^{n} - \frac{1}{2} \Delta t \left( Q_\mathrm{in}^{n+1} + Q_\mathrm{in}^{n} \right) - \frac{1}{2} \Delta t \left( Q_\mathrm{en}^{n+1} + Q_\mathrm{en}^{n} \right),
\end{equation}
\begin{equation}
e_\mathrm{i}^{n+1} = e_\mathrm{i}^{n} - \frac{1}{2} \Delta t \left( Q_\mathrm{ei}^{n+1} + Q_\mathrm{ei}^{n} \right) + \frac{1}{2} \Delta t \left( Q_\mathrm{in}^{n+1} + Q_\mathrm{in}^{n} \right),
\end{equation}
\begin{equation}
e_\mathrm{e}^{n+1} = e_\mathrm{e}^{n} + \frac{1}{2} \Delta t \left( Q_\mathrm{ei}^{n+1} + Q_\mathrm{ei}^{n} \right) + \frac{1}{2} \Delta t \left( Q_\mathrm{en}^{n+1} + Q_\mathrm{en}^{n} \right).
\end{equation}
Since the evolution of the energy densities during the energy exchange timestep corresponds to the evolution of the temperatures, while the kinetic energy term remains constant, Eqn.~\ref{eqn:e_evolve_therm} can be assumed here.
Inserting Eqn.~\ref{eqn:e_evolve_therm} and Eqns.~\ref{eqn:implicitQeiassumption}--\ref{eqn:implicitQenassumption} into each of these equations gives
\begin{equation} \label{eqn:energyexchangelong_n}
\begin{aligned}
K_{\mathrm{en},n} & T_\mathrm{e}^{n+1} 
+ K_{\mathrm{in},n} T_\mathrm{i}^{n+1}
+ \left( 1 - K_{\mathrm{en},n} - K_{\mathrm{in},n} \right) T_\mathrm{n}^{n+1}
= \\
& - K_{\mathrm{en},n} T_\mathrm{e}^{n}
- K_{\mathrm{in},n} T_\mathrm{i}^{n}
+ \left( 1 + K_{\mathrm{en},n} + K_{\mathrm{in},n} \right) T_\mathrm{n}^{n} ,
\end{aligned}
\end{equation}
\begin{equation} \label{eqn:energyexchangelong_i}
\begin{aligned}
K_{\mathrm{ei},i} & T_\mathrm{e}^{n+1} 
+ K_{\mathrm{in},i} T_\mathrm{n}^{n+1}
+ \left( 1 - K_{\mathrm{ei},i} - K_{\mathrm{in},i} \right) T_\mathrm{i}^{n+1}
= \\
& - K_{\mathrm{ei},i} T_\mathrm{e}^{n}
- K_{\mathrm{in},i} T_\mathrm{n}^{n}
+ \left( 1 + K_{\mathrm{ei},i} + K_{\mathrm{in},i} \right) T_\mathrm{i}^{n} ,
\end{aligned}
\end{equation}
\begin{equation} \label{eqn:energyexchangelong_e}
\begin{aligned}
K_{\mathrm{ei},e} & T_\mathrm{i}^{n+1} 
+ K_{\mathrm{en},e} T_\mathrm{n}^{n+1}
+ \left( 1 - K_{\mathrm{ei},e} - K_{\mathrm{en},e} \right) T_\mathrm{e}^{n+1}
= \\
& - K_{\mathrm{ei},e} T_\mathrm{i}^{n}
- K_{\mathrm{en},e} T_\mathrm{n}^{n}
+ \left( 1 + K_{\mathrm{ei},e} + K_{\mathrm{en},e} \right) T_\mathrm{e}^{n} ,
\end{aligned}
\end{equation}
where
\begin{equation}
K_{xy,z} = \frac{\Delta t k_{xy} \left( \gamma_z - 1 \right) }{ 2 n_z k_\mathrm{B} } .
\end{equation}
These equations can be written in matrix form as
\begin{equation}
\mathbf{A} \mathbf{x} = \mathbf{B} ,
\end{equation}
where
\begin{equation} \label{eqn:}
\mathbf{A}
=
\begin{pmatrix}
	A_{11} &
	K_{\mathrm{ei},e} &
	K_{\mathrm{en},e} \\
	K_{\mathrm{ei},i} &
	A_{22} &
	K_{\mathrm{in},i} \\
	K_{\mathrm{en},n} &
	K_{\mathrm{in},n} &
	A_{33}
\end{pmatrix} ,
\end{equation}
\begin{equation} \label{eqn:}
\qquad A_{11} = 1 - K_{\mathrm{ei},e} - K_{\mathrm{en},e} ,
\end{equation}
\begin{equation} \label{eqn:}
\qquad A_{22} = 1 - K_{\mathrm{ei},i} - K_{\mathrm{in},i} ,
\end{equation}
\begin{equation} \label{eqn:}
\qquad A_{33} = 1 - K_{\mathrm{en},n} - K_{\mathrm{in},n} ,
\end{equation}
\begin{equation} \label{eqn:}
\mathbf{x}
=
\begin{pmatrix}
	T_\mathrm{e}^{n+1} \\
	T_\mathrm{i}^{n+1} \\
	T_\mathrm{n}^{n+1} 
\end{pmatrix} ,
\end{equation}
\begin{equation} \label{eqn:}
\mathbf{B}
=
\begin{pmatrix}
	- K_{\mathrm{ei},e} T_\mathrm{i}^{n} - K_{\mathrm{en},e} T_\mathrm{n}^{n} + \left( 1 + K_{\mathrm{ei},e} + K_{\mathrm{en},e} \right) T_\mathrm{e}^{n} \\
	
	- K_{\mathrm{ei},i} T_\mathrm{e}^{n} - K_{\mathrm{in},i} T_\mathrm{n}^{n} + \left( 1 + K_{\mathrm{ei},i} + K_{\mathrm{in},i} \right) T_\mathrm{i}^{n}	\\
	
	- K_{\mathrm{en},n} T_\mathrm{e}^{n} - K_{\mathrm{in},n} T_\mathrm{i}^{n} + \left( 1 + K_{\mathrm{en},n} + K_{\mathrm{in},n} \right) T_\mathrm{n}^{n}
	
\end{pmatrix} .
\end{equation}
At the beginning of the energy-exchange timestep, $\mathbf{A}$ and $\mathbf{B}$ are known and the aim is to calculate $\mathbf{x}$, which we do using Gaussian Elimination.
Once the updated temperatures are known, the updated energy densities are directly calculated.

We have tested our solver using a simpler implicit scheme based on the Backward Euler assumption and Newton iteration and find identical results.
The latter scheme has the advantage that it does not require the assumptions of Eqn.~\ref{eqn:implicitQeiassumption}-\ref{eqn:implicitQenassumption}, but it is much more computationally expensive, since the exchange rates need to be calculated many times per timestep.

\section{Chemical network and solver} \label{appendix:chemicalnetwork}

The chemical reactions in our network are listed in Table~\ref{table:chemicalnetwork}.
Each species varies due to chemical reactions by Eqn.~\ref{eqn:chemicalrate}.
These equations form a stiff system of ordinary differental equations (ODEs) which are impractical to solve using explicit integration methods.
This is mainly because the reaction rates become very rapid in high density gases, meaning explicit integration methods require restrictively small timesteps.
For example, the explicit 5$^\mathrm{th}$ order Runge-Kutte-Fehlberg method given by \citet{cash1990variable} would require timesteps of $\sim10^{-7}$~seconds near the base of our simulations. 
We solve the chemical equations using an implicit multi-step Rosenbrock method.
This class of methods was studied for applications to atmospheric chemistry by \citet{sandu1997benchmarking2} and \citet{sandu1997benchmarking1}, who found that they are generally more favourable than the other methods tested.
The two main advantages of this method are that it is able to take large timesteps even in regions where the reaction rates are high, and that it calculates the timestep length automatically.

Assume \mbox{$\mathbf{n}=\begin{bmatrix} n_1 & n_2 & \ldots & n_N \end{bmatrix}^{T}$} is the number densities of all species, where $N$ is the number of species, and $\mathbf{n}^n$ is this vector at time \mbox{$t^n$}.
The Rosenbrock method is given by
\begin{equation}
\mathbf{n}^{n+1} = \mathbf{n}^{n} + \sum\limits_{i=1}^s b_i \mathbf{k}_i,
\end{equation}
\begin{equation} \label{eqn:rosenbrock_k1}
\mathbf{k}_i = \Delta t \mathbf{f}\left( \mathbf{n}^{n} + \sum\limits_{j=1}^{i-1} \alpha_{ij} \mathbf{k}_j \right) + \Delta t \mathbf{J} \sum\limits_{j=1}^i \gamma_{ij} \mathbf{k}_j,
\end{equation}
where $s$ is the number of steps in the method, \mbox{$\mathbf{f}(\mathbf{n}) = d\mathbf{n}/dt$} (i.e. Eqn.~\ref{eqn:chemicalrate}) is the rate of change of $\mathbf{n}$, and \mbox{$\mathbf{J} = \partial \mathbf{f}/\partial \mathbf{n}$} is the Jacobian of $\mathbf{f}(\mathbf{n})$.
To be clear, when $i=1$, both sums in Eqn.~\ref{eqn:rosenbrock_k1} vanish.
The Jacobian is calculated analytically from Eqns.~\ref{eqn:reactionrate} and \ref{eqn:chemicalrate}.
Eqn.~\ref{eqn:rosenbrock_k1} can be rearranged to give 
\begin{equation} \label{eqn:rosenbrock_k}
\left( \mathbf{I} - \Delta t \gamma_{ii} \mathbf{J} \right) \mathbf{k}_i = \Delta t \mathbf{f}\left( \mathbf{n}^{n} + \sum\limits_{j=1}^{i-1} \alpha_{ij} \mathbf{k}_j \right) + \Delta t \mathbf{J} \sum\limits_{j=1}^{i-1} \gamma_{ij} \mathbf{k}_j,
\end{equation}
where $\mathbf{I}$ is the identity matrix. 
This is a system of linear equations of the form \mbox{$\mathbf{A} \mathbf{k}_i = \mathbf{B}$}, where $\mathbf{A}$ is an NxN matrix and $\mathbf{B}$ is an N element vector.
We solve this system of equations to derive $\mathbf{k}_i$ using Guassian Elimination.
To perform a timestep, we calculate the values of $\mathbf{k}_i$ sequentially and then use them to calculate $\mathbf{n}^{n+1}$.

To determine the appropriate timestep length, we first perform the update using an estimate of \mbox{$\Delta t$} and then estimate the difference between our $\mathbf{n}^{n+1}$ and the exact value.
Since the exact update is not known, we instead estimate this difference for the $i$th species as \mbox{$\mathrm{Est}_i = \tilde{n}_i^{n+1} - n_i^{n+1}$}, where $\tilde{\mathbf{n}}^{n+1}$ is a less accurate estimate for the update. 
If the method for calculating $\mathbf{n}^{n+1}$ has an order of consistency of $p$, then the method for calculating $\tilde{\mathbf{n}}^{n+1}$ should have an order of \mbox{$\tilde{p} = p - 1$}.
This is achieved by calculating $\tilde{\mathbf{n}}^{n+1}$ using Eqn.~\ref{eqn:rosenbrock_k1} with different values of the coefficients $b_i$, such that $\tilde{\mathbf{n}}^{n+1} = \mathbf{n}^{n} + \sum\limits_{i=1}^s \tilde{b}_i \mathbf{k}_i$.
We then estimate the error using
\begin{equation}
\mathrm{Err} = \sqrt{\frac{1}{N_\mathrm{s}} \sum\limits_{i=1}^N \left( \frac{\mathrm{Est}_i}{\mathrm{Tol}_i} \right)^2}
\end{equation}
where $N_\mathrm{s}$ is the number of species and $\mathrm{Tol}_i$ is the error tolerance for the $i$th species, which we assume is given by \mbox{$\mathrm{Tol}_i = \mathrm{aTol} + \mathrm{rTol}_i | n_i^{n+1} |$}.
As in \citet{Grassi14}, we assume \mbox{$\mathrm{aTol} = 10^{-20}$~cm$^{-3}$} and \mbox{$\mathrm{rTol} = 10^{-4}$}.
We then recalculate the desired timestep length using 
\begin{equation}
\Delta t_\mathrm{new} = 0.99 \Delta t \min( 10 , \max(0.1 , 0.9 \mathrm{Err}^{(-1/p)}) ).
\end{equation}
If \mbox{$\mathrm{Err} \ge 1$}, we consider that the timestep has failed and repeat it using \mbox{$\Delta t_\mathrm{new}$} as the estimate for the \mbox{$\Delta t$}; otherwise, we accept our original estimate of $\mathbf{n}^{n+1}$ and use \mbox{$\Delta t_\mathrm{new}$} as the estimate for \mbox{$\Delta t$} on the next timestep.
The extra factor of 0.99 is used to reduce the need to repeat timesteps.
Obviously it is often necessary to reduce \mbox{$\Delta t$} when it is larger than the time that the chemistry should be evolved, especially in the upper atmosphere where chemistry timesteps of several hundred seconds are possible.

The coefficients in the method are $b_i$, $\tilde{b}_i$, $\alpha_{ij}$, and $\gamma_{ij}$.
We use the coefficients derived by \citet{sandu1997benchmarking2} for their `RODAS3' method.
This is a 4-step method, meaning that \mbox{$s=4$}, and is third order, meaning that \mbox{$p=3$}. 
We have tested our implementation of this solver using KROME (\citealt{Grassi14}), which is a freely available package for solving chemical networks and is designed for application to atmospheric and astrophysical problems. 
In all tests, including full atmospheric simulations run using KROME, we find that the solvers gives almost identical results with similar computation times.

\onecolumn 
\begin{longtable}{p{.02\textwidth} p{.3\textwidth} p{.08\textwidth} p{.45\textwidth} p{.02\textwidth} }
\hline
No. & Reaction & Energy & Rate Coefficient & Ref. \\
\hline
\endhead
\hline
\endfoot
\hline
\caption{
Table listing the reactions in our chemical network.
The columns give the reaction number, the reaction, the energy released per reaction, the rate coefficient, and a reference for the reaction.
For photoreactions and non-thermal electron reactions, the reference is for the necessary cross-sections.
The references are:
1. \citet{FoxSung01};
2. \citet{Wakelam12};
3. \citet{Verronen02};
4. \citet{GarciaMunoz07};
5. \citet{Tian08a};
6. \citet{Fox15};
7. \citet{Yelle04};
8. \citet{RichardsVoglozin11};
9. \citet{Rimmer16};
10. \citet{HuebnerMukherjee15};
11. \citet{GreenBarth65};
12. \citet{jusick1967electron};
13. \citet{Jackman77}.
}
\endlastfoot
\\
\multicolumn{2}{l}{Neutral Chemistry:-}\\
1 & N + O$_2$ $\rightarrow$ NO + O & 1.40 eV & 1.5$\times10^{-14}$ $T_\mathrm{gas}$ exp(-3270.0/$T_\mathrm{gas}$) & 1 \\ 
2 & N + NO $\rightarrow$ N$_2$ + O & 2.68 eV & 4.0$\times10^{-11}$ ($T_\mathrm{gas}$/300.0)$^{-0.2}$ exp(-20.0/$T_\mathrm{gas}$) & 2 \\ 
3 & N + CO$_2$ $\rightarrow$ NO + CO & 1.06 eV & 1.7$\times10^{-16}$ & 1 \\ 
4 & N + NO$_2$ $\rightarrow$ N$_2$O + O & 1.81 eV & 3.0$\times10^{-12}$ & 3 \\ 
5 & N + H$_2$ $\rightarrow$ NH + H & -1.06 eV & 1.69$\times10^{-9}$ exp(-18095.0/$T_\mathrm{gas}$) & 4 \\ 
6 & NO + O + M  $\rightarrow$ NO$_2$ + M  & - & 9.0$\times10^{-32}$ (300.0/$T_\mathrm{gas}$)$^{1.5}$  & 3 \\
\end{longtable}
\twocolumn 

\bibliographystyle{aa}
\bibliography{mybib}

\end{document}